\documentclass[aps,physrev,twocolumn,groupedaddress]{revtex4-2}

\usepackage{amsmath}
\usepackage{amssymb}
\usepackage{hyperref}
\usepackage{xcolor}
\usepackage{graphicx}
\usepackage{bigints}
\usepackage{comment}
\usepackage[normalem]{ulem}

\begin{document}


\title{Gravitational light deflection and shadows in a warm plasma}

\author{Barbora Bezd\v{e}kov\'{a}}
\email{bbezdeko@campus.haifa.ac.il}
\affiliation{Department of Physics, Faculty of Natural Sciences, University of Haifa, Haifa 3498838, Israel}
\affiliation{Haifa Research Center for Theoretical Physics and Astrophysics, University of Haifa, Haifa 3498838, Israel}

\author{Volker Perlick}
\email{volker.perlick@uni-bremen.de}
\affiliation{Faculty 1, University of Bremen, 28359 Bremen, Germany}


\date{\today}

\begin{abstract}
Analytical studies of light bending in a dispersive medium near compact objects, e.g., black holes or neutron stars, are most challenged by a suitable definition of the medium. The most realistic model would be a hot magnetized plasma. In such a medium, however, an analytical description of light rays is very difficult. Therefore, usually an isotropic dispersive medium is assumed in analytical calculations. While it is possible to formulate equations for a general refractive index, which some studies do, most attention in the literature is given to the particular case of a cold, non-magnetized electron-ion plasma. Whereas this model covers many astrophysically relevant situations, there are indications that in some cases the plasma temperature is so high that the approximation of a cold plasma is no longer valid. For this reason, we consider in this paper a warm, non-magnetized electron-ion plasma, where the temperature is not set equal to zero but assumed to be small enough, such that relevant equations can be linearized with respect to it. After discussing the general equations for light rays in such a medium on a general-relativistic spacetime, we specify to the axially symmetric and stationary case which includes the spherically symmetric and static case. In particular, we calculate the influence of a warm plasma on the bending angle. In the spherically symmetric and static case, we also calculate the shadow in a warm plasma. We illustrate the general results with a static (respectively corotating) and an infalling warm plasma on Schwarzschild and Kerr spacetimes.
\end{abstract}

\maketitle

\section{Introduction}
Light rays propagating around compact gravitating objects, such as black holes or neutron stars, bend not only due to the gravitational fields of the objects, but possibly also as a result of the effect of a medium which surrounds them. The overall light deflection is thus a combination of effects caused by gravity, refraction and dispersion. In general, it is not an easy task to describe analytically all these effects in realistic astrophysical systems. The main reason for this is that the medium around compact objects is typically a hot magnetized plasma which brings many new effects into the propagation of light \citep[][]{Bittencourt2004}.

Within the geometrical optics approximation, a general analytical description of the ray propagation in an isotropic (i.e., non-magnetized) dispersive medium for relativistic spacetimes was developed by Synge \citep[][]{Synge-1960}. This formalism is based on the assumption that the light propagation is characterized by an index of refraction which depends on the spacetime point and on the frequency of light. Without specifying the index of refraction, Synge demonstrated that then the equations of motion of the light rays can be written in the Hamiltonian form. This formalism has been applied in numerous papers in particular for the case that the medium is a cold non-magnetized electron-ion plasma. In view of applications to astrophysics, light propagation in a cold plasma was first considered by Muhleman et al. \cite{Muhleman-1966,Muhleman-1970}, in combination with the weak-field approximation of gravity, for calculating the travel time and the bending angle of radio rays that pass through the Solar corona. Without the weak-field approximation, Perlick \cite{Perlick-2000} calculated the influence of a cold plasma on the bending angle of light rays in the Schwarzschild spacetime and in the equatorial plane of the Kerr metric. Several  years later, the subject was considerably pushed when Bisnovatyi-Kogan and Tsupko \cite{BK-Tsupko-2009,BK-Tsupko-2010,Tsupko-BK-2013}, Er and Mao \cite{Er-Mao-2014}, and Rogers \cite{Rogers2015} discussed the possibility of actually observing plasma effects on the propagation of light in astrophysics. The location of the photon sphere and the shadow in an unspecified spherically symmetric and static spacetime filled with a cold plasma was first determined by Perlick, Tsupko and Bisnovatyi-Kogan \cite{Perlick-Tsupko-BK-2015}. Whereas this original paper treated the Schwarzschild spacetime and the Ellis wormhole as particular examples, its general results have then be applied to numerous other metrics. Gravitational lensing in the Kerr spacetime in the presence of a cold plasma was discussed in detail by Perlick and Tsupko \cite{Perlick-Tsupko-2017,Perlick-Tsupko-2024}. 

The model of a (non-magnetized) cold electron-ion plasma assumes that the temperature of the electron fluid is negligible and is effectively set to zero. Although such restriction might seem too radical, the (magnetized) cold plasma approximation is efficiently used for the Solar corona, for the Solar wind, and for planetary magnetospheres \citep[e.g.,][]{freeman1985,bagenal1992,czechowski1995}. It is also widely assumed that it is valid in the environment of black holes and other ultracompact objects. However, we know that in galaxy clusters the temperature can be as high as $10 ^9$~K, such that the cold-plasma approximation would not be valid anymore. Moreover, there are some indications that also near supermassive black holes the temperature could come close to such values, see, e.g., \cite{genzel2010,LanzuisiEtAl2019,AkiyamaEtAl2021}.
This motivates us to study lensing in a plasma that cannot be considered as cold.

There are some recent papers that use the original Synge approach and go beyond the cold-plasma approximation. In particular, a formula for the deflection angle in a non-specified static refractive medium around an arbitrary spherically symmetric object was derived by Tsupko \cite{Tsupko-2021}. In a similar manner, Bezd\v{e}kov\'{a} and Bi\v{c}\'{a}k \cite{Bezdekova-2023} obtained a formula for the deflection angle within an axially symmetric stationary spacetime in a general refractive medium. These studies have provided a straightforward way of how to obtain the deflection angles for a broad range of spacetimes and media meeting the required assumptions.
More recently, effects on the deflection angles of a general \emph{moving} medium in spherically symmetric and axially symmetric spacetimes were discussed by Bezd\v{e}kov\'{a} et al. \cite{bezdekova2024} and by Pfeifer et al. \cite{pfeifer2025}, respectively. These studies go beyond the approximation of a cold plasma and discuss how the velocity terms (which are not present in a cold plasma) influence the light deflection. Especially in an axially symmetric spacetime, which was in detail analyzed in \cite{pfeifer2025}, it turns out that in the limit of both a slowly moving medium and a slowly rotating spacetime, effects of these two can completely compensate each other.

Supplementary to previous works, this paper expands the scope of application of the Synge formalism by applying it to the warm-plasma approximation, where the temperature of the electron fluid cannot be neglected, but is low enough that certain expressions can be linearized with respect to it. Besides the fact that in some cases the cold-plasma approximation might not be sufficient, considering a warm plasma allows one to better understand the validity of the cold-plasma approximation, which is its natural limit. We still restrict to the case of a non-magnetized plasma because otherwise the medium would not be isotropic, i.e., the Synge formalism would not be applicable.

The paper is organized as follows. In Section~\ref{wp_approx}, we introduce Synge's formalism and the refractive index in a warm plasma. Moreover, there are two forms of the warm-plasma refractive index (a fractional and an expanded one), which typically appear in the literature. These two are also compared in Section~\ref{wp_approx}. Dealing with a warm plasma in an axially symmetric stationary spacetime is in detail described in Section~\ref{light_ax}. In Section~\ref{def_angl_sh}, we apply the findings from Section~\ref{light_ax} to derive general expressions for deflection angles in axially symmetric stationary and spherically symmetric cases. Additionally, in the spherically symmetric case we also discuss the radius of a black hole shadow. In Section~\ref{examples}, we show how the deflection angles and shadow radii look for particular choices of spacetime and warm plasma radial velocity.
The conclusions of our work are summarized in Section  \ref{sec:conclusions}. The derivation of the refractive index in a warm plasma approximation from the two-fluid model in a kinetic theory is completely performed in Appendix~\ref{sec:warmderiv}. The energy-momentum tensor for such a fluid is rederived in Appendix~\ref{sec:energymom}. In Appendix~\ref{app_C}, we demonstrate how our results for the deflection angle are related to previous works where a different notation was used.

Throughout the paper the metric signature is given as $\{-,+,+,+\}$.
By indices in greek letters we mean $\alpha, \beta =  0,1,2,3$ or $(t,r,\vartheta, \varphi)$, while latin indices denote $i, k = 1,2,3$, resp. $(r,\vartheta, \varphi)$.

\section{The warm-plasma approximation}\label{wp_approx}

To study the ray propagation through an isotropic dispersive medium, the formalism used by Synge \cite{Synge-1960} can be applied. The description relies on the Hamiltonian in the form
\begin{equation}\label{hamiltonian_gen}
\mathcal{H}(x^{\alpha},p_{\alpha})=\frac{1}{2}\left[g^{\beta\delta}p_{\beta}p_{\delta}-(n^2-1)(p_{\gamma}V^{\gamma})^2\right],
\end{equation}
where $x^{\alpha},p_{\alpha}$ are the canonical variables and $n$, $V^\alpha$ characterize the medium by its refractive index and velocity, respectively. The light rays are then determined as the solutions to  Hamilton's equations 
\begin{equation}\label{eq:Ham}
\dot{x}{}^{\alpha} = \dfrac{\partial \mathcal{H}}{\partial p_{\alpha}} \, , \quad
\dot{p}{}_{\alpha} = -\dfrac{\partial \mathcal{H}}{\partial x^{\alpha}} 
\end{equation}
along with the dispersion relation
\begin{equation}\label{eq:disp}
\mathcal{H} = 0 \, .
\end{equation}
Here the overdot denotes differentiation with respect to a (non-affine) parameter which has no particular physical meaning.

The index of refraction must be given as a function of the spacetime point $x^{\alpha}$ and of the photon frequency 
\begin{equation}
\omega(x^{\alpha},p_\alpha)=-p_{\beta}V^{\beta} ( x^{\alpha}) .
\end{equation}
measured by an observer who is at rest in the medium at point $x^{\alpha}$.

The best known example of a dispersive isotropic medium is a \emph{cold plasma} where the index of refraction is given as
\begin{equation}\label{n_cold}
     n^2=1-\frac{\omega_p^2}{\omega^2} \, .
\end{equation}
Here $\omega_p$ denotes the plasma electron frequency which generally reads
\begin{equation}\label{def_omegap}
  \omega_p^2(x^\alpha)=\frac{q^2\, n_e(x^\alpha)}{\epsilon_0m},
\end{equation}
where the charge and the mass of the electron and the vacuum permittivity are denoted by $q$, $m$ and $\epsilon_0$, respectively. 

A cold plasma has the very special property that the velocity $V^{\alpha}$ of the medium drops out from the Hamiltonian which then reads
\begin{equation}\label{hamiltonian_cold}
\mathcal{H}(x^{\alpha},p_{\alpha})=\frac{1}{2}\left[g^{\beta\delta} (x ^{\alpha} )p_{\beta}p_{\delta}-\omega _p (x^{\alpha})^2\right] \, .
\end{equation}

In this paper we want to consider, instead of the often used cold-plasma model, the model of a so-called \emph{warm plasma}, where the influence of the plasma temperature on the propagation of light is taken into account. For a warm plasma, the index of refraction reads
\begin{equation}\label{n_warm}
     n^2=\left[1-\frac{\omega_p^2}{\omega^2}\left(1-\frac{5}{2}\chi\right)\right]\left(1+\frac{\omega_p^2}{\omega^2}\chi\right)^{-1}.
\end{equation}
Here, the dimensionless function $\chi$ is defined as
\begin{equation}\label{def_chi}
  \chi(x^\alpha)=\frac{k_B T(x^\alpha)}{m c^2},
\end{equation}
where $T$ is the temperature, $k_B$ is the Boltzmann constant and $c$ is the vacuum speed of light. It is worth mentioning that $k_B/(m c^2)=1.69\times10^{-10}$~K$^{-1}$ and to gain values of $\chi$ in order of unity, the plasma temperature should hence reach $T\sim10^{10}$~K. This is a very high temperature, but it is known that, e.g., in galaxy clusters the temperature can reach $10^9$~K, so that $\chi$ is not negligibly small in comparison to unity. Additionally, and more relevantly for our work, also near supermassive black holes the temperatures can reach similar values, see, e.g.,\cite{genzel2010,LanzuisiEtAl2019,AkiyamaEtAl2021}.

The index of refraction (\ref{n_warm}) has been well known since decades, see in particular the pioneering papers by Clemmow and Wilson \cite{clemmow56} and by Imre \cite{imre62}. However, it is not easy to find a complete derivation. Therefore, we provide such a derivation in Appendix~\ref{sec:warmderiv}. In this Appendix we consider a collisionless electron-ion plasma. Using kinetic theory and treating an electromagnetic wave as a linear perturbation of the electron fluid, where the unperturbed state is given by the J{\"u}ttner distribution, we derive a formula for the index of refraction for transverse waves, see Eq. (\ref{eq:extrans}). This formula is implicit, in the sense that it cannot be solved for the index of refraction, because the latter occurs both inside and outside of an integral. The step from this formula, which is exact as long as the assumption of a collisionless plasma is satisfied, to the warm-plasma approximation (\ref{n_warm}) is subtle. We show in Appendix~\ref{sec:warmderiv} that this approximation results from linearizing with respect to $\chi$ not the original Hamiltonian (\ref{hamiltonian_gen}), but rather another Hamiltonian that results from multiplying the original one with a nowhere vanishing function. Here one makes use of the fact that multiplication of the Hamiltonian with a nowhere vanishing function leaves the solutions to Hamilton's equations (\ref{eq:Ham}) with (\ref{eq:disp}) invariant up to parametrization.

It is important to realize that the warm-plasma approximation is \emph{not} a linearization of $n^2$ with respect to $\chi$. Clearly, the latter would require to expand the denominator in (\ref{n_warm}) resulting in
\begin{equation}\label{n_warm_exp}
     n^2=1-\frac{\omega_p^2}{\omega^2}\left(1-\chi\left(\frac{3}{2}+\frac{\omega_p^2}{\omega^2}\right)\right).
\end{equation}
This is indeed also a legitimate approximation which holds for sufficiently small $\chi$. However, as will be discussed in Appendix  
\ref{sec:warmderiv}, the warm-plasma approximation is still valid for some values of $\chi$ where (\ref{n_warm_exp}) already fails. 

The exact formula (\ref{eq:extrans}) implies that the index of refraction can only take values between 0 and 1. By contrast, the 
warm-plasma formula allows values $n > 1$. More specifically, we read from Eq.~(\ref{n_warm}) that for $\chi > 2/3$ this formula yields $n > 1$ for all $\omega$. This clearly demonstrates that the warm-plasma approximation is not valid for $\chi > 2/3$. Actually, as discussed in Appendix \ref{sec:warmderiv}, this approximation breaks down already at considerably lower values of $\chi$.

\begin{figure}[h!]
  \centering
  \includegraphics[width=0.45\textwidth]{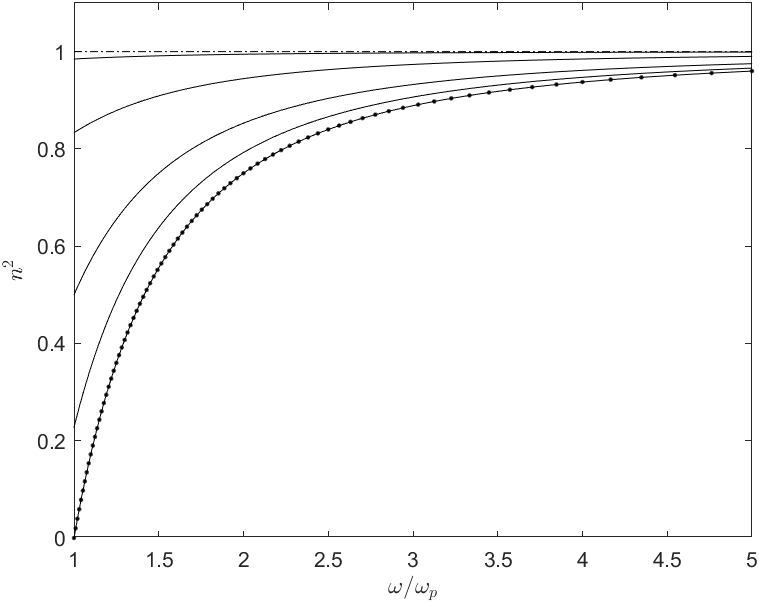}
  \caption{Refractive index squared of the warm plasma as a function of $\omega/\omega_p$ for different values of $\chi$. The cold plasma case ($\chi=0$) is shown by the dotted curve. The plain solid curves show situations when $\chi=$ 0.1, 0.25, 0.5, and 0.65 (from the cold plasma case bottom up).}\label{warm_cold}
\end{figure}

Profiles of refractive indices shown in Fig.~\ref{warm_cold} were drawn for $\chi=0$ (cold plasma case, dotted curve) and $\chi=$ 0.1, 0.25, 0.5, and 0.65 (solid curves). The limiting value $n^2=1$ is plotted by the dash-dotted line and it is seen that it is almost achieved when $\chi=0.65$. In general, it holds that for the same $\omega/\omega_p$ the refractive index in a warm plasma is larger than in a cold plasma.

In a cold plasma, according to Eq. (\ref{n_cold}), waves can propagate only with frequencies $\omega > \omega _p$ because $n^2$ cannot be negative. By contrast, in a warm plasma with index of refraction given by Eq. (\ref{n_warm}) waves can propagate if $\omega^2>\omega_p^2\left(1-\frac{5}{2}\chi\right)$, i.e., the cut-off frequency is lower than in a cold plasma with the same $\omega _p$. In Appendix \ref{sec:warmderiv} we compare this cut-off frequency according to the warm-plasma approximation  with the numerically determined cut-off frequency according to the exact formula (\ref{eq:extrans}) and also with the cut-off frequency according to the alternative approximation (\ref{n_warm_exp}). We will see that Eq. (\ref{n_warm}) gives, indeed, a better approximation of the exact value than Eq. (\ref{n_warm_exp}).

\begin{figure}[h!]
\centering
\includegraphics[width=0.45\textwidth]{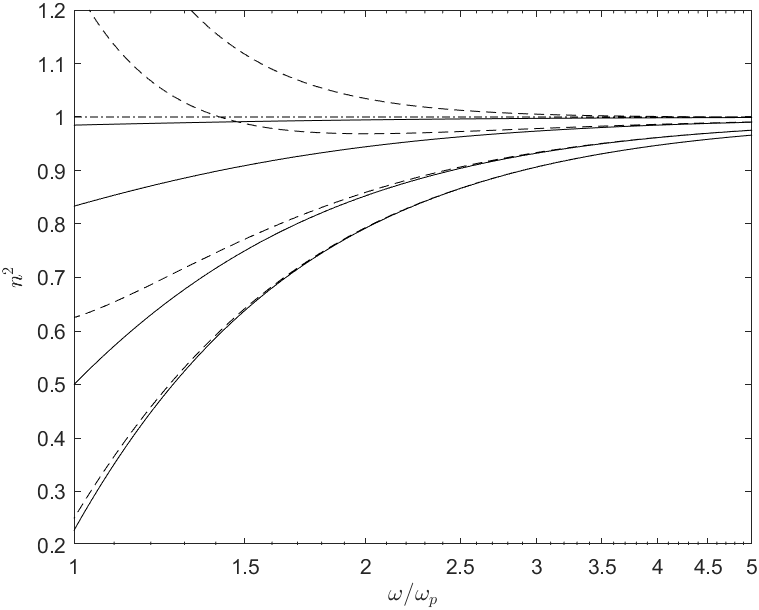}
\caption{Comparison of refractive indices for a warm plasma calculated when fractional and expanded forms were used. The solid curves correspond to formula (\ref{n_warm}), while dashed curves shown results when expression (\ref{n_warm_exp}) was applied. Situations when $\chi=$ 0.1, 0.25, 0.5, and 0.65 are shown (from bottom up). Notice a logarithmic scale of the $x$-axis.}\label{frac_exp}
\end{figure}

Fig.~\ref{frac_exp} compares the refractive indices for the same nonzero values of $\chi$ used in Fig.~\ref{warm_cold}, which were obtained when expressions (\ref{n_warm}) and (\ref{n_warm_exp}) were applied (solid and dashed curves, respectively). The plot demonstrates that for sufficiently small values of $\chi$ (up to $\sim0.1$)
the expanded formula can be applied with approximately the same precision as the fractional form.

\section{Light rays in a warm plasma on an axially symmetric 
and stationary spacetime}\label{light_ax}

We now consider a warm plasma on an axially symmetric and 
stationary spacetime
\[
g_{\mu \nu} dx^{\mu} dx^{\nu} =
-A \,  c^2 dt^2 +B \, dr^2 
\]
\begin{equation}   
+
D \, \mathrm{sin} ^2 \vartheta \,  \big( d \varphi- P \, c \, dt \big) ^2
+ F \, d \vartheta ^2 \, ,
\label{eq:axistat}
\end{equation}
where the metric coefficients $A$,$B$, $D$, $F$, and $P$ are functions
of $r$ and $\vartheta$.
The energy-momentum tensor of the electron fluid reads, as derived in 
Appendix~\ref{sec:energymom}, 
\begin{equation}
T_{\mu \nu} = \dfrac{m^2 \omega _p^2}{\mu _0 q^2}
\Bigg( \Big( 1 + \dfrac{5}{2} \chi \Big) V_{\mu} V_{\nu} 
+ \chi \, g_{\mu \nu} \Bigg)
\, ,
\end{equation}
where $\omega _p$ and $\chi$ are functions of $r$ and $\vartheta$; $\mu _0$ is the permeability of vacuum. We assume 
that the four-velocity of the fluid is of the form
\begin{equation}
V^{\mu} \partial _{\mu} = 
\sqrt{\dfrac{1+u^2 B}{A}} 
\big( \partial _t + P \, c \, \partial _{\varphi} \big)
- u \,  c \, \partial _r\,,  
\label{eq:V}
\end{equation}
where $u$ is a function of $r$ and $\vartheta$ that is arbitrary
except for the condition
\begin{equation}
    \dfrac{1+u^2 B}{A} \ge 0 \, .
\label{eq:norm}
\end{equation}
On the domain where this condition holds, $V^{\mu} \partial _{\mu}$ is timelike 
and normalized according to $g_{\mu \nu} V^{\mu} V^{\nu} = - c^2$.  Note that 
(\ref{eq:norm}) is certainly true on the domain where $A$ and $B$ are positive. The 
three functions $\omega _p$, $\chi$, and $u$ characterize the plasma.

For a stationary situation it is natural to assume that
the conservation laws of charge and energy hold for the electron
fluid and for the ion fluid separately. If the metric coefficients 
are given, this gives us two equations for
the three functions $\omega _p$, $\chi$, and $u$ have to satisfy, namely
\begin{equation}
0 =\partial _{\mu} \big( \sqrt{|g|} \, j^{\mu} \big) 
= \partial _r \big( \sqrt{|g|} \, j^r \big)
\, ,
\label{eq:Con1}
\end{equation} 
\begin{equation}
0 =\partial _{\mu} \big( \sqrt{|g|} \, T^{\mu}{}_t \big) 
= \partial _r \big( \sqrt{|g|} \, V^r V^t g_{tt} \big)\,,
\label{eq:Con2}
\end{equation}
where 
\begin{equation}
g = -A\, B\, D \, F \, c^2 \, \mathrm{sin} ^2 \vartheta
\end{equation}
is the determinant of the metric and 
$j^{\mu} \sim \omega _p^2 V^{\mu}$ is the electric current
density of the electron fluid. After pulling the factor 
$c^2 \mathrm{sin} ^2 \vartheta$ out of the $\partial _r$
derivative and dividing by this quantity, (\ref{eq:Con1})
implies that
\begin{equation}
C_u  = \sqrt{A \, B  \, D \, F} \, \omega _p^2 \, u
\label{eq:C1}
\end{equation}
is independent of $r$. Using this result, Eq. (\ref{eq:Con2}) requires that
\begin{equation}
C_{\chi} = \Big( 1 + \dfrac{5}{2} \, \chi \Big) 
\, \sqrt{\big( 1 + u^2 B \big) A}
\label{eq:C2}
\end{equation}
is also independent of $r$. Note that $C_u$ and $C_{\chi}$ may be functions
of $\vartheta$. 

If the spacetime is asymptotically flat, i.e.,
\begin{equation}
A \to 1 \, , \: \;
B \to 1 \, , \: \;
\dfrac{D}{r^2} \to 1 \, , \: \;
\dfrac{F}{r^2} \to 1 
\; \: \mathrm{if} \: r \to \infty
\, ,
\label{eq:as}
\end{equation}
Eqs. (\ref{eq:C1}) and (\ref{eq:C2}) require that
\begin{equation}
\dfrac{C_u}{r^2 \omega _p^2\, u} \to 1 
\, , \quad
\dfrac{C_{\chi}}{\sqrt{1+u^2}} \to 1+ \dfrac{5}{2} \chi _{\infty}
\, \quad \,
\mathrm{if} \: r \to \infty\,,
\label{eq:C1as}
\end{equation}
where $\chi _{\infty}$ is the temperature at infinity which may depend on 
$\vartheta$. Note that (\ref{eq:C1as}) implies that $\omega _p^2\, u$ falls 
off like $r^{-2}$ and that $C_{\chi} \ge 1$.

With $u$ chosen appropriately, the plasma model suggested here may very well
be valid inside an ergoregion and also on both sides of a horizon. It can,
however, not be valid near a curvature singularity where $A \to - \infty$ and
$AB$ stays positive, such as, e.g., in the Schwarzschild spacetime at $r \to 0$. 
Then, Eq. (\ref{eq:C2}) necessarily fails to hold on a certain
neighborhood of the curvature singularity. 

With any choice of $C_u$ and $C_{\chi}$, Eqs. (\ref{eq:C1}) and (\ref{eq:C2}) 
give us two equations for the three functions $u$, $\omega _p$, and $\chi$. As long as we have no third equation, one of the three functions,
e.g. $\omega _p$, can be freely chosen.

In order to get a third equation, one might think of requiring the validity of an equation of state.
This is what Michel \cite{Michel1972} did who 
was the first to consider accretion of a perfect fluid in the full formalism 
of general relativity. (Actually, Michel restricted to spherically symmetric and
static spacetimes which is more special than the situation considered here.)
Michel had two conservation laws, equivalent to our equations (\ref{eq:C1}) and (\ref{eq:C2}), and required in addition a polytropic equation of state,
\begin{equation}
p = K \, \varepsilon ^{\gamma},
\label{eq:pol_eos}
\end{equation}
where $K$ and in principle also $\gamma$ may depend on the temperature. Here, however, we are in a different situation. In contrast to Michel we have a specific plasma model, based in particular on the assumptions that the electron fluid is collisionless and in the unperturbed state given by the Juettner distribution. As shown in Appendix~\ref{sec:energymom}, in this case the equation of state (\ref{eq:pol_eos}) always holds with $\gamma =1$ and
\begin{equation}
K = \chi / \big(1+3 \chi /2 + O(\chi ^2) \big) \, ,
\label{eq:K}
\end{equation}
i.e., the equation of state places no additional conditions on $\omega _p$, $\chi$ or $u$. Following this line of thought, we may choose $\omega _p$ freely as a function of the coordinates $r$ and $\vartheta$. Then the conservation laws (\ref{eq:C1}) and (\ref{eq:C2}) determine $u$ and $\chi$ as functions of $r$ and $\vartheta$, where we are still free to choose $C_u$ and $C_{\chi}$, and the equation of state (\ref{eq:pol_eos}) holds with $\gamma = 1$ and $K$ given by (\ref{eq:K}). This is the procedure we will use in all our examples below. Note that this line of argument also applies to the cold-plasma case, where (\ref{eq:pol_eos}) holds with $K= 0$. Also in this case we are free to choose $\omega _p$ at will. This is what is done in virtually all papers on lensing in a cold plasma, where in the spherically symmetric and static case usually a power law is chosen for $\omega _p^2$.

As a mathematical curiosity, we mention that in some cases the equation of state (\ref{eq:pol_eos}) may also hold with $\gamma \neq 1$. If we insert the expressions for energy density and pressure from Appendix~\ref{sec:energymom}, Eqs. (\ref{eq:density}) and (\ref{eq:pressure}), this equation of state may be rewritten as
\begin{equation}
\chi \, \Big(\dfrac{m^2}{\mu_0 q^2} \Big)^{1-\gamma }
\,\Big( 1+ \dfrac{3}{2} \chi + O(\chi ^2 )\Big) ^{- \gamma}
= K \, \big( \omega _p^2 \big) ^{\gamma -1}
\, .
\label{eq:poly}
\end{equation}
As we observed already, this equation is always satisfied with $\gamma = 1$ and $K$ given by (\ref{eq:K}). Let us investigate if it can also hold with some given $\gamma \neq 1$ and $K \neq 0$, where the latter is prescribed as a function of $\chi$ which has to go to 0 for $\chi \to 0$. Then (\ref{eq:poly}) can be solved for $\omega _p^2$, i.e., it allows us to express $\omega _p^2$ in terms of $\chi$. Thereupon the two conservation laws (\ref{eq:C1}) and (\ref{eq:C2}) give us two equations for $\chi$ and $u$. If these equations admit a solution with $\chi (r, \vartheta ) \ge 0$ and $u(r , \vartheta ) \ge 0$, this gives us an infalling or corotating plasma where the electron fluid satisfies the equation of state (\ref{eq:pol_eos}) with the prescribed $\gamma \neq 1$. The corresponding plasma density $\omega _p (r , \vartheta )^2$ is then given by (\ref{eq:poly}). One has to check further whether the warm-plasma approximation is valid.

We will not follow this rather contrived procedure in the following. The important observation is that we can choose $\omega _p ^2$ freely, and that the equation of state always holds with $\gamma = 1$ and $K$ given by (\ref{eq:K}), disregarding the fact that in some particular cases it might hold simultaneously with some other value of $\gamma$.

\section{Deflection angle and shadow}\label{def_angl_sh}
We will now demonstrate how one can calculate the deflection (bending) angle of light in the situation described in the previous section. We will perform these calculations first for the general axially symmetric and stationary situation and then specify to the spherically symmetric and static situation. In the latter case we will also calculate the location of the photon sphere and the angular radius of the shadow. In the axially symmetric and stationary case, analytical calculations of the shadow cannot be presented here because they are possible only in the presence of a Carter constant. For light rays in a warm plasma, the conditions for the existence of a Carter constant have not yet been worked out.

\subsection{The axially symmetric and stationary case}
To write down the Hamiltonian, we have to determine the inverse metric which can be read directly from Eq. (\ref{eq:axistat}). We find that, 
for a general dispersive and isotropic medium, the Hamiltonian (\ref{hamiltonian_gen}) takes the form
\begin{gather}
\mathcal{H}(x^{\alpha},p_{\alpha})=
\frac{1}{2}
\left[
-\dfrac{\big( p_t+ P \, c \, p_{\varphi} \big) ^2}{c^2 A}  + \dfrac{p_r^2}{B} 
+ \dfrac{p_{\varphi}^2}{D \, \mathrm{sin} ^2 \vartheta}
\right.
\nonumber
\\[0.2cm]
\left.
 + \dfrac{p_{\vartheta}^2}{F}  
-(n^2-1)\omega^2\right].
\label{hamiltonian_axistat}
\end{gather}
We remind the reader that the metric coefficients $A$, $B$, $D$, $F$, and $P$ are functions of $r$ and $\vartheta$.

Let us now consider a spacetime and a plasma such that a light ray that starts tangentially to the equatorial plane, $\vartheta = \pi /2$, remains in this plane. This is certainly the case if the spacetime and the plasma are symmetric with respect to transformations $\vartheta \mapsto \pi = \vartheta$. Then, we can restrict our consideration to the equatorial plane and hence, with the choice $\vartheta = \pi /2$ and $d \vartheta = 0$, the Hamiltonian~(\ref{hamiltonian_axistat}) reduces to
\begin{equation}
\mathcal{H}(x^{\alpha},p_{\alpha})=
\frac{1}{2}\left[\frac{p_r^2}{B}+ \frac{p_\varphi^2}{D}
-\frac{\left(p_t+P\,c \, p_\varphi\right)^2}{c^2A}-(n^2-1)\omega^2\right].
\end{equation}
From now on the metric coefficients $A$, $B$, $D$, and $P$ are to be taken at $(r, \vartheta = \pi/ 2 )$.

In a corotating medium, i.e., when $u=0$, in the equatorial plane the index of refraction 
$n$ is a function of $r$, $p_t$, and $p_{\varphi}$, but independent of $p_r$, and
\begin{equation}
\omega=-\frac{1}{c}p_\mu V^{\mu}=-\sqrt{\dfrac{1}{c^2A}}\big(p_t + P \, c \, p_{\varphi} \big).
\end{equation}
This allows us to write the Hamiltonian as
\begin{equation}
\mathcal{H}(x^{\alpha},p_{\alpha})=\frac{1}{2}\left[\frac{p_r^2}{B}+ \frac{p_\varphi^2}{D}-n^2\dfrac{\big(p_t + P \, c \, p_{\varphi} \big)^2}{c^2A}\right].
\end{equation}
Then the dispersion relation $\mathcal{H}=0$ gives us a quadratic equation for $p_r$ which can be easily solved and returns
\begin{equation}
p_r=\pm\sqrt{B}\sqrt{n^2\dfrac{\big(p_t + P \, c \, p_{\varphi} \big)^2}{c^2A}-\frac{p_\varphi^2}{D}}.
\end{equation}
The corresponding equations of motion read
\begin{align}
    \dot{r}=
\frac{\partial \mathcal{H}}{\partial p_r}&=\frac{p_{r}}{B},
\\
\dot{\varphi}=\frac{\partial \mathcal{H}}{\partial p_\varphi}&= \frac{p_{\varphi}}{D}-n^2\dfrac{P\,\big(p_t + P \, c \, p_{\varphi} \big)}{cA}.
\end{align}
The orbit equation for light rays in the equatorial plane can now be derived by forming the quotient of these two equations which returns
\begin{widetext}
\begin{equation}
\label{def_angl_axis}
  \frac{d\varphi}{dr}
  = \frac{B}{p_{r}}\left(\frac{p_{\varphi}}{D}-n^2\dfrac{P\,\big(p_t + P \, c p_{\varphi} \big)}{cA}\right)\\
  =\pm\sqrt{\frac{B}{D}}\left[
  \frac{
  n^2D\left(1-P^2 \, c^2 \frac{p_{\varphi}^2}{p_t^2}+n^2\dfrac{D\,P^2 }{A}\big(1 + P \, c \frac{p_{\varphi}}{p_t} \big)^2\right)
  }{
  c^2A\left(\frac{p_{\varphi}}{p_t} -n^2\dfrac{D\,P}{c\,A}\big(1 + P \, c \frac{p_{\varphi}}{p_t} \big)\right)^2
  }
  -1\right]^{-1/2} \, . 
\end{equation}
\end{widetext}

This equation gives us the deflection angle: We just have to solve for $d \varphi$ and to integrate over a light ray that comes in from infinity, passes through a minimum radius $R$ and then goes out to infinity again.

In the case that $u\ne0$, the frequency depends on $p_r$, i.e.,
\begin{equation}
  \omega=-\frac{1}{c}p_\mu V^{\mu}=-\sqrt{\dfrac{1+u^2B}{c^2A}}\big(p_t + P \, c \, p_{\varphi} \big)+u\,p_r.
\end{equation}
Therefore, one has to specify how the index of refraction $n$ depends on $\omega$ before one can solve the dispersion relation $\mathcal{H}=0$ for $p_r$. With the choice of the refractive index (\ref{n_warm}) for a warm plasma, the Hamiltonian reads
\begin{gather}
\mathcal{H}(x^{\alpha},p_{\alpha})=\frac{1}{2}
\left[\frac{p_r^2}{B}+ \frac{p_\varphi^2}{D}\right.\nonumber\\
\left.-\frac{\left(p_t+P\,c \, p_\varphi\right)^2}{c^2A}+\frac{\omega_p^2\omega^2\left(1-\frac{3}{2}\chi\right)}{\omega^2+\omega_p^2\chi}\right].
\end{gather}
In this case $\mathcal{H} =0$ gives us a quartic equation for $p_r$. There are four in general complex solutions from which we have to select the two relevant real ones, one for ingoing and one for outgoing light rays. They can be written explicitly with one of the well-known solution formulas for quartic equations, but the resulting expressions are rather awkward and will not be given here. Note that the branches for ingoing and outgoing rays are not in general symmetric with respect to the point of the closest approach. With the two relevant expressions for $p_r$ found, the deflection angle can then be calculated in analogy to the case where $u = 0$. 

We will illustrate the procedure in the following section with the Kerr spacetime.

\subsection{The spherically symmetric and static case}\label{subsec:sphstat}
We now specialize to the spherically symmetric case, i.e., we assume that $P = 0$ and that $A$, $B$, $D$, $F$, $\omega _p$, $\chi$, and $u$ are functions of $r$ only. We first consider the case of a static plasma, $u = 0$. Then 
the orbit equation (\ref{def_angl_axis}) simplifies to
\begin{gather}
  \frac{d\varphi}{dr}=
  \pm\sqrt{\frac{B(r)}{D(r)}}\left[\frac{p_t^2}{c^2 p_{\varphi}^2}h^2(r)-1\right]^{-1/2},
\end{gather}
where
\begin{equation}\label{h_Tsupko}
 h^2(r)=\frac{D(r)}{A(r)}n^2(r, \omega(r)).
\end{equation}
Here $n$ has to be inserted as a function of $r$ and $\omega (r)= -p_t / (c \sqrt{A(r)})$. This corresponds to the orbit equation derived for this situation by Tsupko~\cite{Tsupko-2021}, see Eq.~(21) there. The deflection angle is then given as
\begin{equation}\label{def_angle_Tsupko}
\alpha= 2\int_{R}^{\infty}\sqrt{\frac{B(r)}{D(r)}}\left(\frac{h^2(r)}{h^2(R)}-1\right)^{-1/2}dr-\pi,
\end{equation}
cf. Eq. (26) in Ref.~\cite{Tsupko-2021}. The radius coordinate $R$ of the point of the closest approach is determined by the equation $dr/d\varphi=0$.

Plugging the refractive index (\ref{n_warm}) into (\ref{h_Tsupko}) leads to
\begin{equation}\label{h_warm}
 h^2(r)=\frac{D(r)}{A(r)}\left[1-\frac{\omega_p^2(r)A(r) }{\omega _0^2+\omega_p^2(r)\chi(r) A(r)} \left(1-\frac{3}{2}\chi(r)\right)\right],
\end{equation}
where 
\begin{equation}
\omega _0 = - p_t /c \, .
\end{equation}
Note that choosing $\chi=0$ gives Eq.~(17) of \cite{Perlick-Tsupko-BK-2015}.

In the case of a spherically symmetric metric ($P=0$) and a static plasma ($u=0$) it is straight-forward to calculate the angular radius of the shadow in a warm plasma, following the approach of Tsupko \cite{Tsupko-2021}. This requires to calculate
the angle $\alpha_R$ between the direction of a light ray and the radial direction at the position of the observer
which is given as
\begin{equation}
  \sin^2\alpha_R=\frac{h^2(R)}{h^2(r_O)},
\end{equation}
where $r_O$ denotes the radial coordinate of the observer. Since the shadow is effectively bounded by past-oriented rays that spiral towards an unstable photon sphere, the angular radius $\alpha_{\mathrm{sh}}$ of the shadow can be derived from
\begin{equation}
  \sin^2\alpha_{\mathrm{sh}}=\frac{h^2(r_{ph})}{h^2(r_O)}.
\end{equation}
The radius $r_{ph}$ of the circular light orbits has to be found from the relation
\begin{equation}
  \left.\frac{d}{dr}h^2(r)\right|_{r=r_{ph}}=0.
\end{equation}
What is new here in comparison to the cold plasma case which was considered in \cite{Perlick-Tsupko-BK-2015} is an additional term proportional to the temperature gradient. Indeed, using the definition (\ref{h_Tsupko}) with the index of refraction from (\ref{n_warm}) returns
\begin{equation}\label{r_ph_cond}
\dfrac{D'(r)}{D(r)}-\dfrac{A'(r)}{A(r)}  + \dfrac{(n^2)'(r)}{n^2} = 0,
\end{equation}
where
\begin{widetext}
\begin{equation}
 (n^2)'(r)=\frac{-\omega_0^2\left[(\omega_p^2)' (r) A(r)+\omega_p^2 (r) A'(r)\right] \left(1-\frac{3}{2}\chi (r) \right)+\omega_p^2 (r) A(r)\chi' (r)\left(\frac{3}{2}\omega_0^2+\omega_p^2 (r) A(r)\right)}{\left(\omega_0^2+\omega_p^2 (r) A(r)\chi (r)\right)^2}.
\end{equation}
\end{widetext}
Here both the plasma frequency and the temperature are assumed to be prescribed functions of $r$, i.e., $\omega_p^2(r)$, $\chi(r)$.

The bending angle and the shadow can also be calculated if the plasma is in radial motion, i.e., when $ u \neq 0$. However, then the frequency depends on $p_r$,
\begin{equation}
\omega=-\frac{1}{c}p_\mu V^{\mu}=-\dfrac{p_t}{c \, \sqrt{A}} + u \, p_r ,
\end{equation}
and with 
\begin{equation} 
\mathcal{H}(x^{\alpha},p_{\alpha})
=\frac{1}{2}\left[\frac{p_r^2}{B}
+ \frac{p_\varphi^2}{D}- \dfrac{p_t^2}{c^2A}
+\frac{\omega_p^2\omega^2\left(1-\frac{3}{2}\chi\right)}{\omega^2+\omega_p^2\chi}\right]
\end{equation}
the dispersion relation $\mathcal{H}=0$ still gives us a quartic equation for $p_r$, so the situation is not much simpler than in the axially symmetric stationary case. Again, we have to select the two relevant solutions of the quartic equation, one for ingoing and one for outgoing rays. These solutions have to be inserted into the orbit equation 
\begin{equation}
  \frac{d\varphi}{dr}
  = \frac{B}{p_{r}}\left(\frac{p_{\varphi}}{D}-n^2\dfrac{p_t\,P}{c\,A}\right).
\end{equation}
Integration over the ingoing and the outgoing branch then gives us the bending angle.

The location of the photon sphere and the angular radius of the shadow can be determined in the following way. With the parameters $C_u$, $\omega _c$, $C_{\chi}$, $k$, and $\omega _0 = - p_t /c$ fixed we solve the cubic equation $\partial \mathcal{H}/\partial p_r=0$ for $p_r$. Again, we do not write down the analytic solution here because it is rather awkward. This gives us the minimum radius $r=R$ of light rays with constants of motion $p_{\varphi}$ and $p_t$. Reinserting this expression into the dispersion relation $\mathcal{H} = 0$ then gives us the constant of motion $p_{\varphi}$ of a light ray with constant of motion $p_t$ and minimum radius $R$. Setting the derivative of this $p _{\varphi}$ with respect to $R$ equals to zero determines the critical value of $p_{\varphi}$ and the radius coordinate $r_{\mathrm{ph}}$ of the photon sphere. 

For calculating the angular radius $\alpha _{\mathrm{sh}}$ of the shadow we have to recall that the boundary of the shadow corresponds to light rays that spiral towards the photon sphere, i.e., to light rays whose constant of motion $p_{\varphi}$ is equal to its critical value which was just calculated by solving $dp_\varphi/dR = 0$. From elementary geometry we find that for a static observer at $r = r_O$ it holds that
\[
    \mathrm{tan} \, \alpha _{\mathrm{sh}} 
= 
    \sqrt{\dfrac{D(r)}{B(r)}} \dfrac{\dot{\varphi}}{\dot{r}} \Big| _{r = r_O} 
=    
    \sqrt{\dfrac{D(r)}{B(r)}} 
    \dfrac{\partial \mathcal{H} / \partial p_{\varphi}}{\partial \mathcal{H}/\partial p_r} \Big| _{r = r_O}
\]
\begin{equation}
=
    \dfrac{1}{\sqrt{D(r) \, B(r)}} \, \dfrac{p_{\varphi}}{\partial \mathcal{H} / \partial p_r} \Big| _{r = r_O};
\end{equation}
see Fig.\,7 in Perlick and Tsupko \cite{Perlick-Tsupko-2022}. Here we have to insert into the expression for $\partial \mathcal{H}/\partial p_r$ first the solution to the third-order equation for $p_r$ and then into the entire expression the critical value of $p _{\varphi}$. In this way we get an analytical expression for the angular radius of the shadow for light rays for an observer at $r_O$, in dependence of the parameters $C_u$, $\omega _c$, $C_{\chi}$, $k$, and $\omega _0$.   

We will illustrate the procedure in the second example below, see Subsection~\ref{infall_Schw}.      

\section{Examples}\label{examples}
\subsection{Example 1: Plasma at rest on Schwarzschild spacetime}

We consider the Schwarzschild spacetime with mass parameter $M$, 
which is of the form of Eq. (\ref{eq:axistat}) with 
\begin{equation}
A = B^{-1} = 1 - \dfrac{2M}{r} \, , \quad
D = F = r^2 \, , \quad P = 0\, .
\label{eq:Schw}
\end{equation}
Then the conservation laws (\ref{eq:C1}) and (\ref{eq:C2}) require
\begin{equation}
    C_u = r^2 \omega _p^2 u
\label{eq:CuS}
\end{equation}
and
\begin{equation}
    C_{\chi} = \Big( 1 + \dfrac{5}{2} \, \chi \Big) \sqrt{ 1 - \dfrac{2M}{r} + \dfrac{C_u^2}{r^4 \omega _p^4}} \, . 
\label{eq:CchiS}
\end{equation}
We see that the case of a cold plasma, $\chi = 0$, is included with
$C_{\chi} = 1$ and 
\begin{equation}
\omega _p ^2 = \dfrac{C_u}{\sqrt{2M} \, r^{3/2}}
\, , \quad
u = \sqrt{\dfrac{2M}{r}}
\, .
\end{equation}
We recognize that this is a well-known result if we remember that $\chi = 0$ 
implies that the electron fluid is a dust:
It was shown already by Michel \cite{Michel1972}
that the density of a dust infalling from rest onto a Schwarzschild black hole is 
proportional to $r^{-3/2}$. This follows from Eqs. (8), (9) and (10) in Michel's
paper if specified for a dust, $P = 0$ and $\epsilon = 0$, and assuming infall from 
rest, $u\to 0$ for $r \to \infty$.

Here we are of course interested in the case that $\chi \neq 0$. We follow the line
of thought outlined at the end of Sec. \ref{light_ax}, i.e., we prescribe the 
plasma density and then get the other relevant quantities from the conservation
laws (\ref{eq:C1}) and (\ref{eq:C2}). Recall that then the equation of state 
(\ref{eq:pol_eos}) always holds with $\gamma = 1$ and $K$ given by (\ref{eq:K}).
We will assume a power-law for the plasma density,
\begin{equation}
\omega _p (r) ^2 = \omega _c ^2 \Big( \dfrac{M}{r} \Big) ^k \, , 
\label{eq:omegapk}
\end{equation}
where $\omega _c$ is a constant with the dimension of 
a frequency and $k$ is a dimensionless constant.
Then the conservation laws (\ref{eq:CuS}) and (\ref{eq:CchiS})
determine $u$ and $\chi$, 
\begin{equation}
    u = \dfrac{C_u r^{k-2}}{\omega _c^2 M^k}
    \, , 
\label{eq:uS}
\end{equation}
\begin{equation}
    \chi = \dfrac{2}{5} \Bigg( 
    C_{\chi}  \Big( 1 - \dfrac{2M}{r} + \dfrac{C_u^2 r^{2k-4}}{\omega _c^4 M^{2k}}
    \Big)^{-1/2} - 1  
    \Bigg) \, ,
\label{eq:chiS}
\end{equation}
where $C_u$ and $C_{\chi}$ can be chosen at will. In principle, they could depend
on $\vartheta$, but we will restrict to the case that they are constant. 

Note that the conservation laws (\ref{eq:CuS}) and (\ref{eq:CchiS}) can hold for 
a cold plasma, $\chi =0$, only if $k = 3/2$. Therefore, a power-law (\ref{eq:omegapk}) 
with $k \neq 3/2$ cannot be strictly true for a cold plasma unless one gives up 
the conservation law of energy or of charge for the electron fluid (or both). To be 
sure, such a power-law with $k \neq 3/2$ has been assumed in several papers for a cold
plasma, see e.g. \cite{BK-Tsupko-2009,Rogers2015,Perlick-Tsupko-BK-2015}. We do not 
say that there is anything wrong with this. We just want to point out 
that such a power-law ansatz with $k \neq 3/2$ is mathematically inconsistent with the
conservation laws if $\chi =0$. Our equations for a warm plasma allow us to calculate
the temperature in this situation. If it comes out so low that its influence on the 
trajectories of light rays can be neglected, the use of the power-law ansatz for a 
cold plasma is justified as a valid approximation. Of course, the power-law ansatz is
always compatible with assuming $\chi = 0$ if one gives up one of the conservation
laws, i.e., if one allows for external forces or for recombination or ionization 
processes. In this paper, however, we always assume that the conservation
laws are valid.

While in the next section we will consider the case of an infalling plasma, $C_u > 0$, in this section we 
restrict to the case of a plasma at rest, $C_u = 0$. Then (\ref{eq:uS}) requires $u =0$, i.e., the four-velocity
of the plasma is  
\begin{equation}
V^{\mu} \partial _{\mu} = \Big( 1 - \dfrac{2M}{r} \Big) ^{-1/2} \partial _t \, .
\end{equation}
This is, of course, possible only in the domain $2M< r < \infty$ to which we restrict in the following.
Then, (\ref{eq:chiS}) requires 
\begin{equation}
\chi =  \dfrac{2}{5} \Bigg( 
C_{\chi} \, \left( 1- \dfrac{2M}{r}\right)^{-1/2} 
- 1 \Bigg) \, .
\label{chi_C2}
\end{equation}

Here $C_{\chi}$ is a constant that can be chosen arbitrarily, but we have to 
require $C_{\chi} \ge 1$ if we want (\ref{chi_C2}) to hold for arbitrarily
large $r$ because $\chi$ cannot be negative. But then, with $C_{\chi}$ restricted 
in this way, we observe that $\chi$ must be strictly bigger than zero
at any finite value of $r$, i.e., that this model does not allow for a 
cold-plasma limit. 

Note that (\ref{chi_C2}) requires the temperature to go 
to infinity if the horizon at $r = 2M$ is approached. This is easily 
understood: energy conservation of the 
electron fluid means that there are no external forces acting on 
this fluid. Moreover, $u=0$ can hold only if the gravitational attraction
of the central mass is balanced by a pressure gradient. If the horizon 
is approached, the gravitational attraction becomes infinite, so the 
pressure must also go to infinity. As the plasma satisfies the ideal-gas equation, as demonstrated in Appendix~\ref{sec:energymom},
the temperature must go to infinity as well. 
In particular, this line of argument demonstrates that on the Schwarzschild
spacetime energy conservation of the electron fluid implies that our 
assumption of a static plasma, $u=0$, cannot hold for a warm plasma 
arbitrarily close to the horizon. We have already seen that for a cold
plasma it cannot hold anywhere.

\begin{figure}[h!]
    \centering
    \includegraphics[width=1.0\linewidth]{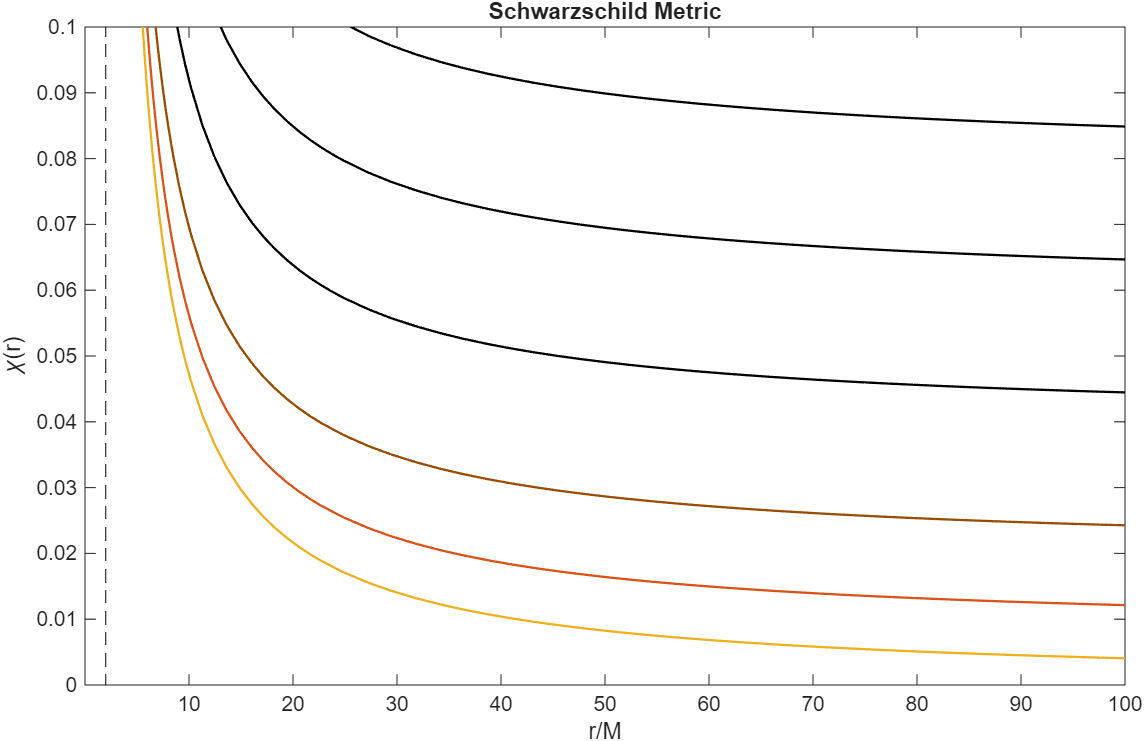}
    \caption{Radial profiles of temperature $\chi$ for various choices
of $C_\chi$ with $u(r)=0$ in the Schwarzschild spacetime. Values of $C_\chi$ span between 1 and 1.2, namely, from the bottom yellow curve up, $C_{\chi}$ equals 1, 1.02, 1.05, 1.1, 1.15, and 1.2. The dashed vertical line shows $r=2M$.}
    \label{fig:chi_prof}
\end{figure}

The temperature $\chi$ is plotted, as a function of $r$ according to (\ref{chi_C2}), 
for various values of $C_{\chi}$ in Fig. \ref{fig:chi_prof}.

\begin{figure}[h!]
    \centering
    \includegraphics[width=0.9\linewidth]{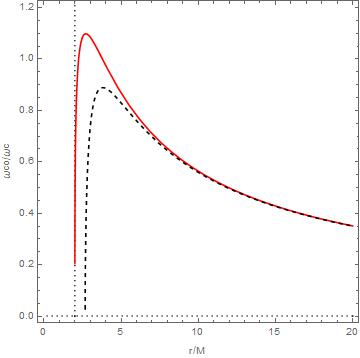}
    \caption{Cut-off frequency $\omega _{\mathrm{co}}$ as a function of $r$,
    calculated with the exact formula (\protect\ref{eq:exactco}) (solid, red) and
    with the warm-plasma approximation (\protect\ref{eq:frco}) (dashed, black). We 
    have chosen $k = 1.45$ and $C_{\chi} =1$.}
    \label{fig:Schwrestomegaco}
\end{figure}

We now calculate the cut-off frequency, once with the exact formula (\ref{eq:exactco}) 
and once with the warm-plasma approximation (\ref{eq:frco}). The result is shown in 
Fig. \ref{fig:Schwrestomegaco}. We see that, for the chosen values of $k$ and $C_{\chi}$, 
the warm-plasma approximation is good for about $r > 8M$, but that this approximation becomes quite poor 
at lower radius coordinates. 

To see which light rays can actually travel through the plasma, we have to 
relate the cut-off frequency to the photon frequency 
\begin{equation} 
\omega = - \dfrac{1}{c} \, p_{\mu} V^{\mu} = \omega _0 \Big( 1 - \dfrac{2M}{r} \Big) ^{-1/2} \, ,
\end{equation}
measured by static observers (i.e., by observers at rest with respect to the 
plasma). Here $\omega _0 = - p_t /c$ denotes the frequency constant of the 
considered light ray. With the values $k = 1.45$ and $C_{\chi} =1$, which were
chosen for the plot in Fig. \ref{fig:Schwrestomegaco}, we see that the inequality
$\omega > \omega _{\mathrm{co}}$ is satisfied for $\omega _0 > \omega _c$ on the 
entire domain $2M<r<\infty$, i.e., light rays with such a frequency constant
$\omega _0$ can travel through all points of this domain.

As the temperature goes to infinity if the horizon is approached, it is clear that
the warm-plasma approximation cannot hold arbitrarily close to $2M$. In
order to see how far this approximation is valid, we calculate the index of
refraction $n(r)$, once numerically from the exact formula (\ref{eq:exactn})
and once for the warm-plasma approximation~(\ref{n_warm}) which is 
derived in Appendix \ref{sec:warmderiv} as Eq. (\ref{eq:frn}). In the latter case,
we can write the index of refraction analytically as
\begin{equation}
n(r) ^2 = \dfrac{1- \dfrac{\omega _c^2}{\omega _0^2} \Big( \dfrac{M}{r} \Big) ^k 
\Big( 1 - \dfrac{2M}{r} \Big)
\Big( 1 - \dfrac{5}{2} \chi (r) \Big) }{1+ \dfrac{\omega _c^2}{\omega _0^2} \Big( \dfrac{M}{r} \Big) ^k
\Big( 1 - \dfrac{2M}{r} \Big) \chi (r)},  
\label{eq:Schwn}
\end{equation}
with $\chi (r)$ derived from (\ref{chi_C2}).

\begin{figure}[h!]
    \centering
    \includegraphics[width=0.9\linewidth]{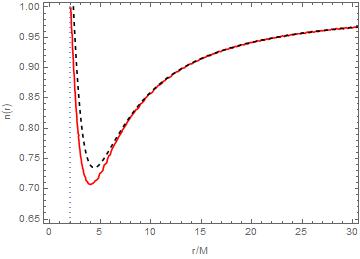}
    \caption{Index of refraction $n(r)$ for a plasma density      
    (\protect\ref{eq:omegapk}) with $k = 1.45$, temperature parameter
    $C_{\chi} = 1$ and for light rays with frequency constant $\omega _0^2
    = \omega _c ^2 /10$ from the exact formula (solid, red) and from
    the warm-plasma approximation (dashed,black).}
    \label{fig:Schwrestn}
\end{figure}

The result is plotted
in Fig. \ref{fig:Schwrestn}. We see that the warm-plasma approximation 
for $n$ is good for big $r$ and also near the horizon. The reason can be read from 
(\ref{eq:frn}): As $\chi$ occurs only multiplied with $\omega _p^2/\omega ^2$, the
warm-plasma approximation for $n$ is good if $\chi\, \omega _p^2/\omega ^2$
is small in comparison to 1. For big $r$ this is true because there 
$\chi$ is small, and near the horizon it is true because there 
$\omega ^2 = \omega _0^2 (1-2M/r)^{-1}$ is big. In the region between, 
roughly, $2.2 M$ and $8M$, the function $\chi\, \omega _p^2/\omega ^2$ is not small in comparison to 1. 

\begin{figure}[h!]
    \centering
    \includegraphics[width=0.9\linewidth]{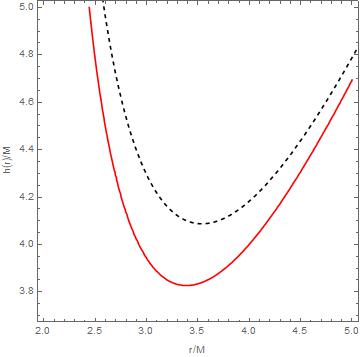}
    \caption{Function $h(r)$ for a plasma density      
    (\protect\ref{eq:omegapk}) with $k = 1.45$, temperature parameter
    $C_{\chi} = 1$ and for light rays with frequency constant $\omega _0^2
    = \omega _c ^2 /10$ from the exact formula (solid, red) and from
    the warm-plasma approximation (dashed,black).}
    \label{fig:Schwresth}
\end{figure}

So we see that for the plasma distribution
considered in this example the warm-plasma approximation is good for 
calculating weak deflection angles but not for treating light rays that 
come close to the photon sphere. In particular, this approximation does 
not give good values for the location of the photon sphere. To demonstrate 
this more clearly, we plot the function 
\begin{equation}
 h(r)= r \Big(1- \dfrac{2M}{r} \Big) ^{-1/2}  \, n(r)
\end{equation}
which gives us the radius coordinate of the photon sphere via the
equation $h'(r_{\mathrm{ph}}) = 0$. We do this again once with
the exact formula (\ref{eq:exactn}) for the index of refraction, and
once with the warm-plasma approximation (\ref{n_warm}). The result is 
plotted in Fig. \ref{fig:Schwresth}. We see that 
the warm-plasma approximation locates the photon sphere near $r = 3.6M$ 
while actually it is close to $3.4M$. This corroborates our earlier 
observation that, with the chosen parameters, the warm-plasma approximation 
is valid only for $r \gtrsim 8M$. 

Correspondingly, we can use (\ref{def_angle_Tsupko}) for the deflection
angle $\alpha$, which for a medium at rest in the Schwarzschild metric reads
\begin{gather}
\alpha + \pi = 
\\[0.2cm]
2\bigintsss_{R}^{\infty}\sqrt{\frac{1}{r(r-2M)}} 
\left(
\frac{r^3 n(r)^2 \big(R- 2M \big) }{R^3 n(R)^2 \big(r- 2M \big) }-1
\right)^{-1/2}
dr.
\nonumber
\end{gather}
However, inserting the warm-plasma approximation (\ref{eq:Schwn}) for
the index of refraction is justified only as long as 
the minimum radius coordinate $R$ is bigger than $\approx 8M$.

\subsection{Example 2: Infalling plasma on Schwarzschild spacetime}\label{infall_Schw}

\begin{figure}[h!]
\centering
\includegraphics[width=0.45\textwidth]{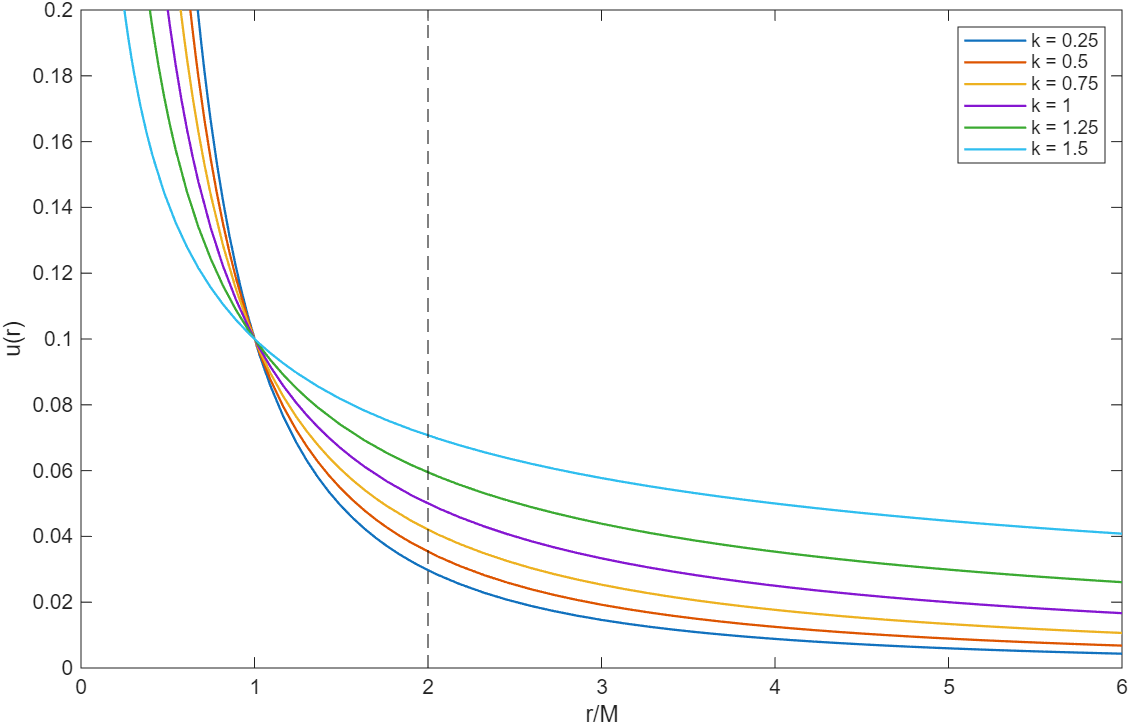}
\caption{Velocity radial component as a function of $r$ for various choices of $k$. 
Remaining parameter values were chosen to give $C_u / \omega _c^2 = M^2/10$.}\label{ur_prof}
\end{figure}

We consider again a plasma with electron density (\ref{eq:omegapk})
on the Schwarzschild spacetime (\ref{eq:Schw}), and we assume again 
that the conservation laws of charge and energy hold for the electron 
fluid. This time, however, we consider the case that the plasma is 
in radial inward motion, i.e., we assume that
Eq. (\ref{eq:uS})
holds
with  a constant $C_u  > 0$, see Fig. \ref{ur_prof}.

Then, the temperature $\chi$ is determined
by (\ref{eq:chiS}) with $C_u >0$ and $C_{\chi} \ge 1$.
It reads
\begin{equation}
1 + \dfrac{5}{2} \, \chi (r) =
C_{\chi} \, \Bigg( 1- \dfrac{2M}{r} +\dfrac{C_u^2}{\omega _c^4 M^4}
\Big( \dfrac{M}{r} \Big)^{4-2k} \Bigg)^{-1/2} \,.
\label{eq:chiinf}
\end{equation}
The model is physically viable on the domain 
where the four-velocity of the electron fluid is timelike and $\chi \ge 0$. 

We see that both conditions are satisfied on the entire domain 
$0 < r < \infty$ if and only if $k = 3/2$ and  $C_u^2/ \omega _c^4 = 2 M^4$. 
In this case, the temperature is a constant, $\chi = (2/5) (C_{\chi} - 1)$. In particular,
this gives us a cold-plasma model if we choose $C_{\chi}  =1$.

For $k > 3/2$, the model is not valid for big $r$ because then
$\chi$ comes out negative. As a consequence, we cannot consider light rays 
that come in from infinity in this case, i.e., the bending angle is not
defined. Therefore, we do not consider this case in the following.

\begin{figure}[h!]
\centering
\includegraphics[width=1\linewidth]{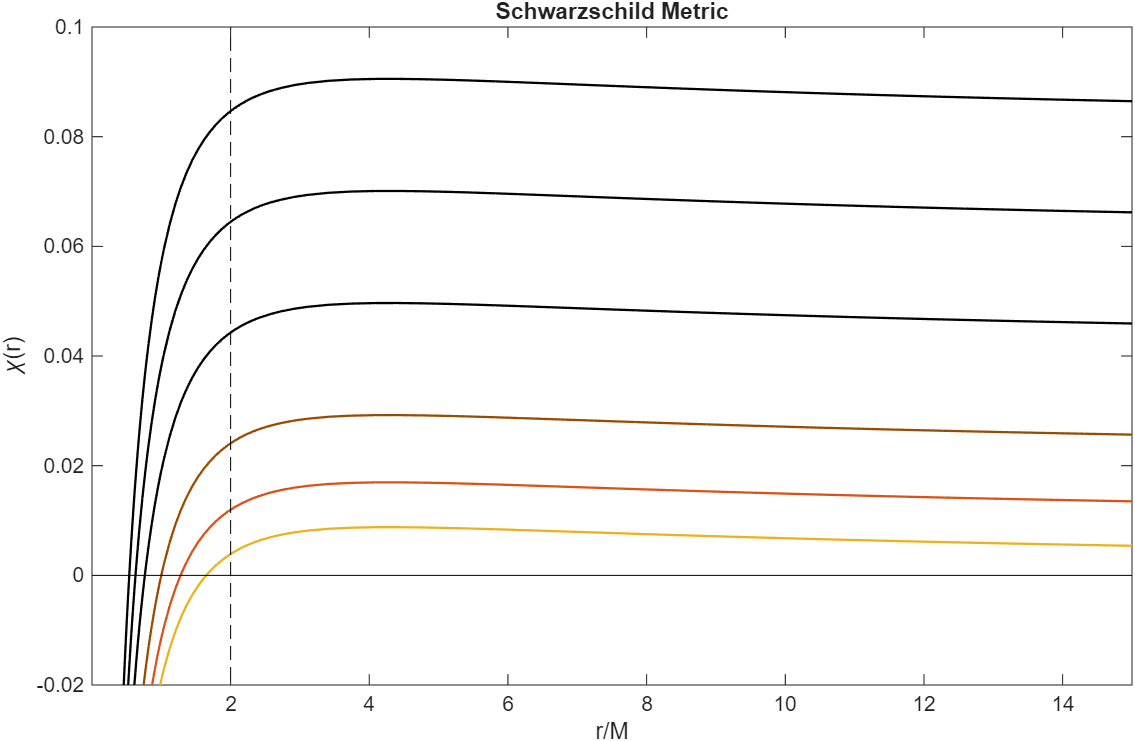}
\caption{Radial profiles of the temperature function $\chi$ for various choices
of $C_\chi$ with $C_u/\omega_c^2=1.45 \, M^2$ and $k=1.45$ in the Schwarzschild spacetime. 
Values of $C_\chi$ span between 1 and 1.2 (namely, from the bottom yellow curve up, they read 
1, 1.02, 1.05, 1.1, 1.15, and 1.2). The dashed vertical line shows $r = 2M$. }\label{chi_Vr_Schw}
\end{figure}

If $0 < k < 3/2$, then $\chi \geq 0$ holds everywhere outside 
the horizon, provided that $C_{\chi} \ge 1$, with $\chi (r) \to (2/5) (C_{\chi} -1 )$ 
for $r \to \infty$. However, $\chi$ is negative near the central singularity, 
see Fig. \ref{chi_Vr_Schw}. Nonetheless, we may use such a profile for calculating 
the bending angle and the calculation of the photon sphere if we assume 
that inside the horizon some other density profile has been matched. The special 
form of this density profile is of no interest for our considerations.

We keep in mind that the choices of $k=3/2$, $C_u = \sqrt{2} \, \omega _c^2 M^2$ 
and $C_{\chi}= 1$ give a cold plasma for which the equations for a warm plasma are equivalent
to the exact equations.  Therefore, it is clear that the warm-plasma approximation will be 
still valid if $k$, $C_u$, and $C_{\chi}$ are chosen sufficiently close to these values. In
order to find out what exactly ``sufficiently close'' means, we now calculate the index of
refraction as  a function of the frequency $\omega$, once with the warm-plasma approximation, 
for which we have the analytical equation (\ref{n_warm}), and then with the exact equation
(\ref{eq:extrans}). The latter has to be solved numerically. We use an iterative method, 
beginning with the warm-plasma expression as the zeroth-order approximation. Inserting this 
expression under the integral in (\ref{eq:extrans}) gives us a first-order approximation for 
$n$; then inserting this first-order approximation under the integral in (\ref{eq:extrans}) 
gives us a second-order approximation, and so on. In all the examples below
where results are plotted according to the exact equation, we did the calculation up to
the second order and we found that the first-order approximation is already indistinguishable,
in the plots, from the next one.

\begin{figure}[h!]
    \centering
    \includegraphics[width=0.9\linewidth]{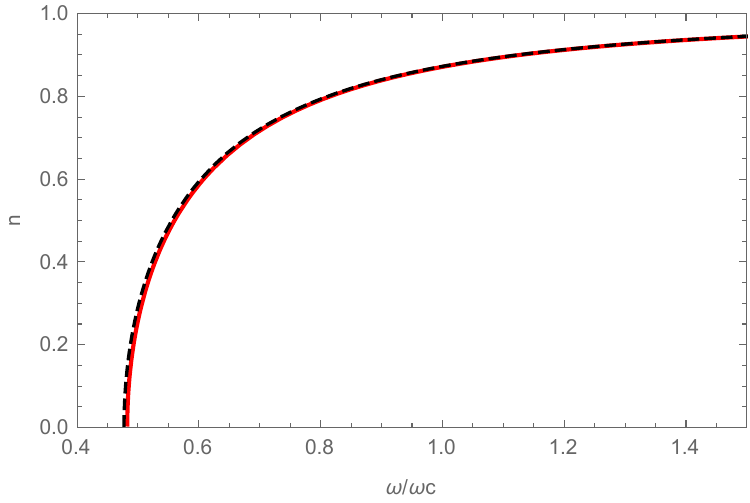}
    
    \includegraphics[width=0.9\linewidth]{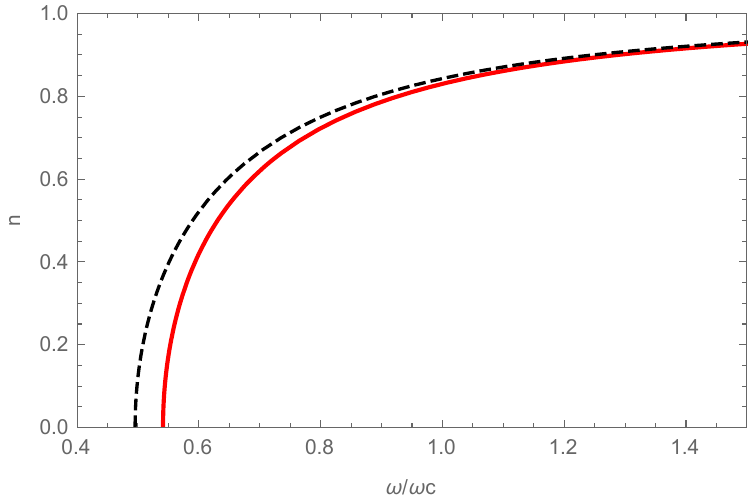}
    \caption{ Index of refraction $n$ as a function of frequency $\omega$ at $r = 2.5 M$,
    according to the warm-plasma approximation (dashed, black) and according to the exact
    equation (solid, red). In both plots we have chosen $C_u = \sqrt{2} \, \omega _c^2 M^2$.
    The upper plot is for $C_{\chi} = 1.1$ and $k=1.45$, where the warm-plasma 
    approximation is accurate enough. The bottom plot, however, is for 
    $C_{\chi} = 1$ and $k=1$ where the warm-plasma approximation is not 
    accurate enough.}
    \label{fig:Schwinfallingn}
\end{figure}

In Fig.~\ref{fig:Schwinfallingn} we plot the index of refraction $n$ as a function of the 
frequency $\omega$ at $r = 2.5 M$, which is close to the point where the temperature $\chi$
has its maximum, see Fig.~\ref{chi_Vr_Schw}.
In the upper panel of Fig.~\ref{fig:Schwinfallingn}, the parameters $C_\chi$, $C_u$, and $k$ are chosen close 
to the cold-plasma values, so the warm-plasma approximation gives good values for $n$. 
In the bottom panel,
however, the parameters are chosen such that the warm-plasma approximation is no longer 
valid: While the exact cut-off frequency is 
$\omega _{\mathrm{co}} \approx 0.55 \, \omega _c$, the warm-plasma approximation places the 
cut-off frequency near $0.50 \, \omega _c$. 
At the cut-off frequency, where the index of refraction takes the value 0, the 
warm-plasma approximation gives the absurdly wrong value of $n \approx 0.4$. So we conclude 
that, if we keep the value of $C_u = \sqrt{2} \, \omega _c^2 M^2$ fixed, the
warm-plasma approximation is applicable for $1 \le C_{\chi} \lesssim  1.1.$ and $1.45 
\lesssim k \leq 1.5$; for values outside this range this approximation would give erroneous 
results if the frequency comes close to the cut-off frequency.

\begin{figure}[h!]
\centering
\includegraphics[width=0.45\textwidth]{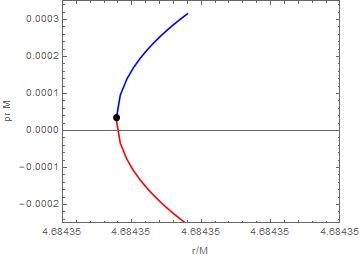}
\caption{The functions $p_r^-$ (bottom, red) and $p_r^+$ (top,blue) along a light ray with 
impact parameter $b = 6 M$ in a plasma with $C_{\chi}=1$, $C_u = \sqrt{2} \, \omega _c^2 M^2$
and $k = 1.45$. The black dot marks the turning point where $\dot{r}=0$.}
\label{fig:pr}
\end{figure}

The frequency measured by an observer who is comoving with the electron fluid is now given by the equation
\begin{equation}
    \omega = - \dfrac{1}{c} \, p_{\mu} V^{\mu} =
    \sqrt{\dfrac{1- \dfrac{2M}{r} + u(r)^2}{\Big( 1-\dfrac{2M}{r} \Big) ^2}} \, \omega _0 + u(r) p_r,
\label{eq:omegainf}
\end{equation}
where $\omega _0 = - p_t /c$ is the frequency constant of the light ray under consideration.
In contrast to the case $u(r) = 0$, which was treated in the previous example, the
frequency $\omega$, and therefore the index of refraction, now depends on $p_r$. 
Therefore, we have to determine $p_r$ along each light ray by solving the
equation $\mathcal{H} =0$ for $p_r$.
In the warm-plasma approximation, we can write the Hamiltonian in the equatorial plane as
\begin{gather}
    \mathcal{H} = \frac{1}{2}\left[ 
    \Big(1- \dfrac{2M}{r} \Big) p_r^2 + \dfrac{p_{\varphi} ^2}{r^2} 
    - \dfrac{p_t^2}{\Big( 1 - \dfrac{2M}{r} \Big)c^2} \right.
    \nonumber
    \\[0.2cm]
   \left. +
    \dfrac{
    \omega ^2 \, \omega _p(r)^2 \Big( 1 - \dfrac{3}{2} \chi (r) \Big)
    }{
    \omega ^2 + \omega _p(r)^2 \, \chi (r)
    } \right]\, .
    \label{eq:HSchwinfalling}
\end{gather}
Inserting (\ref{eq:omegainf}) into the dispersion relation $\mathcal{H}=0$ gives
us a quartic equation for $p_r$ that admits four in general complex solutions.
These solutions can be determined analytically, with one of the well-known
solution methods for quartic equations, but we do not write them out here
because they are rather awkward. In our calculations, we used MATHEMATICA for 
producing these analytical solutions. For a light ray that comes in from infinity,
goes through a minimum radius $R$ and then escapes to infinity again, exactly two of 
the four solutions, let us call them $p_r^-$ and $p_r^+$, are real on the interval 
$R< r < \infty$. Note that the two branches are not symmetrical, $p_r^- \neq - p_r^+$,
although the difference between $p_r^-$ and $-p_r^+$ is small in the domain where the 
warm-plasma approximation is valid. At the turning point ($r=R$, where $\dot{r} = 0$)
$p_r$ is positive, see Fig. \ref{fig:pr}.

The deflection angle is to be determined by integrating the equation
\begin{equation}
d \varphi =
\Bigg( \dfrac{\partial \mathcal{H}}{\partial p_r} \Bigg) ^{-1}
\Bigg( \dfrac{\partial \mathcal{H}}{\partial p_{\varphi}} \Bigg) \, dr \, ,
\label{eq:drdphiinf}
\end{equation} 
which follows from Hamilton's equations, over the light ray.
Here, the derivatives of $\mathcal{H}$ on the right-hand side have to be calculated 
after inserting (\ref{eq:omegainf}) into the Hamiltonian. Then, the 
integration has to be done first over the part of the light ray where 
$p_r = p_r^-$ and then over the part where $p_r = p_r^+$.
Each light ray is labelled by the 
frequency constant $\omega _0$  and the minimum radius $R$. The 
impact parameter $b = -c \, p _{\varphi}/p_t$ is related to these two
constants of motion by the condition that $\dot{r} = 
\partial \mathcal{H} / \partial p_r$ vanishes at $r=R$.

\begin{figure}[h!]
\centering
\includegraphics[width=0.45\textwidth]{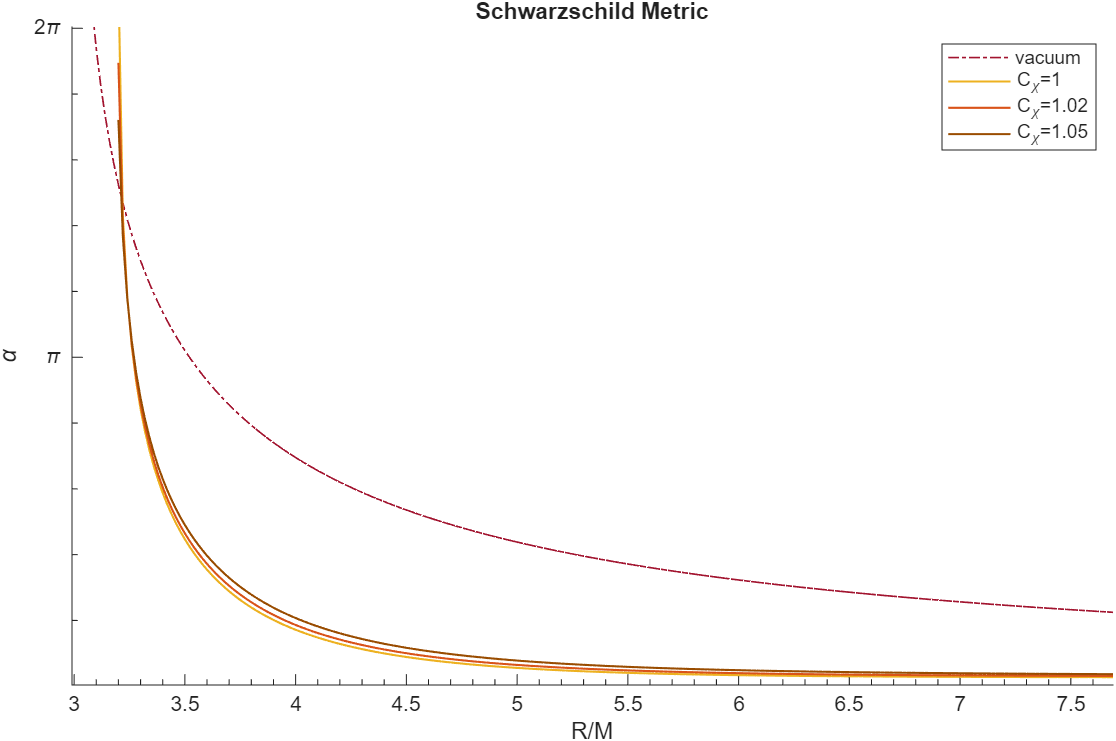}
\caption{Deflection angle in an infalling warm plasma on the Schwarzschild spacetime,
with various $C_\chi$, as a function of the radius coordinate $R$ (the point of the closest approach)
in comparison with vacuum (dashed purple curve). 
The values of $C_\chi$ were chosen as 1, 1.02, and 1.05, while $C_u/\omega_c^2=\sqrt{2} \, M^2$ 
and $k=1.5$. Note that for this choice of parameters, the case with $C_\chi=1$ corresponds to a cold plasma.}
\label{angle_Schw1}
\end{figure}

Fig. \ref{angle_Schw1} shows plots of the deflection angle in the warm-plasma
approximation, where we made sure that the parameter values have 
been chosen such that this approximation is applicable.
The photon sphere is determined by the value of $R$ where
the deflection angle diverges. For the sake of comparison, we also plot
the case of light rays in vacuum and in a cold plasma. Note that, for 
the parameters chosen in Fig.~\ref{angle_Schw1}, the cold-plasma limit 
$\chi = 0$ is incompatible with the energy conservation law (\ref{eq:chiinf}).\ref{eq:chiinf}
In other words, a cold plasma can exist on the Schwarzschild spacetime with
the chosen density and velocity profiles only if the electron fluid is acted 
upon by external forces.

From Fig.~\ref{angle_Schw1} we read that for big $R$ the deflection angle in the plasma is smaller
than in vacuum. For small $R$, however, it is bigger which has the effect that the photon sphere
in the plasma is at a bigger radius coordinate than in vacuum. The fact that, for a sufficiently 
high $\omega _p$, such a crossing of the plasma curve and the vacuum curve occurs, was already 
observed for a cold plasma on the Schwarzschild spacetime by Perlick and Tsupko \cite{Perlick-Tsupko-2024}.

\begin{figure}[h!]
\centering
\includegraphics[width=0.45\textwidth]{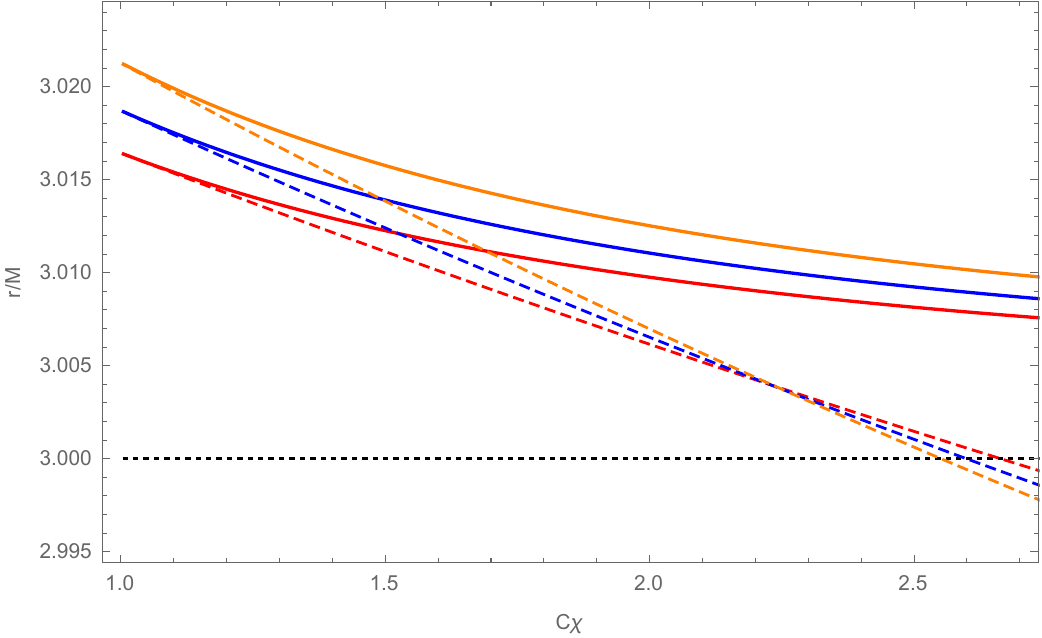}
\caption{Radius of the photon sphere as a function of the temperature constant $C_{\chi}$,
calculated with the exact formula (solid) and with the warm-plasma approximation (dashed).
All plots are for $C_u = \sqrt{2} \, \omega _c^2 M^2$ and $\omega _0 = \omega _c$. For $k$
we have chosen, from the bottom up, $k= 1.5$ (red), $k= 1.45$ (blue), and $k= 1.4$ (orange). 
The red curve is valid for a plasma of constant temperature, where $C_{\chi}=1$ corresponds 
to a cold plasma, i.e., $\chi = 0$. Recall that for $\chi = 0$ both the warm-plasma
approximation (\protect\ref{eq:frn}) and the exact equation (\protect\ref{eq:extrans}) reduce to
the cold-plasma case.
}
\label{fig:InfallingPhotonSphere}
\end{figure}

For calculating the location of the photon sphere in the infalling warm plasma, 
we proceed as outlined in Section~\ref{subsec:sphstat}.
In Fig. \ref{fig:InfallingPhotonSphere} we plot $r = r_{\mathrm{ph}}$ as a function of the 
temperature constant $C_{\chi}$. For the sake of comparison, we also plot the results from the exact 
equation, which we determined numerically with the above-mentioned iterative procedure. The figure 
confirms our earlier observation that the warm-plasma approximation is good if the parameter values 
are close to those of a cold plasma, but that they can become quite incorrect if we go too far
away from these values. In particular, we see that for too high temperature constants $C_{\chi}$ 
the warm plasma approximation gives us values for the radius of the photon sphere that are
below the value for vacuum light rays, $r = 3M$, while the actual values are above this value.

\begin{figure}[h!]
\centering
\includegraphics[width=0.45\textwidth]{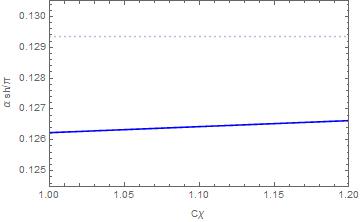}
\caption{
Angular radius of the shadow, for an observer at $r_O= 12 \, M$ on 
Schwarzschild spacetime, in an infalling plasma with $C_u =\sqrt{2} \, \omega _c ^2 M^2 $, 
$k = 3/2$, $\omega _0 = \omega _c$. The dotted line indicates the shadow radius in
vacuum for an observer at the same position. 
}\label{fig:Schwshadow}
\end{figure}

\begin{figure}[h!]
\centering
\includegraphics[width=0.45\textwidth]{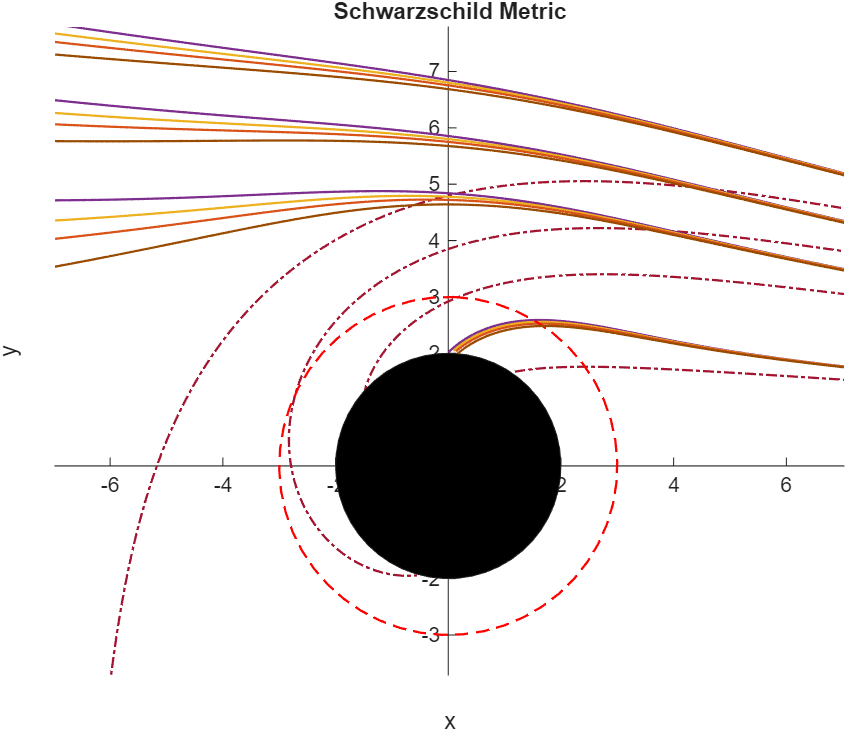}
\caption{Ray trajectories around a Schwarzschild black hole in vacuum (dashed purple curves), cold plasma (solid violet curves) and in a warm plasma (remaining solid curves) in the equatorial plane. The values of the corresponding plasma parameters used to construct the plot are $C_u/\omega_c^2=1.45 \, M^2$  and $k=1.45$, and the colors are the same as in Fig.~\ref{angle_Schw1}. 
The dashed red circle shows the photon sphere in vacuum. The impact parameters of rays equal $2 \, M$, $4 \, M$, $5 \, M$, and $6 \, M$.}\label{traj_Schw1}
\end{figure}

Moreover, we can calculate the angular radius of the shadow in an infalling warm plasma 
on the Schwarzschild spacetime with the method outlined in Section \ref{subsec:sphstat}. 
The result is shown in Fig.~\ref{fig:Schwshadow}. We 
have chosen the plasma parameters $C_u = \sqrt{2} \omega _c^2 M^2$ and $k = 3/2$
which give a constant temperature, in particular $\chi = 0$ for $C_{\chi} = 1$. We see
that in the domain where the warm-plasma approximation is valid (i.e., $1 \le C_{\chi} 
\lesssim 1.2$) the influence of a non-zero temperature on the shadow radius is quite small.
We also notice that the shadow in the plasma is smaller than in vacuum. For a cold plasma,
it was already observed by Perlick, Tsupko and Bisnovatyi-Kogan \cite{Perlick-Tsupko-BK-2015} 
that this is true for a plasma density that satisfies a power law with $k < 2$. The angular radius of the shadow slightly increases with increasing $C_\chi$, i.e., it gets closer to the vacuum value. The same tendency can be read from the shape of the trajectories, shown in Fig.~\ref{traj_Schw1}, where we plot the trajectories of light rays in an infalling warm plasma on 
the Schwarzschild spacetime. We see that, unsurprisingly, the deflection becomes stronger if
the temperature parameter $C_{\chi}$ is increased while keeping the impact parameter constant.

\subsection{Example 3: Stationarily rotating plasma on Kerr spacetime}

We consider the Kerr spacetime, which is of the form of 
Eq. (\ref{eq:axistat}) with 
\[
A = 1 - \dfrac{2Mr}{\rho ^2}+ 
\dfrac{4M^2r^2a^2 \mathrm{sin} ^2 \vartheta}{\rho ^2 \big( (r^2+a^2) \rho ^2 +
2Mra^2 \mathrm{sin} ^2 \vartheta \big)} \, , 
\]
\[
B = \dfrac{\rho ^2}{\Delta} \, , \quad
F = \rho ^2 \, ,
\]
\[
D = \dfrac{1}{\rho ^2} \big( (r^2+a^2) \rho ^2 +
2Mra^2 \mathrm{sin} ^2 \vartheta \big) \, , 
\]
\begin{equation}
P = \dfrac{2Mra}{
(r^2+a^2) \rho ^2 + 2Mra^2 \mathrm{sin} ^2 \vartheta},
\label{eq:Kerr}
\end{equation}
where $\rho^2 = r^2 + a^2 \mathrm{cos} ^2 \vartheta$ 
and $\Delta = r^2 + a^2 - 2Mr$. Here $M$ and $a$ are, respectively, the mass parameter and the 
spin parameter both of which have the dimension of a length. We assume that the plasma 
density is symmetric with respect to the equatorial plane
and in the equatorial plane of the form
\begin{equation}
\omega _p (r) ^2 = \omega _c ^2 \Big( \dfrac{M}{r} \Big) ^k,
\end{equation}
with constants $\omega _c$ and  $k$. 
We remind the reader that then the equation of state (\ref{eq:pol_eos}) is 
satisfied with $\gamma = 1$ and $K$ given by (\ref{eq:K}).

In this section, we assume that the plasma is corotating with the black hole in the sense
that the four-velocity $V^{\mu} \partial _{\mu}$ of the electron fluid satisfies 
(\ref{eq:V}) with $u = 0$.
In the black-hole case, $a^2 \le M^2$, 
we restrict the following consideration to the domain of outer 
communication, i.e., to the region outside the outer horizon, 
$r > M + \sqrt{M^2-a^2}$, where the four-velocity 
\begin{equation}
V^{\mu} \partial _{\mu} = 
\dfrac{1}{\sqrt{A}} 
\big( \partial _t + P \, c \, \partial _{\varphi} \big)
\label{eq:VKerrco}
\end{equation}
is timelike. 

We assume again that charge conservation and energy 
conservation hold for the electron fluid. With $u=0$, 
(\ref{eq:C1}) is satisfied with $C_u = 0$, while (\ref{eq:C2})
requires that the temperature in the equatorial plane
is given by the equation
\begin{equation}
1 + \dfrac{5}{2} \, \chi (r) =
C_{\chi} \, \left(1 - \dfrac{2M(r^2+a^2)}{r(r^2+a^2) +
2Ma^2}\right)^{-1/2} 
\label{eq:Kerrcochi}
\end{equation}
with a constant $C{\chi}$. As $\chi$ cannot be negative, we have 
to require $C_{\chi} \ge 1$. Then we may assume that Eq.~(\ref{eq:Kerrcochi})
holds for arbitrarily large $r$. 

\begin{figure}[h!]
\centering
\includegraphics[width=0.45\textwidth]{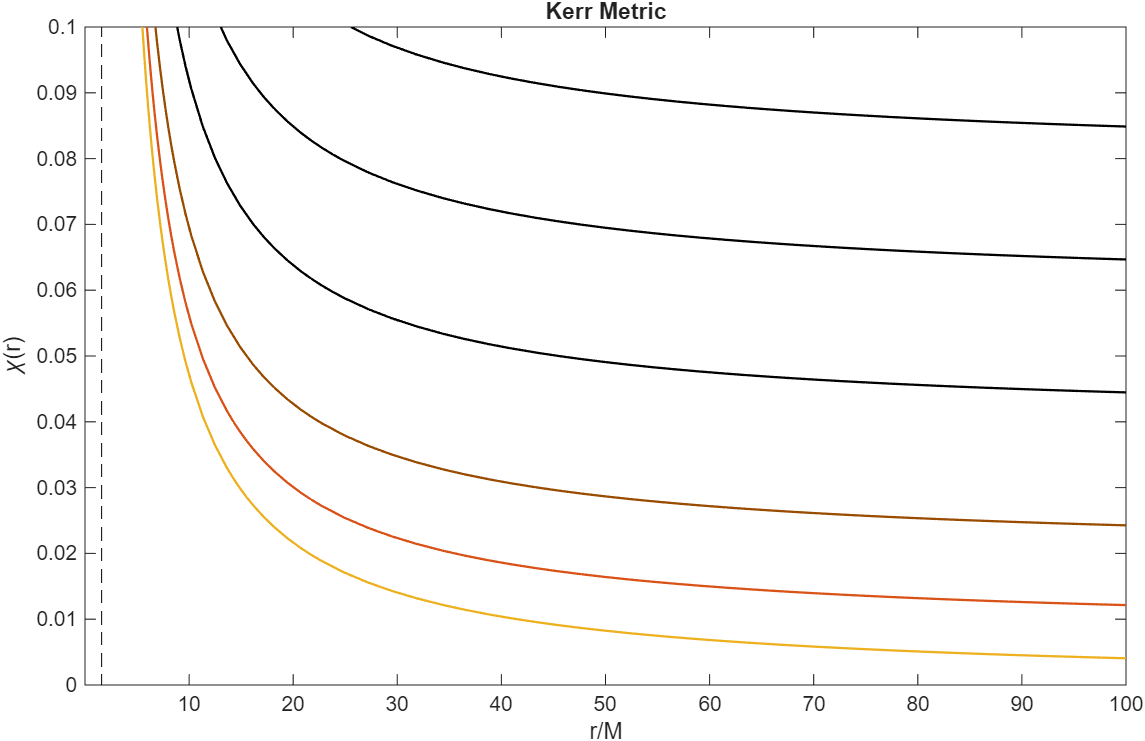}
\caption{Radial profiles of temperature $\chi$ for various choices 
of $C_\chi$ with $u(r) = 0$ in the Kerr spacetime with $a=0.8\,M$. 
Values of $C_\chi$ span between 1 and 1.2, namely, from the bottom 
yellow curve up, $C_\chi$ equals 1, 1.02, 1.05, 1.1, 1.15, and 1.2. 
The dashed vertical line shows the horizon radius.}
\label{fig:chi_prof_Kerr_new}
\end{figure}

From (\ref{eq:Kerrcochi}) we read that $\chi$ goes to infinity if the 
horizon at $M + \sqrt{M^2-a^2}$ is approached, cf. Fig.
\ref{fig:chi_prof_Kerr_new}. Therefore, the warm-plasma
approximation cannot hold arbitrarily close to the horizon. 

\begin{figure}[h!]
    \centering
    \includegraphics[width=0.9\linewidth]{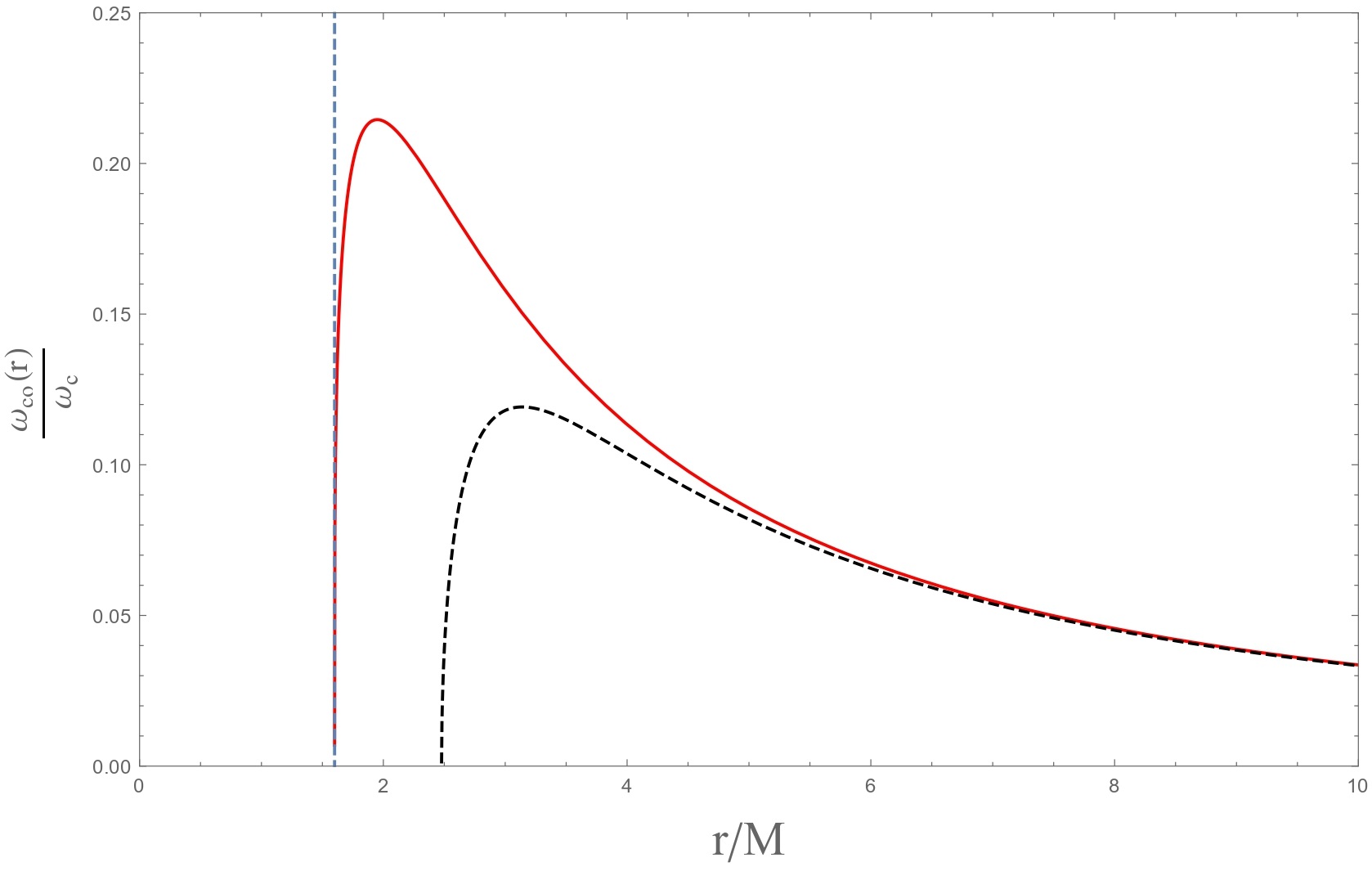}
    \caption{Cut-off frequency $\omega _{\mathrm{co}} (r)$ divided by $\omega _c $,
    calculated with the exact formula (\protect\ref{eq:exactco}) (solid, red) and
    with the warm-plasma approximation (\protect\ref{eq:frco}) (dashed, black), on a
    Kerr spacetime with $a = 0.8 \, M$. The vertical dashed line marks the horizon at 
    $M + \sqrt{M^2-a^2}$. We have chosen $k = 1.45$ and $C_{\chi} =1$.}
    \label{fig:Kerrcoromegaco}
\end{figure}

To see how close to the horizon we can go with the warm-plasma
approximation, we calculate the cut-off frequency, once with the 
exact formula (\ref{eq:exactco}) and once with the warm-plasma 
approximation (\ref{eq:frco}). The result is shown in 
Fig. \ref{fig:Kerrcoromegaco}. Similar to what we have seen
in the Schwarzschild spacetime for a plasma with $u =0$, we
observe again that for the chosen values of $k$ and $C_{\chi}$, 
the warm-plasma approximation is good only for $r \gtrsim 8M$.
Hence, we cannot use it for light rays that are strongly bent.  
In particular, when calculating the deflection angle with the help 
of (\ref{def_angl_axis}), we can use the warm-plasma
approximation of $n$ only for 
light rays whose point of the closest approach is at $r \gtrsim 8 M$.

\subsection{Example 4: Infalling plasma on Kerr spacetime}

We consider again the Kerr spacetime (\ref{eq:Kerr}), but this time with a plasma 
that is falling towards the center, i.e., the four-velocity has
the form of (\ref{eq:V}) with $u(r)>0$. We restrict again to the
equatorial plane, assuming that the plasma functions are such that
a light ray stays in this plane if it starts tangential to it. 

For the plasma density in the equatorial plane, we assume again
a power law (\ref{eq:omegapk}) with a positive $k$. Then the 
conservation laws of energy and charge, Eqs. (\ref{eq:C1}) and 
(\ref{eq:C2}), determine the radial velocity and the temperature, respectively, as
\begin{equation}
u(r) = 
\dfrac{C_u}{\omega _c^2 M^2} \Big( \dfrac{M}{r} \Big)^{2-k}
\, ,
\end{equation}
\begin{equation}
\chi (r) = \dfrac{2}{5} \left( 
C_{\chi} \sqrt{\dfrac
{1+\dfrac{a^2}{r^2} + \dfrac{2 \, m \, a^2}{r^3}
}{
1+\dfrac{C_u^2}{\omega _c^4 M^4} \Big( \dfrac{M}{r} \Big)^{4-2k}  - \dfrac{2 \, M}{r}  + \dfrac{a^2}{r^2}
}
}
- 1 \right) 
\label{eq:chiKerr}
\end{equation}
with constants $C_u$ and $C_{\chi}$. For an infalling plasma we must have $C_u > 0$. Moreover, we choose $C_{\chi} \ge 1$ and $0<k<3/2$ to make sure that the temperature is positive for arbitrarily large $r$. For $C_u/\omega _c^2 < 4 \, M^2$, the temperature is then positive and bounded on the domain of outer communication, $M + \sqrt{M^2-a^2} < r < \infty$.  

In contrast to the Schwarzschild case, in a Kerr spacetime with $a \neq 0$ we cannot choose $C_u$, $C_{\chi}$ and $k$ such that the temperature is a constant. In particular, the cold-plasma case is not included. In other words, our power-law ansatz for the plasma density is incompatible with the assumptions that the conservation laws hold and that the temperature is a constant.    

\begin{figure}[h!]
\centering
\includegraphics[width=0.5\textwidth]{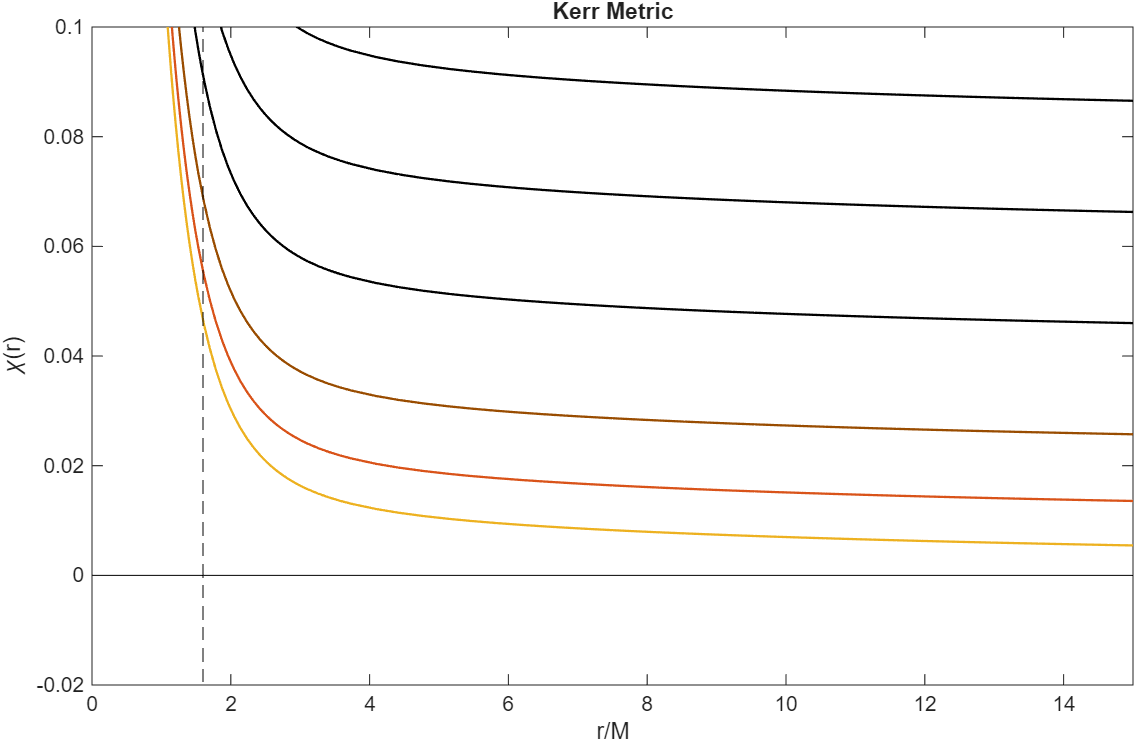}
\caption{Radial profiles of the temperature function $\chi$ for various choices of $C_\chi$ with $C_u/\omega_c^2=1.45 \, M^2$ and $k=1.45$ in the Kerr spacetime with $a=0.8M$.  Values of $C_\chi$ span between 1 and 1.2 (namely, from the bottom yellow curve up, they read 1, 1.02, 1.05, 1.1, 1.15, and 1.2). The dashed vertical line shows $r = r_H=1.6M$.}
\label{fig:chi_prof_Kerr_new_Vr}
\end{figure}

Fig. \ref{fig:chi_prof_Kerr_new_Vr} shows the temperature as a function of $r$ on a Kerr spacetime with $a = 0.8 \, M$, for various values of $C_{\chi}$.

\begin{figure}[h!]
\centering
\includegraphics[width=0.45\textwidth]{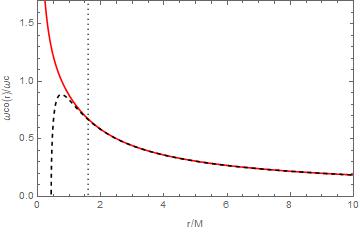}
\caption{Cut-off frequency in an infalling plasma on Kerr spacetime with $a = 0.8 M$, $k = 1.45$, $C_u / \omega _c^2 = 1.45 \, M^2$ and $C_{\chi} =1$, calculated with the exact formula (\protect\ref{eq:exactco}) (red, solid) and with the warm-plasma approximation (\protect\ref{eq:frco}) (black, dashed). The black dotted line marks the horizon.}
\label{fig:Kerrinfomegaco}
\end{figure}

Before calculating and plotting the deflection angle of light rays in an infalling plasma on the Kerr spacetime, we want to check whether the warm-plasma approximation is applicable. To that end we calculate the cut-off frequency, once with the exact formula  (\ref{eq:exactco}) and once with the warm-plasma approximation (\ref{eq:frco}) for the same parameters $a = 0.8 M$, $k = 1.45$, $C_u / \omega _c^2 = 1.45 \, M^2$ and $C_{\chi} =1$ for which we then want to calculate the deflection angle. Fig.~\ref{fig:Kerrinfomegaco} demonstrates that the approximation is very good on the domain of outer communication, i.e., on the entire domain which is relevant for the deflection angle.

\begin{figure}[h!]
\centering
\includegraphics[width=0.45\textwidth]{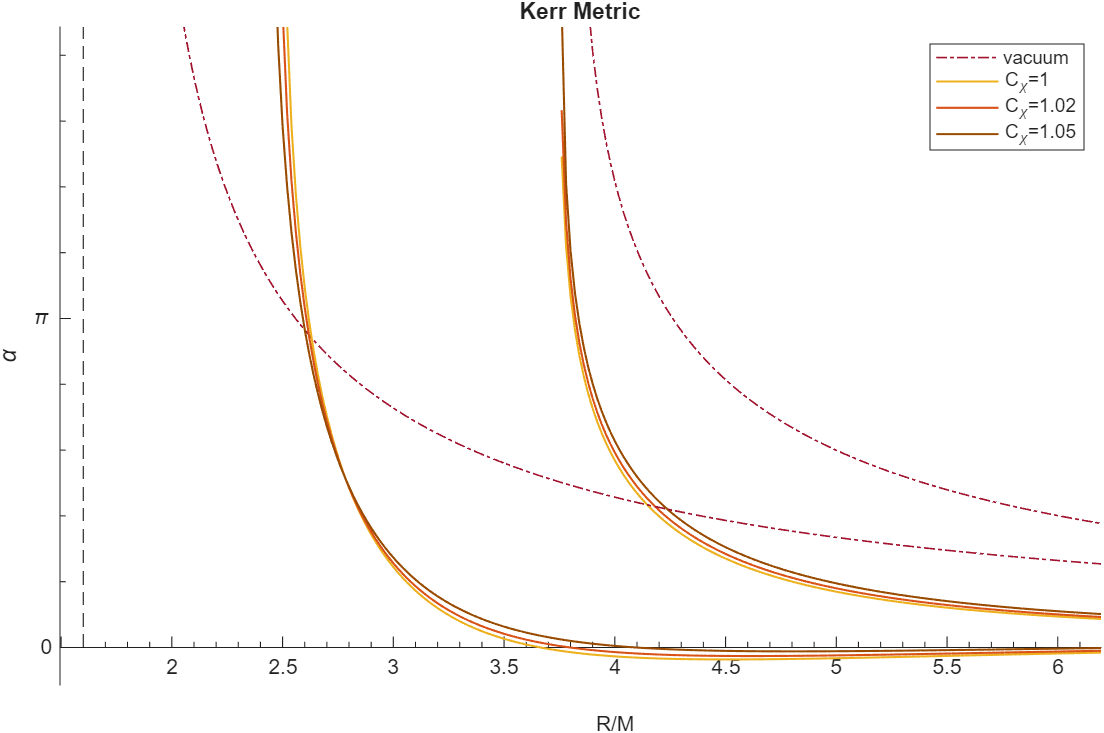}
\caption{Deflection angle in an infalling warm plasma on the Kerr  spacetime with $a=0.8 \, M$, with various $C_\chi$, as a function of the radius coordinate $R$ (the point of the closest approach) in comparison with vacuum (dashed purple curve). The values of $C_\chi$ were chosen as 1, 1.02, and 1.05, while $C_u/\omega_c^2=\sqrt{2} \, M^2$ and $k=1.5$. The dashed vertical line shows $r=r_H=1.6M$. We remind the reader that in a Kerr spacetime with $a \neq 0$ the parameters $C_{\chi}$, $C_u$ and $k$ cannot be chosen such that $\chi$ is a constant, see (\protect\ref{eq:chiKerr}), so in particular the cold-plasma case is not included. }\label{angle_Kerr1}
\end{figure}

Having convinced ourselves that the warm-plasma approximation is applicable, we plot the deflection 
angle for various values of the temperature parameter $C_{\chi}$ in Fig.~\ref{angle_Kerr1} for co-rotating
and for counter-rotating rays. The values of $R$ where the bending angle diverges gives the location of 
the circular light rays. In the counter-rotating case the deflection in the plasma is always
smaller than in vacuum. Correspondingly, the counter-rotating circular light ray in the plasma is 
at a smaller radius value than in vacuum. By contrast, in the co-rotating case the plasma curves intersect
the vacuum case, i.e., for sufficiently small $R$ the bending in the plasma is stronger than in vacuum.
Correspondingly, the co-rotating circular light ray in the plasma is at a bigger radius value than in
vacuum. This phenomenon has already been observed for a cold plasma on the Kerr spacetime by Perlick and 
Tsupko \cite{Perlick-Tsupko-2024}. Also, we see in Fig.~\ref{angle_Kerr1} that for co-rotating rays there 
is a regime where the bending angle takes negative values which indicates that the ray is repelled rather 
than attracted. Again, this has already been observed for a cold plasma on the Kerr spacetime by Perlick 
and Tsupko \cite{Perlick-Tsupko-2024}.

\begin{figure}[h!]
\centering
\includegraphics[width=0.45\textwidth]{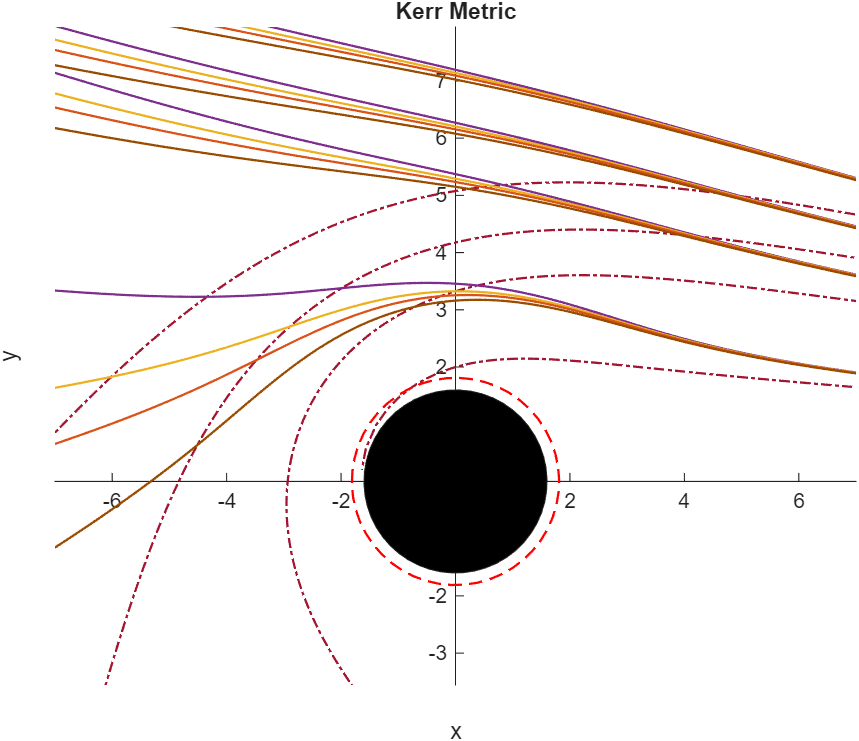}
\caption{Trajectories of co-rotating light rays around a Kerr black hole with $a=0.8 \, M$ in vacuum (dashed purple curves), cold plasma (solid violet curves) and in a warm plasma (remaining solid curves) in the equatorial plane. The values of the corresponding plasma parameters used to construct the plot are $C_u/\omega_c^2=1.45 \, M^2$  and $k=1.45$, and the colors are the same as in Fig.~\ref{angle_Schw1}. The dashed red circle shows the co-rotating circular light ray in vacuum. The impact parameters of rays equal $2 \, M$, $4 \, M$, $5 \, M$, and $6 \, M$. Here we have included, for the sake of comparison, the cold-plasma case with the same $\omega_p$ as for the other curves, but the reader should keep in mind that this case can be realized only if there is an external force acting on the electron fluid.}\label{traj_Kerr1}
\end{figure}

Fig.~\ref{traj_Kerr1} shows the trajectories of a few co-rotating light rays in a warm plasma on the 
Schwarzschild spacetime. This picture confirms the observation discussed above that because of the plasma some of 
these rays are repelled from the center.

\section{Conclusions}\label{sec:conclusions}

In this paper, we have discussed light propagation on a general-relativistic spacetime,
in particular in the neighborhood of a black hole, in the presence of a warm plasma. 
The motivation for considering a warm plasma, rather than a cold plasma which has been
treated in numerous papers before, was twofold. Firstly, there is some indication that 
near supermassive black holes the temperature may be so high that the cold-plasma 
approximation is no longer valid. Secondly, working out the formalism for a warm plasma
gives important new insights into the validity of the cold-plasma approximation. 

Let us discuss the former motivation first. In view of observations, the 
results of this paper are relevant in situations where the plasma 
density is so high that light rays cannot be treated as vacuum light 
rays (i.e., lightlike geodesics) anymore, and where the temperature 
is so high that the cold-plasma approximation is no longer applicable. 
More precisely, the first condition requires that $\omega _p^2/\omega ^2$ 
is not negligible in comparison to unity and the second condition 
requires that $\chi$ is not negligible in comparison to unity. 

As to the first condition, we have to take into account that observations 
of (supermassive) black holes and their immediate neighborhood are usually 
done at frequencies of $\omega = 86 \, \mathrm{GHz}$ 
(see e.g. Issaoun et al. \cite{IssaounEtAl2019})
or higher. It is 
widely believed that observations at lower frequencies are not 
possible because the images would be washed out by scattering. If
we take this value of $\omega = 86 \, \mathrm{GHz}$ for a numerical
estimate of the relevant effects, we see that we would need a plasma
(electron) density of about $10^{19} \mathrm{m}^{-3}$ to make 
$\omega _p^2/\omega ^2$ approximately equal to $0.1$ which would 
reduce the square of the index of refraction from its vacuum
value $n^2 = 1$ to $n^2 \approx 0.9$ which is certainly measurable.
To be sure, a plasma density of $10^{19} \mathrm{m}^{-3}$ is extremely
high, but it is believed to occur near some stellar mass black
holes and maybe even near some supermassive black holes, see, e.g., Jiang et 
al. \cite{JiangEtAl2019,MallickEtAl2022}. In the case of 
supermassive black holes, this could have a measurable effect on
the shadow and on bending angles, while in the case of a stellar 
mass black hole, where it is certainly impossible to get a resolved
image of the black hole and its immediate neighborhood, it might
nevertheless have a measurable impact on microlensing observations.
For a theoretical investigation of how a (cold) plasma influences
microlensing we refer to Tsupko and Bisnovatyi-Kogan 
\cite{Tsupko-BK-2020-microlensing}. 

As to the second condition, we have already observed in Sec. II that 
$\chi$ being non-negligible in comparison to unity 
requires a temperature of $10^8 \, \mathrm{K}$ or higher. Whereas this
is a very high temperature, it may occur near some stellar mass and 
even near some supermassive black holes, see, e.g., 
\cite{genzel2010,LanzuisiEtAl2019,AkiyamaEtAl2021}. Temperatures above 
$10^8 \mathrm{K}$ correspond to values of $\chi$ above $0.02$ which, 
according to Fig.21, have a measurable impact on the cut-off frequency 
and, thereby, on the index of refraction if $\omega _p^2/\omega ^2$ 
is not negligibly small. 

As to the second motivation of this work,
it is crucial to notice that in a cold plasma the Hamiltonian for the
light rays involves only the density of the electron fluid (in its rest frame), while in
a warm plasma it also involves the velocity of the electron fluid. In this paper, we have
seen that the validity of the warm-plasma approximation puts some restrictions on the 
velocity of the electron fluid, in particular if we require that the conservation
laws of energy and charge should hold for the electron fluid alone. (This is a reasonable
assumption in a stationary situation.) As the cold-plasma approximation cannot be valid if
the warm-plasma approximation is not, this observation implies that for a cold plasma the
velocity of the electron fluid is not arbitrary: Although this velocity does not appear in
the equations for a cold plasma, just by assuming that the cold-plasma approximation is valid,
one implicitly restricts the velocity of the electron fluid. Of course, this does not mean 
that there is anything wrong with the numerous papers that considered light propagation in
a cold plasma. 
One just has to be aware of the fact that by prescribing a certain plasma density
and assuming that the effect of the temperature on the trajectories of light rays
is negligibly small, one implicitly puts restrictions onto the velocity of the 
electron fluid, unless one allows for violations of the conservation laws.


Throughout the paper we considered a plasma without an electromagnetic background field and 
we excluded ionization and recombination processes. If these conditions were not satisfied, 
it would not be justified to assume that the conservation laws for energy and charge would 
hold for the electron fluid alone: An electromagnetic background field would not only modify the dispersion relation, but also exert an external
force onto the electron fluid and ionization or recombination processes would change its charge.
In view of applications to astrophysics, including an electromagnetic background  field is the
most desirable generalization of the formalism considered here. The index of refraction for light in a magnetized warm plasma is known. It was derived by Imre \cite{imre62} for the case of a homogeneous magnetized plasma on Minkowski spacetime. This result can be carried over to general relativity by invoking the principle of minimal coupling. However, then the index of refraction depends not only on the spacetime point and on the frequency, but also on the spatial direction, i.e., the medium is no longer isotropic, so Synge's formalism does not apply. In the axisymmetric and stationary case, it might be possible to choose the electromagnetic field such that a Carter constant still exists. Then some analytical calculations, e.g., of the shadow, would be possible. We leave this generalization for a future work.

In our examples, we have assumed that the underlying spacetime is given by the Schwarzschild or 
the Kerr metric. As both are vacuum solutions of Einstein's field equations, in these examples
we have implicitly assumed that the plasma is not self-gravitating, i.e., we have assumed that
the gravitational field produced by the plasma is negligibly small in comparison to the 
gravitational field produced by the central object. We emphasize, however, that our general
equations apply also to a self-gravitating plasma, because Einstein's field equations were not
used when deriving them.

We have taken some pains to make sure that we apply the warm-plasma approximation only in cases
where it is indeed applicable. For that purpose we have derived, in Appendix~\ref{sec:warmderiv},
an exact equation for the index of refraction in a collisionless non-magnetized plasma without 
any restrictive assumptions on the temperature, see Eq.~(\ref{eq:extrans}). This allowed us to 
give a derivation of the warm-plasma approximation which, in our view, goes beyond the derivations 
that are available in the literature: We have seen that the warm-plasma approximation is the 
result of a linearization with respect to the temperature, not of the index of refraction and 
not of the standard Hamiltonian according to Synge formalism, but rather of a rescaled Hamiltonian 
that has the same solution curves as the standard one, just with another parametrization. By comparing
with the exact equation (\ref{eq:extrans}), we were able to determine the range of validity of the 
warm-plasma approximation in all our example calculations. While our calculations of bending
angles, shadows and other lensing features with the warm-plasma approximation are fully analytic,
evaluation of the exact formula (\ref{eq:extrans}) requires time-consuming numerical work. Therefore,
using the warm-plasma approximation, where applicable, is of a great advantage. Since the warm-plasma 
approximation provides additional aspects about the influence of a medium on ray trajectories around 
compact objects, it is worthwhile to explore further systems with this type of plasma.

For analytically calculating the influence of a warm plasma on the shadow we had to restrict to
the spherically symmetric and static case. For extending these calculations to rotating black holes 
one has to determine the necessary and sufficient conditions for separability of the 
Hamilton-Jacobi equation for light rays in a warm plasma.  
In a cold plasma, these conditions were found on the Kerr spacetime by Perlick and Tsupko 
\cite{Perlick-Tsupko-2017} and on an arbitrary axially symmetric and stationary spacetime by 
Bezd{\v{e}}kov{\'a} et al. \cite{Bezdekova-2022}. Another important future goal is to calculate 
the influence of a warm plasma on image distortion, magnification and distance measures. For 
this purpose, it would be necessary to derive generalized Sachs equations for light bundles 
in such a medium. In a cold plasma, the corresponding results have been worked out by 
Schulze-Koops et al. \cite{Schulze-Koops-Perlick-Schwarz-2017}, also see S{\'a}ren{\'y} 
and Balek \cite{Sareny-2019}.

\begin{acknowledgments}
We would like to thank Oleg Tsupko for helpful discussions.
\end{acknowledgments}

\begin{widetext}

\appendix
\section{The dispersion relation for light propagation in a warm plasma}
\label{sec:warmderiv}

We work on Minkowski spacetime in standard inertial coordinates 
$x=(x^0=ct,\vec{x})$. Einstein's summation convention is used for 
greek indices $\mu , \nu, \rho , {\, \dots} = 0,1,2,3$ and latin indices 
$i, j, k,{\, \dots 1,2,3}$. We use SI units, where the vacuum speed of 
light is given as $c^2 = (\mu _o \varepsilon _0 )^{-1}$ with 
$\varepsilon _0$ and $\mu _0$ denoting the permittivity and the 
permeability of vacuum, respectively.  We, frequently and tacitly,  
lower (greek) indices with the Minkowski metric $(\eta _{\mu \nu}) 
= \mathrm{diag} (-1,1,1,1)$ and we raise them with the inverse 
Minkowski metric $(\eta ^{\mu \nu}) = \mathrm{diag} (-1,1,1,1)$.    

To model a plasma, we consider Maxwell's equations with the vacuum 
constitutive law and a two-fluid source, one fluid modelling the ions
and the other fluid modelling the electrons. With the electromagnetic 
field tensor (Faraday tensor) denoted $F_{\mu \nu}$, 
Maxwell's equations read
\begin{equation}
\partial _{[\mu} F_{\nu \sigma ]} (x) = 0 \, , \quad
\partial _{\mu} F ^{\mu \nu} (x)= \mu _0
\Big( j^{\nu} (x) + J^{\nu} (x) \Big) \, .
\label{eq:Maxwell}
\end{equation}
Here $j^{\nu}$ and $J^{\nu}$ denotes the 4-current density 
of the electron fluid and of the ion fluid, respectively.

Using kinetic theory, we model the electron fluid in terms of a 
distribution function $f$ on the one-particle phase space which 
is parametrized by the coordinates 
$(x^0,x^1,x^2,x^3,p_0,p_1,p_2,p_3)
=(x^0=ct, \vec{x}, p_0, \vec{p\,})$. 
As the electrons are restricted to the mass shell
\begin{equation}
p_{\mu}p^{\mu} = -m^2 c^2 
\end{equation}
where $m$ denotes the electron mass,
we have for future-oriented trajectories
\begin{equation}
-p_0 = p^0  = \sqrt{m^2 c^2 + \mathrm{p}^2} \, , \quad
\mathrm{p} = \sqrt{\big| \vec{p \,} \big| ^2} \, .
\end{equation}
Therefore, $f$ is a function of $(x, \vec{p\,})=(x^0,\vec{x}, \vec{p\, })$ 
only. The 4-current density of the electron fluid can be written as
\begin{equation}
j^{\nu} (x) = q \bigintssss _{\mathbb{R}{}^3} 
f ( x , \vec{p\,} ) \, \dfrac{p^{\nu}}{m} \, 
\dfrac{d^3 \vec{p\,}}{\sqrt{1+ \mathrm{p}{}^2/(m^2c^2)}},
\label{eq:f}
\end{equation}
and $f$ is normalized such that
\begin{equation}
N (x^0) 
= 
\bigintssss _{\mathcal{V}} \bigintssss_{\mathbb{R}^3}
f(x, \vec{p\,} ) 
\sqrt{1 + \mathrm{p}^2/(m^2c^2)} \, 
\dfrac{d^3 \vec{p\,} \, d^3 \vec{x}}{\sqrt{1+ \mathrm{p}{}^2/(m^2c^2)}}
=
\bigintssss _{\mathcal{V}} \bigintssss_{\mathbb{R}^3}
f(x^0,\vec{x}, \vec{p\,} ) \,
d^3 \vec{p\,} \,  d^3 \vec{x} 
\label{eq:N}
\end{equation} 
gives the number of electrons at time $t = x^0/c$ in the volume 
$\mathcal{V}$. Here $d^3 \vec{p} / \sqrt{1 + \mathrm{p}^2/(m^2c^2)}$ 
is the invariant volume form on the mass shell and the factor of 
$\sqrt{1 + \mathrm{p}^2/(m^2c^2)}$ in the numerator of (\ref{eq:N}) 
has to be applied because the worldline of a particle with momentum 
$\mathrm{p}$ crosses the hypersurface $x^0 = \mathrm{constant}$ 
in Minkowski spacetime at a corresponding angle.  
We assume that the temperature is low enough for the electron fluid 
to be considered as collisionless, so $f$ has to satisfy the collisionless
Boltzmann equation (also known as the Liouville equation or, for
the case of charged particles considered here, as the Vlasov equation),
\begin{equation}    
0 = \dfrac{df (x, \vec{p\,})}{dt} =
\dfrac{dx^{\mu}}{dt} 
\dfrac{\partial f (x , \vec{p\,})}{\partial x^{\mu}} 
+\dfrac{dp^j}{dt} 
\dfrac{\partial f (x, \vec{p\,})}{\partial p^j} 
\, .
\label{eq:Liouville0}
\end{equation}
Along each individual electron trajectory we have
\begin{equation}
\dfrac{dx^{\mu}}{dt} 
=
\dfrac{d\tau}{dt} 
\dfrac{dx^{\mu}}{d\tau} 
=
\dfrac{d\tau}{dt} 
\, \dfrac{p^{\mu}}{m} \, ,
\end{equation}
\begin{equation}
\dfrac{dp^{\mu}}{dt} 
=
\dfrac{d\tau}{dt} 
\dfrac{dp^{\mu}}{d\tau} 
=
\dfrac{d\tau}{dt} 
 \, q \, F^{\mu \nu} \dfrac{p_{\nu}}{m}
\end{equation}
where $\tau$ is proper time and $q(<0)$ is the electron charge. 

Now (\ref{eq:Liouville0}) can be rewritten as
\begin{equation}    
0 = p^{\mu} 
\dfrac{\partial f (x , \vec{p\,})}{\partial x^{\mu}} 
+q F^{j \rho} (x) p_{\rho} 
\dfrac{\partial f (x , \vec{p\,})}{\partial p^{j}} 
\, .
\label{eq:Liouville}
\end{equation}
The equations (\ref{eq:Maxwell}), (\ref{eq:f}), (\ref{eq:Liouville})
together with equations for the ion fluid analogous to  (\ref{eq:f})
and (\ref{eq:Liouville}) determine our dynamical system.

We want to consider the situation that we have a homogeneous 
background field with a perturbation given by a plane-harmonic wave, 
and we want to linearize all equations with respect to the perturbation. 
We mark the background field by an upper index $(0)$ and the 
perturbation by an upper index $(1)$. We assume that the background  
electromagnetic field vanishes, i.e., that the background plasma 
is non-magnetized. We further assume that the frequency of the
perturbation is so high that only the electrons are influenced by it
whereas the ions, whose inertia is much bigger, stay put. Hence
\begin{equation}
F_{\mu \nu} (x)= F_{\mu \nu} ^{(0)}
+ F_{\mu \nu} ^{(1)} (x)\, , \quad
F_{\mu \nu} ^{(0)} = 0 \, , \quad
F_{\mu \nu} ^{(1)} (x)= 
\mathcal{F}_{\mu \nu} e^{i k_{\rho}x^{\rho}}
\, ,
\end{equation}
\begin{equation}
j_{\mu} (x)= j_{\mu}^{(0)} + j_{\mu}^{(1)} (x)\, ,
\end{equation}
\begin{equation}
J_{\mu} = J_{\mu}{^{(0)}} + J_{\mu}^{(1)} \, , \quad  
J_{\mu}^{(1)} = 0   
\, .
\end{equation}
The electron distribution function is assumed to be of the form
\begin{equation}
f (x , \vec{p\,} ) = f ^{(0)} (\mathrm{p}) +
f ^{(1)} (x , \vec{p\,} ) \, ,
\end{equation}
with a background distribution $f^{(0)}$ that depends only on
$\mathrm{p} = \sqrt{| \vec{p \,} |^2}$. Then the zeroth order
equations require
\begin{equation}
J_{\mu}^{(0)} = -j_{\mu}^{(0)} = 
- q \, c \bigintssss _{\mathbb{R}^3} f^{(0)} ( \mathrm{p} )
\dfrac{p_{\mu} \, d^3 \vec{p\,}}{\sqrt{m^2 c^2 + \mathrm{p}^2}} \, ,
\label{eq:j0}
\end{equation}   
and the first order equations require
\begin{equation}
k_{[\mu}\mathcal{F}_{\nu \rho]} = 0 \, ,
\label{eq:kF}
\end{equation}
\begin{equation}
i k_{\mu} \mathcal{F}^{\mu \nu} e^{i k_{\sigma} x^{\sigma}}
= \mu _0 \, j^{\nu (1)} (x)= \mu _0 \, q \, c  
\bigintssss _{\mathbb{R}^3} f^{(1)} (x, \vec{p\,}) 
\dfrac{p^{\nu} \, d^3 \vec{p}}{\sqrt{m^2 c^2 + \mathrm{p}^2}} \, ,
\label{eq:F1}
\end{equation}   
\[
0 = p^{\mu} 
\dfrac{\partial f ^{(1)} (x , \vec{p\,})}{\partial x^{\mu}} 
+q \, \mathcal{F}^{j \rho} p_{\rho} 
\dfrac{d f ^{(0)}(\mathrm{p})}{d\mathrm{p}} \dfrac{p_j}{\mathrm{p}} 
e^{i k_{\sigma} x^{\sigma}}
\]
\begin{equation}
=p^{\mu} 
\dfrac{\partial f ^{(1)} (x , \vec{p\,})}{\partial x^{\mu}} 
-q \, \mathcal{F}^{j 0} p_j 
\dfrac{\sqrt{m^2c^2 + \mathrm{p}{}^2}}{\mathrm{p}} 
\dfrac{d f ^{(0)}(\mathrm{p})}{d\mathrm{p}} 
e^{i k_{\sigma} x^{\sigma}}
\, .
\end{equation}
Integration of the last equation yields
\begin{equation}
f ^{(1)} (x , \vec{p\,})
=
\dfrac{
-i \, q \, \mathcal{F}^{j 0} p_j \sqrt{m^2c^2+\mathrm{p}^2}
}{
k_{\rho}p^{\rho} \, \mathrm{p}
}
\dfrac{d f^{(0)} ( \mathrm{p})}{d \mathrm{p}}
e^{i k_{\sigma} x^{\sigma}}
\, .
\label{eq:f1}
\end{equation}
From (\ref{eq:F1}) and (\ref{eq:f1}) we find
\begin{equation}
k_{\mu} \mathcal{F}^{\mu \nu} 
= - \mu _0  \, q^2 \, c  
\bigintssss _{\mathbb{R}^3}  
\dfrac{d f^{(0)} ( \mathrm{p})}{d \mathrm{p}}
\,
\dfrac{
\mathcal{F}^{j 0} p_j 
}{
k_{\rho}p^{\rho} \, 
}
\dfrac{p^{\nu}}{\mathrm{p}} \, d^3 \vec{p} \, .
\label{eq:kF2}
\end{equation}
If $f^{(0)}$ is given, $\mathcal{F}{}_{\mu \nu}$ can be
determined from (\ref{eq:kF}) and (\ref{eq:kF2}). We will 
do this as far as possible with $f^{(0)}$ unspecified, before
choosing $f^{(0)}$ to be the J{\"u}ttner distribution. If
$\mathcal{F}{}_{\mu \nu}$ has been found,  (\ref{eq:F1}) 
determines $j^{\nu (1)}$, whereas $J_{\mu}^{(0)}$
and $j_{\mu}^{(0)}$ are given by (\ref{eq:j0}).

We now decompose (\ref{eq:kF}) and (\ref{eq:kF2})
into temporal and spatial parts, writing  
\begin{equation}
\mathcal{F}{}^{0j} = \dfrac{\mathcal{E}{}^j}{c} 
\, , \quad
\vec{\mathcal{E}} = (\mathcal{E}^1,\mathcal{E}^2,\mathcal{E}^3) 
\, , \quad
\mathcal{E} = \sqrt{\big| \vec{\mathcal{E}} \big| ^2}
\, ,
\end{equation}
\begin{equation}
\mathcal{F}{}^{\ell j} = \varepsilon ^{\ell j k} \mathcal{B}{}_k 
\, , \quad
\vec{\mathcal{B}} = (\mathcal{B}_1,\mathcal{B}_2,\mathcal{B}_3) \, ,
\, , \quad
\mathcal{B} = \sqrt{\big| \vec{\mathcal{B}} \big| ^2}
\, ,
\end{equation}
\begin{equation}
(k_0,k_1,k_2,k_3) = \big(-\omega /c , \vec{k\,} \big) \, , \quad
\mathrm{k} = \sqrt{\big| \vec{k \,} \big| ^2 }
\, ,
\end{equation}
where $\varepsilon ^{\ell j k}$ is the totally antisymmetric Levi-Civita
symbol. Then (\ref{eq:kF}) yields
\begin{equation}
\vec{k} \cdot \vec{\mathcal{B}} = 0 \, , \quad
\vec{k} \times \vec{\mathcal{E}} = \omega \, \vec{\mathcal{B}}
\, ,
\label{eq:kBE0}
\end{equation}
while the spatial part of (\ref{eq:kF2}) can be rewritten as
\begin{equation}
\dfrac{\omega}{c^2} \, \vec{\mathcal{E}} +
\vec{k} \times \vec{\mathcal{B}} 
= - \mu _0  \, q^2 \, c 
\bigintsss _{\mathbb{R}^3}  
\dfrac{d f^{(0)} ( \mathrm{p})}{d \mathrm{p}}
\,
\dfrac{
\big( \vec{\mathcal{E}} \cdot \vec{p\,} \big) \, \vec{p} \; d^3 \vec{p} 
}{
\Big( 
\omega \,  \sqrt{m^2c^2+ \mathrm{p}{}^2} 
-
c \, \vec{k} \cdot \vec{p} 
\Big) 
\mathrm{p}
}
\label{eq:kBE1}
\end{equation}
The temporal part of (\ref{eq:kF2}) gives no further information because
transvecting (\ref{eq:kF2}) with $k_{\nu}$ results in the identity $0=0$,
so $\vec{\mathcal{E}}$ and $\vec{\mathcal{B}}$ are determined
by (\ref{eq:kBE0}) and (\ref{eq:kBE1}). Taking the scalar product
of (\ref{eq:kBE1}) with $\vec{\mathcal{E}}$ and using (\ref{eq:kBE0}) 
results in 
\begin{equation}
\Bigg(  \dfrac{\omega}{c^2} - \dfrac{\mathrm{k} ^2}{\omega} \Bigg) 
\mathcal{E} ^2
+
\dfrac{1}{\omega} \big( \vec{k} \cdot \vec{\mathcal{E}\,} \big)^2 
= 
- \mu _0  \, q^2 \, c 
\bigintsss _{\mathbb{R}^3}  
\dfrac{d f^{(0)} ( \mathrm{p})}{d \mathrm{p}}
\,
\dfrac{
\big( \vec{\mathcal{E}} \cdot \vec{p\, } \big)^2 \, d^3 \vec{p} 
}{
\Big(
\omega \, \sqrt{m^2c^2+ \mathrm{p}{}^2} 
-
c \, \vec{k} \cdot \vec{p}
\Big) 
\mathrm{p}
}
\, .
\label{eq:kE1}
\end{equation}
We evaluate the integral in spherical polar coordinates with
\begin{equation}
\vec{k} = \begin{pmatrix} \, 0 \, \\ 0 \\ \mathrm{k} \end{pmatrix} 
\, , \quad
\vec{\mathcal{E}} = 
\mathcal{E} \, 
\begin{pmatrix} 
\, \mathrm{sin} \, \alpha \, \\ 0 \\ \, \mathrm{cos} \, \alpha \,  
\end{pmatrix} 
\, , \quad
\vec{p} = 
\mathrm{p} \, 
\begin{pmatrix} 
\, \mathrm{sin} \, \vartheta  \, \mathrm{cos} \, \varphi \,
\\
\, \mathrm{sin} \, \vartheta  \, \mathrm{sin} \, \varphi \,
\\ 
\, \mathrm{cos} \, \vartheta  \,   
\end{pmatrix} 
\, .
\end{equation}
Here setting $\alpha =0$ gives the longitudinal modes and setting $\alpha = \pi /2$ 
gives the transverse modes. As we linearized all equations with respect to the 
perturbations, a general wave is a linear combination of longitudinal and transverse
modes. With this representation in polar coordinates, (\ref{eq:kE1}) reads 
\[
\Bigg( \dfrac{\omega ^2}{c^2 \, \mathrm{k}{}^2} - 1 \Bigg) 
\mathcal{E} ^2
+
\mathcal{E}{}^2 \, \mathrm{cos}^2 \alpha
= 
\]
\begin{equation}
- \dfrac{\mu _0  \, q^2 \, c \, \omega }{k^2} 
\bigintsss _0 ^{\infty} \bigintsss _0 ^{\pi} \bigintsss _0 ^{2 \pi}  
\dfrac{d f^{(0)} ( \mathrm{p})}{d \mathrm{p}}
\,
\dfrac{\mathcal{E}{}^2 
\Big( 
\mathrm{sin} \, \alpha \, \mathrm{sin} \, \vartheta \, \mathrm{cos} \, \varphi
+ 
\mathrm{cos} \, \alpha \, \mathrm{cos} \, \vartheta
\Big) ^2
\, \mathrm{sin} \, \vartheta \, \mathrm{p}{}^3 
d \varphi \, d \vartheta \, d \mathrm{p} 
}{
\Big(
\omega \, \sqrt{m^2c^2+ \mathrm{p}{}^2} 
-
c \, \mathrm{k}  \, \mathrm{p} \, \mathrm{cos} \, \vartheta
\Big) 
}
\, .
\label{eq:kE2}
\end{equation}
By (\ref{eq:kBE0}), $\mathcal{E}=0$ implies $\mathcal{B} = 0$; as 
we exclude, of course, the case that there is no electromagnetic wave,
we may divide (\ref{eq:kE2}) by $\mathcal{E}{}^2$ without losing any 
information. After performing the integration over $\varphi$ this equation 
then reduces to
\begin{equation}
\dfrac{\omega ^2}{c^2 \, \mathrm{k}{}^2} 
-
\mathrm{sin} ^2 \alpha 
= 
- \dfrac{\mu _0  \, q^2 \, \pi}{k^2 \, m} 
\bigintss _0 ^{\infty}    
\dfrac{d f^{(0)} ( \mathrm{p})}{d \mathrm{p}}
\bigintss _0 ^{\pi}
\dfrac{
\Big( 
\mathrm{sin} ^2 \alpha 
+
(2 \, - \, 3 \, \mathrm{sin} ^2 \alpha ) \mathrm{cos} ^2 \vartheta
\Big)
\, \mathrm{sin} \, \vartheta \
d \vartheta  
}{
\Big(
 \mathrm{cosh} \, w( \mathrm{p} )
-
\dfrac{c \, \mathrm{k} }{\omega}  \, \mathrm{sinh} \, w( \mathrm{p} )
\, \mathrm{cos} \, \vartheta 
\Big) 
}
\, \mathrm{p}{}^3 \, d \mathrm{p},
\label{eq:kE4}
\end{equation}
where $w ( \mathrm{p} )$ is defined by the equation
\begin{equation}
\mathrm{sinh} \, w( \mathrm{p} ) = \dfrac{\mathrm{p}}{m \, c} \, .
\label{eq:defw}
\end{equation}
With the substitution $\xi = \mathrm{cos} \, \vartheta$ this can be rewritten as
\begin{equation}
\dfrac{\omega ^2}{c^2 \, \mathrm{k}{}^2} 
-
\mathrm{sin} ^2 \alpha 
= 
- \dfrac{\mu _0  \, q^2 \, \pi \, m^2 \, c^3}{k^2} 
\bigintss _0 ^{\infty}    
\dfrac{d f^{(0)} ( \mathrm{p})}{d \mathrm{p}}
\bigintss _{-1} ^1
\dfrac{
\Big( 
\mathrm{sin} ^2 \alpha 
+
(2 \, - \, 3 \, \mathrm{sin} ^2 \alpha ) \xi ^2
\Big)
\, d \xi  
}{
\Big(
\mathrm{cosh} \, w( \mathrm{p} )
-
\dfrac{c \, \mathrm{k} }{\omega}  \, \mathrm{sinh} \, w( \mathrm{p} )
\, \xi 
\Big) 
}
\, \mathrm{sinh}{}^3 w ( \mathrm{p} ) \, d \mathrm{p}.
\label{eq:kE5}
\end{equation}
With
\begin{equation}
\bigintssss _{-1} ^1\dfrac{d \xi}{b-a \xi} =\dfrac{1}{a} 
\, \mathrm{ln} \dfrac{b+a}{b-a} \, ,
\quad
\bigintssss _{-1} ^1\dfrac{\xi ^2 \, d \xi}{b-a \xi} =-\dfrac{b}{a^3} 
\Bigg(2\, a-b \, \mathrm{ln} \dfrac{b+a}{b-a} \Bigg)
\, ,
\end{equation}
integration over $\xi$ results in
\[
\dfrac{k^3}{\mu _0  \, q^2 \, \pi \, m^2 \, c^2\, \omega}
\Bigg(
\dfrac{\omega ^2}{c^2 \, \mathrm{k}{}^2} 
-
\mathrm{sin} ^2 \alpha 
\Bigg)
= 
\]
\[
2 \, (2- 3 \, \mathrm{sin} ^2 \alpha ) \, 
\dfrac{\omega}{c \, \mathrm{k}}
\bigintsss _0 ^{\infty}    
\dfrac{d f^{(0)} ( \mathrm{p})}{d \mathrm{p}}
\mathrm{cosh} \, w ( \mathrm{p} ) \,
\mathrm{sinh} \,  w ( \mathrm{p} ) \, dp
\]
\begin{equation}
+ 
\bigintsss _0 ^{\infty}    
\dfrac{d f^{(0)} ( \mathrm{p})}{d \mathrm{p}}
\Bigg(
\, \mathrm{sinh} ^2 w ( \mathrm{p} ) \, \mathrm{sin} ^2 \alpha 
-
\dfrac{\omega ^2}{c^2 \, \mathrm{k} ^2}
( 2 - 3 \, \mathrm{sin} ^2 \alpha ) \, \mathrm{cosh} ^2 w ( \mathrm{p} ) 
\Bigg)
\,
\Bigg(
\mathrm{ln}
\dfrac{ 
1 + \frac{c \, \mathrm{k}}{\omega} 
 \, \mathrm{tanh} \, w ( \mathrm{p} )
}{
1 - \frac{c \, \mathrm{k}}{\omega} 
 \, \mathrm{tanh} \, w ( \mathrm{p} )
}
\Bigg)
\, d \mathrm{p}.
\label{eq:kE6}
\end{equation}
We now specify $f^{(0)}$ to be the J{\"u}ttner distribution,
\begin{equation}
f^{(0)} ( \mathrm{p} ) =
\dfrac{
\nu ^{(0)} \, \beta \, e^{- \beta \sqrt{1 + \mathrm{p}^2 / (m^2 c^2 )}}
}{ 
4 \pi m^3 c^3 K_2 ( \beta )
}
 \, , \quad
\beta = \dfrac{m \, c^2}{k_B T},
\label{eq:Juettner}
\end{equation}
where $T$ is the temperature, $k_B$ is the Boltzmann constant and 
\begin{equation}
K_s ( \beta ) = \dfrac{ 2^{s-1} \, (s-1)! \, \beta ^s}{(2s-1)!}
\int _0 ^{\infty} e^{- \beta \, \mathrm{cosh} \, u} 
\mathrm{sinh}^{2s} u \, du
\label{eq:defKs}
\end{equation}
is the modified Bessel function of the second kind. From  the normalization 
condition (\ref{eq:N}) we find that the constant 
\begin{equation}
\nu ^{(0)} = 4 \pi \int _0 ^{\infty} f^{(0)} ( \mathrm{p} )
\, \mathrm{p}^2 d \mathrm{p}
\end{equation}
is the number density of electrons in the background state. 
Differentiating (\ref{eq:Juettner}) yields
\begin{equation}
\dfrac{df^{(0)} ( \mathrm{p} )}{d \mathrm{p}} = 
\dfrac{-
\nu ^{(0)} \, \beta^2 \, \mathrm{p}}
{
4 \pi m^5 c^5 K_2 ( \beta ) \sqrt{1 + \mathrm{p}^2 / (m^2 c^2 )}
}
\,
e^{- \beta \, \sqrt{1 + \mathrm{p}^2/(m^2c^2)}} 
\, .
\end{equation}
By inserting this expression into (\ref{eq:kE6}) we find
\[
\dfrac{4 \, \, k^3 \, m \, c \, K_2 ( \beta )}{\mu _0  \, q^2 \, \, \omega \, \nu ^{(0)} \, \beta^2}
\,
\Bigg(
\dfrac{\omega ^2}{c^2 \, \mathrm{k}{}^2} 
-
\mathrm{sin} ^2 \alpha 
\Bigg)
= 
\]
\[
-
2 \, (2- 3 \, \mathrm{sin} ^2 \alpha ) \, 
\dfrac{\omega}{c \, \mathrm{k}}
\bigintsss _0 ^{\infty}    
e^{- \beta \, \mathrm{cosh} \,  w} 
\,
\mathrm{sinh} ^2  w  \, \mathrm{cosh} \, w \,  dw
\]
\[
+
\mathrm{sin} ^2 \alpha 
\bigintsss _0 ^{\infty}    
\!\!\!\!
e^{- \beta \, \mathrm{cosh} \, w} 
\,
\Bigg(
\mathrm{ln}
\dfrac{ 
1 + \frac{c \, \mathrm{k}}{\omega} 
 \, \mathrm{tanh} \, w
}{
1 - \frac{c \, \mathrm{k}}{\omega} 
 \, \mathrm{tanh} \, w
}
\Bigg)
\, \mathrm{sinh} ^3 w \,    d w
\]
\begin{equation}
+
\dfrac{\omega ^2}{c^2 \, \mathrm{k} ^2}
\big( 2 - 3 \mathrm{sin} ^2 \alpha \big) 
\bigintsss _0 ^{\infty}    
\!\!\!\!
e^{- \beta \, \mathrm{cosh} \, w} 
\,
\Bigg(
\mathrm{ln}
\dfrac{ 
1 + \frac{c \, \mathrm{k}}{\omega} 
 \, \mathrm{tanh} \, w
}{
1 - \frac{c \, \mathrm{k}}{\omega} 
 \, \mathrm{tanh} \, w
}
\Bigg)
\mathrm{cosh} ^2 w \, \mathrm{sinh} \, w \,   d w
\end{equation}
If we introduce the plasma frequency
\begin{equation} 
\omega _p ^2 = \dfrac{\mu _0 \, c^2 \, q^2 \, \nu ^{(0)}}{m} 
\end{equation}
and the index of refraction
\begin{equation}
n = \dfrac{c \, k}{\omega}
\, ,
\end{equation}
this equation reads
\[
\dfrac{4 \, n^3 K_2(\beta) \omega ^2}{\beta ^2 \omega _p^2}
\, (1 -n^2 \, \mathrm{sin}^2 \alpha )
=
n^2 \, \mathrm{sin} ^2 \alpha 
\bigintssss _0 ^{\infty} e^{-\beta \, \mathrm{cosh} \, w}
\Big( \mathrm{ln} 
\dfrac{1+n \, \mathrm{tanh} \, w}{1-n \, \mathrm{tanh} \, w}
\Big)
\big( \mathrm{cosh}^2 w - 1 \big) \, \mathrm{sinh} \, w  \, d w
\]
\[
+
(2- 3 \, \mathrm{sin} ^2 \alpha )
\bigintssss _0 ^{\infty}
e^{-\beta \, \mathrm{cosh} \, w}
\Big( \mathrm{ln}
\dfrac{1+n \, \mathrm{tanh} \, w}{1-n \, \mathrm{tanh} \, w}
- 2 \, n \, \mathrm{tanh} \, w
\Big)
\mathrm{sinh} \, w \, \mathrm{cosh}^2 w 
\, dw
\]
\[
=
- n^2 \, \mathrm{sin} ^2 \alpha 
\bigintssss _0 ^{\infty} e^{-\beta \, \mathrm{cosh} \, w}
\Big( \mathrm{ln} 
\dfrac{1+n \, \mathrm{tanh} \, w}{1-n \, \mathrm{tanh} \, w}
\Big)
\mathrm{sinh} \, w  \, d w
\]
\[
+ 
\bigintssss _0 ^{\infty}
e^{-\beta \, \mathrm{cosh} \, w}
\Bigg( \big( 2- 3 \, \mathrm{sin} ^2 \alpha \big)
\Big( \mathrm{ln}
\dfrac{1+n \, \mathrm{tanh} \, w}{1-n \, \mathrm{tanh} \, w}
- 2 \, n \, \mathrm{tanh} \, w
\Big)
+
n^2 \, \mathrm{sin} ^2 \alpha 
\Big( \mathrm{ln}
\dfrac{1+n \, \mathrm{tanh} \, w}{1-n \, \mathrm{tanh} \, w}
\Big)
\Bigg)
\mathrm{sinh} \, w \, \mathrm{cosh}^2 w 
\, dw
\]
\[
=
 \dfrac{n^2}{\beta} \, \mathrm{sin} ^2 \alpha 
\bigintssss _0 ^{\infty} \dfrac{d e^{-\beta \, \mathrm{cosh} \, w}}{dw}
\Big( \mathrm{ln} 
\dfrac{1+n \, \mathrm{tanh} \, w}{1-n \, \mathrm{tanh} \, w}
\Big)
d w
\]
\[
-
\dfrac{d^2}{d \beta ^2} 
\left(
\dfrac{1}{\beta}
\bigintssss _0 ^{\infty}
\dfrac{de^{-\beta \, \mathrm{cosh} \, w}}{dw}
\Bigg( \big( 2- 3 \, \mathrm{sin} ^2 \alpha \big)
\Big( \mathrm{ln}
\dfrac{1+n \, \mathrm{tanh} \, w}{1-n \, \mathrm{tanh} \, w}
- 2 \, n \, \mathrm{tanh} \, w
\Big)
+
n^2 \, \mathrm{sin} ^2 \alpha 
\Big( \mathrm{ln}
\dfrac{1+n \, \mathrm{tanh} \, w}{1-n \, \mathrm{tanh} \, w}
\Big)
\Bigg)
dw
\right)
\]
\[
=
- \dfrac{n^2}{\beta} \, \mathrm{sin} ^2 \alpha 
\bigintssss _0 ^{\infty} 
\dfrac{
e^{-\beta \, \mathrm{cosh} \, w} \, 2 \, n \, dw
}{
\big( 1 - n^2 \mathrm{tanh}^2 w \big) \, \mathrm{cosh}^2 w
}
\]
\begin{equation}
+
\dfrac{d^2}{d \beta ^2} 
\left(
\dfrac{1}{\beta}
\bigintssss _0 ^{\infty}
e^{-\beta \, \mathrm{cosh} \, w}
\Bigg( \big( 2- 3 \, \mathrm{sin} ^2 \alpha \big)
\Big(
\dfrac{2 \, n }{
\big( 1 - n^2 \mathrm{tanh}^2 w \big) \, \mathrm{cosh}^2 w
}
- \dfrac{2 \, n}{\mathrm{cosh} ^2 w}
\Big)
+ 
\dfrac{
2 \, n^3 \, \mathrm{sin} ^2 \alpha 
}{
\big( 1 - n^2 \mathrm{tanh}^2 w \big) \, \mathrm{cosh}^2 w
}
\Bigg)
dw
\right)
\, ,
\end{equation}
hence
\[
\dfrac{2 \, n^2 \, K_2(\beta) \omega ^2}{\beta ^2 \omega _p^2}
\, (1 -n^2 \, \mathrm{sin}^2 \alpha )
=
- \dfrac{n^2}{\beta}  \, \mathrm{sin} ^2 \alpha
\bigintssss _0 ^{\infty} 
\dfrac{
e^{-\beta \, \mathrm{cosh} \, w} \, dw
}{
\mathrm{cosh}^2 w - n^2 \mathrm{sinh}^2 w}
\]
\begin{equation}
+
\bigintsss _0 ^{\infty}
\dfrac{
e^{-\beta \, \mathrm{cosh} \, w}
\Big(
\dfrac{1}{\beta} \mathrm{cosh} ^2 w
+ \dfrac{2}{\beta ^2} \mathrm{cosh} \, w
+ \dfrac{2}{\beta^3}
\Big)
\Big(
\big( 2- 3 \, \mathrm{sin} ^2 \alpha \big) \,  n^2 \, \mathrm{tanh}^2 w
+  n^2 \, \mathrm{sin} ^2 \alpha 
\Big) \, dw
}{
\mathrm{cosh}^2 w - n^2 \mathrm{sinh}^2 w}
\, .
\end{equation}
After dividing by $n^2/\beta$, this reduces to 
\[
\dfrac{2 \, K_2(\beta) \omega ^2}{\beta \, \omega _p^2}
\, (1 -n^2 \, \mathrm{sin}^2 \alpha )
=
- \mathrm{sin} ^2 \alpha
\bigintssss _0 ^{\infty} 
\dfrac{
e^{-\beta \, \mathrm{cosh} \, w} \, dw
}{
1 + (1-n^2)\, \mathrm{sinh}^2 w
}
\]
\[
+
\bigintsss _0 ^{\infty}
\dfrac{
e^{-\beta \, \mathrm{cosh} \, w}
\Big(
\mathrm{cosh} ^2 w
+ \dfrac{2}{\beta} \mathrm{cosh} \, w
+ \dfrac{2}{\beta^2}
\Big)
\Big(
\big( 2- 3 \, \mathrm{sin} ^2 \alpha \big) \, \mathrm{tanh}^2 w
+ \mathrm{sin} ^2 \alpha 
\Big) \, dw
}{
1 + (1-n^2)\, \mathrm{sinh}^2 w
}
\]
\[
=
2 \, \mathrm{cos} ^2 \alpha 
\bigintsss _0 ^{\infty}
\dfrac{
e^{-\beta \, \mathrm{cosh} \, w}
\Big(
\mathrm{cosh} ^2 w
+ \dfrac{2}{\beta} \mathrm{cosh} \, w
+ \dfrac{2}{\beta^2}
\Big)
\, \mathrm{tanh}^2 w
\, dw
}{
1 + (1-n^2)\, \mathrm{sinh}^2 w
}
\]
\[
+
\mathrm{sin} ^2 \alpha 
\bigintsss _0 ^{\infty}
\dfrac{
e^{-\beta \, \mathrm{cosh} \, w}
\Bigg(
-1 +\big(1-\mathrm{tanh}^2 w \big) 
\Big(
\mathrm{cosh} ^2 w
+ \dfrac{2}{\beta} \mathrm{cosh} \, w
+ \dfrac{2}{\beta^2}
\Big)
\Bigg) \, dw
}{
1 + (1-n^2)\, \mathrm{sinh}^2 w
}
\]
\[
=
2 \, \mathrm{cos} ^2 \alpha 
\bigintsss _0 ^{\infty}
\dfrac{
e^{-\beta \, \mathrm{cosh} \, w}
\Big(
\mathrm{cosh} ^2 w
+ \dfrac{2}{\beta} \mathrm{cosh} \, w
+ \dfrac{2}{\beta^2}
\Big)
\, \mathrm{sinh}^2 w
\, dw
}{
\Big( 1 + (1-n^2)\, \mathrm{sinh}^2 w \Big)
\, \mathrm{cosh}^2 w
}
\]
\begin{equation}
+
\dfrac{2}{\beta} \, \mathrm{sin} ^2 \alpha 
\bigintsss _0 ^{\infty}
\dfrac{
e^{-\beta \, \mathrm{cosh} \, w}
\Big(
\mathrm{cosh} \, w
+ \dfrac{1}{\beta}
\Big)
\, dw
}{
\Big( 1 + (1-n^2)\, \mathrm{sinh}^2 w \Big)
\, \mathrm{cosh}^2 w  
}
\, .
\label{eq:exactn}
\end{equation}
This gives us, in implicit form, an exact expression for the index of 
refraction $n$ as a function of $\omega$. 

From now on it is convenient to work, rather than with $\beta$, with its inverse
\begin{equation}
    \chi = \beta ^{-1} = \dfrac{k_B T}{m \, c^2} \, .
\end{equation}
Then (\ref{eq:exactn}) simplifies for longitudinal modes ($\alpha = 0$) to
\begin{equation}
\chi \, K_2(\chi ^{-1}) \, \dfrac{\omega ^2}{\omega _p^2}
=
\bigintsss _0 ^{\infty}
\dfrac{
e^{- (\chi ^{-1} \mathrm{cosh} \, w )}
\Big(
\mathrm{cosh} ^2 w
+ 2 \, \chi \, \mathrm{cosh} \, w
+ 2 \, \chi ^2
\Big)
\, \mathrm{sinh}^2 w
\, dw
}{
\Big( 1 + (1-n^2)\, \mathrm{sinh}^2 w \Big)
\, \mathrm{cosh}^2 w
}
\label{eq:exlong}
\end{equation}
and for transverse modes ($\alpha = \pi /2$) to 
\begin{equation}
(1 -n^2 ) \, K_2(\chi ^{-1}) \, \dfrac{\omega ^2}{\omega _p^2}
= 
\bigintsss _0 ^{\infty}
\dfrac{
e^{-( \chi ^{-1} \mathrm{cosh} \, w )}
\Big(
\mathrm{cosh} \, w
+ \chi
\Big)
\, dw
}{
\Big( 1 + (1-n^2)\, \mathrm{sinh}^2 w \Big)
\, \mathrm{cosh}^2 w  
}
\, .
\label{eq:extrans}
\end{equation}
Note that for $\chi \to 0$, (\ref{eq:extrans}) gives indeed the index
of refraction of a cold plasma, $n^2=1- \omega _p^2/\omega ^2$, as can be verified with the help of asymptotic formulas for the modified Bessel function.

The index of refraction $n$ can take values only between 0 and 1, because otherwise
the integrals in (\ref{eq:exlong}) and (\ref{eq:extrans}) do not converge. For the 
transverse modes, according to (\ref{eq:extrans}) the index of refraction 
increases monotonically from  0 to 1 if the frequency $\omega$ increases from a 
cut-off frequency $\omega _{co} (\chi)$ to infinity, where $\omega _{co}$ 
is given by 
\begin{equation}
\omega _{co} (\chi )^2 = \dfrac{\omega _p^2}{K_2 (\chi ^{-1} )}
\int _0 ^{\infty} 
\dfrac{
e^{-(\chi ^{-1} \mathrm{cosh} \, w)}
\Big( \mathrm{cosh} \, w + \chi \Big) \, dw
}{  
\mathrm{cosh} ^4 w
}
\, .
\label{eq:exactco}
\end{equation}
Eqs. (\ref{eq:exlong}) and (\ref{eq:extrans}) are exact for all $\chi$ (but 
note that we have assumed that the electron fluid is collisionless which will
not be justified for very high temperatures unless the density is very low).
 
If we want to go beyond the approximation of a cold plasma, but still
assuming that the temperature is not too high, we may linearize the relation 
between $n$ and $\omega$ with respect to $\chi$. To that end we substitute 
the integration variable $w$ by a new one, $z$, defined by    
\begin{equation} 
\mathrm{cosh} \, w  = 1 + \chi \, z \, , \quad
\mathrm{sinh} \, w  = \sqrt{2 \, \chi \, z} \, \sqrt{1+ \chi \, \dfrac{z}{2}} \, , \quad
\mathrm{sinh} \, w \, dw = \chi \, dz
\, .
\end{equation}
Then (\ref{eq:exactn}) takes the following form:
\[
\chi \, K_2(\chi ^{-1}) \dfrac{\omega ^2}{\omega _p^2}
\, (1 -n^2 \, \mathrm{sin}^2 \alpha )
=
\mathrm{cos} ^2 \alpha 
\bigintsss _0 ^{\infty}
\dfrac{
e^{-(\chi^{-1})} \, e^{-z} 
\Big(
1+ 2 \, \chi \, z + 2 \, \chi \Big)
\, \sqrt{2z} \, \sqrt{1+ \chi \, \dfrac{z}{2}}\, \sqrt{\chi}^{\, 3}
\Big( 1 + O \big( \chi ^2 \big) \Big) \, dz
}{
\Big( 1 + 2 \, (1-n^2)\, \chi \, z \Big)
\, \Big( 1+ 2 \, \chi \, z \Big) 
}
\]
\begin{equation}
+
\mathrm{sin} ^2 \alpha 
\bigintsss _0 ^{\infty}
\dfrac{
e^{-( \chi ^{-1})}  \, e^{-z} 
\Big(
1+ \chi \, z
+ \chi
\Big)
\, \sqrt{\chi}^{\, 3} \, \Big( 1 + O \big( \chi ^2 \big) \Big) \, dz
}{
\Big( 1 + 2 \, (1-n^2)\, \chi \, z \Big)
\, \Big( 1 + 2 \, \chi \, z \Big) 
\, \sqrt{2z} \, \sqrt{1+ \chi \, \dfrac{z}{ 2 }}
}
\, ,
\end{equation}
hence
\[
\dfrac{e^{-(\chi ^{-1})} \, K_2(\chi ^{-1}) \, \sqrt{2} \, \omega ^2}{\sqrt{\chi} \, \omega _p^2}
\, (1 -n^2 \, \mathrm{sin}^2 \alpha )
=
2 \, \mathrm{cos} ^2 \alpha 
\bigintsss _0 ^{\infty}
e^{-z} 
\Big(
1+ 2 \, \chi \, z + 2 \chi 
+ \chi \, \dfrac{z}{4}  - 2 \, (1-n^2)\, \chi \, z \
- 2 \, \chi \, z + O \big( \chi ^2 \big)\Big)
\, \sqrt{z} \, dz
\]
\begin{equation}
+
\mathrm{sin} ^2 \alpha 
\bigintsss _0 ^{\infty}
e^{-z} 
\Big(
1+ \chi \, z + \chi 
- 2 \, (1-n^2)\, \chi \, z- 2\, \chi \, z
-\chi \, \dfrac{z}{ 4}+ O \big( \chi ^2 \big) 
\Big) \, \dfrac{dz}{\sqrt{z}}
\end{equation}
With the asymptotic series expansion of the modified Bessel functions
\begin{equation}
K_s ( \chi ^{-1} ) = \sqrt{\dfrac{\pi \, \chi }{2}}\, e^{- (\chi ^{-1})} 
\Big( 1 + \dfrac{4 s^2 - 1}{8} \, \chi  + O ( \chi ^2 ) \Big) \, ,
\label{eq:asyKs}
\end{equation}
this can be rewritten as
\[
\dfrac{\sqrt{\pi} \, \omega ^2}{\omega _p^2}
\, \big(1 -n^2 \, \mathrm{sin}^2 \alpha \big)
\Big( 1 + \dfrac{15}{8} \, \chi \Big) + O ( \chi  ^2 ) 
\]
\begin{equation}
=
2 \, \mathrm{cos} ^2 \alpha  
\Bigg( \Big( 1 + 2 \, \chi  \Big) \dfrac{\sqrt{\pi}}{2}
+ \chi  \Big(2 \, n^2  
-\dfrac{7}{4} 
\Big) \, \dfrac{3 \, \sqrt{\pi}}{4} \Bigg)
+
\mathrm{sin} ^2 \alpha 
\Bigg( 
\Big(1 + \chi \Big) \, \sqrt{\pi}
+ \chi  \Big(2 \, n^2 - \dfrac{13}{4} \Big)
\dfrac{\sqrt{\pi}}{2} \Bigg)
\, ,
\end{equation}
hence
\begin{equation}
\dfrac{\omega ^2}{\omega _p^2}
\, \big(1 -n^2 \, \mathrm{sin}^2 \alpha \big)
\Big( 1 + \dfrac{15}{8} \, \chi  \Big) + O ( \chi ^2 ) 
=
\mathrm{cos} ^2 \alpha  
\Bigg(1 + \chi 
\Big( 2+3 \, n^2  
-\dfrac{21}{8} 
\Big) \Bigg)
+
\mathrm{sin} ^2 \alpha 
\Bigg( 
1 + \chi 
\Big( 1+ n^2 - \dfrac{13}{8} \Big)
\Bigg),
\end{equation}
and thus
\begin{equation}
\dfrac{\omega ^2}{\omega _p^2}
\, \big(1 -n^2 \, \mathrm{sin}^2 \alpha \big)
=
\mathrm{cos} ^2 \alpha  
\Bigg(1 + \chi
\Big( 3 \, n^2 - \dfrac{5}{2} 
\Big) \Bigg)
+
\mathrm{sin} ^2 \alpha 
\Bigg( 
1 + \chi 
\Big( n^2 - \dfrac{5}{2} \Big)
\Bigg)
+ O ( \chi ^2 ). 
\end{equation}
So, if linearized with respect to $\chi$, the relation between $n$ and 
$\omega$ reads for longitudinal modes ($\alpha = 0$)
\begin{equation}
\dfrac{\omega ^2}{\omega _p^2}
=
1 + \dfrac{1}{\beta}
\Big( 3 \, n^2 + \dfrac{5}{2} 
\Big) + O ( \chi ^2 ) 
\end{equation}
and for transverse modes ($\alpha = \pi /2$)
\begin{equation}
\dfrac{\omega ^2}{\omega _p^2}
\, \big(1 -n^2  \big)
=
1 + \chi
\Big( n^2 - \dfrac{5}{2} \Big)
+ O ( \chi ^2 )  \, .
\label{eq:nsolve}
\end{equation}
Solving the last equation for $1-n^2$ results in
\begin{equation}
1 -n^2 
=
\dfrac{\omega _p^2}{\omega ^2}
\Bigg( 1- \chi 
\Big( \dfrac{\omega _p^2}{\omega ^2} + \dfrac{3}{2} \Big)
\Bigg)
+ O ( \chi ^2 )  \, .
\end{equation}
Inserting this expression into the Hamiltonian
\begin{equation}
\mathcal{H} = \dfrac{1}{2} \left[ g^{\mu \nu} p_{\mu} p_{\nu}
+ \Big(1-n^2 \Big) \omega ^2 \right]   
\end{equation}
results in 
\begin{equation}
H=
\dfrac{1}{2} \Bigg\{ g^{\mu \nu} p_{\mu} p_{\nu}
+ \omega _p^2 \Bigg[ 1
- \chi \left( \dfrac{\omega _p^2}{\omega ^2} 
+ \dfrac{3 }{2} \right)
+ R(\chi )  \Bigg] \Bigg\},
\end{equation}
where $R(\chi)$ is a term of order $O(\chi ^2)$. Dropping 
this term gives us an approximation for the Hamiltonian $H$ that is 
correct to within linear approximation with respect to $\chi$ or,
what is the same, with respect to $\chi \, \omega _p^2/\omega ^2$.
The corresponding approximate index of refraction is 
\begin{equation}
n^2 
\approx
1-\dfrac{\omega _p^2}{\omega ^2} \Bigg(
1- \chi 
\Big( \dfrac{\omega _p^2}{\omega ^2} + \dfrac{3}{2} \Big)
\Bigg)
\, .
\label{eq:linn}
\end{equation}
In this approximation, the cut-off frequency $\omega _{co}$
(corresponding to $n=0$) is given by
\begin{equation}
\omega _{co} (\chi )^2  = \dfrac{\omega _p^2}{2}
\Bigg(
1-\dfrac{3}{2} \, \chi + \sqrt{1- 7 \chi  +\dfrac{9}{4} \, \chi ^2}
\Bigg)
\, .
\label{eq:linco}
\end{equation}
However, this is not the only way in which the linearization with respect 
to $\chi$ can be achieved. There is a certain ambiguity because the 
Hamiltonian can be multiplied with a function on phase space that is 
arbitrary except for the fact that it should have no zeros. Then the light
rays remain unchanged up to parametrization. E.g., we could switch to
the Hamiltonian 
\begin{equation}
\tilde{\mathcal{H}} = \Big( 1 + \dfrac{\omega _p ^2}{\omega ^2} \, \chi \Big) \mathcal{H}
\, .
\end{equation}
As (\ref{eq:nsolve}) can be rewritten as 
\begin{equation}
\big( 1 -n^2 \big) 
\Big( 1 + \dfrac{\omega _p^2}{\omega ^2} \, \chi\Big)
=
\dfrac{\omega _p^2}{\omega ^2}
\Big( 1- \dfrac{3}{2} \, \chi  \Big)
+ O ( \chi ^2 )  \, ,
\end{equation}
the linearization of the Hamiltonian $\tilde{\mathcal{H}}$ with respect to $\chi$ results in
\begin{equation}
\tilde{\mathcal{H}} = \dfrac{1}{2} \Bigg( 
\Big( 1 + \dfrac{\omega _p^2}{\omega ^2} \, \chi\Big)
g^{\mu \nu} p_{\mu} p_{\nu}
+\omega _p^2 
\Big( 1- \dfrac{3}{2} \, \chi \Big)
+ \tilde{R} ( \chi )
\Bigg)
\, ,
\label{eq:tildeH}
\end{equation}
where $\tilde{R} (\chi )$ is again a term of order $O(\chi ^2 )$,
but different from $R(\chi )$.  According to (\ref{eq:tildeH}) the 
original Hamiltonian $\mathcal{H}$ is now represented as
\begin{equation}
\mathcal{H} = \dfrac{1}{2} \Bigg(
g^{\mu \nu} p_{\mu} p_{\nu}
+\dfrac{
\omega _p^2 
\Big( 1- \dfrac{3}{2} \, \chi \Big)
}{
\Big( 1 + \dfrac{\omega _p^2}{\omega ^2} \, \chi \Big)
}
+ 
\dfrac{
\tilde{R} ( \chi )
}{
\Big( 1 + \dfrac{\omega _p^2}{\omega ^2} \, \chi \Big)
}
\Bigg)
\, .
\end{equation}
Dropping the $\tilde{R} (\chi )$ term gives an approximation that
is different from the one above. The corresponding approximate
index of refraction is now
\begin{equation}
n^2 \approx
\dfrac{1- \dfrac{\omega _p^2}{\omega ^2}
\Big( 1- \dfrac{5}{2} \, \chi  \Big)
}{
\Big( 1 + \dfrac{\omega _p^2}{\omega ^2} \, \chi \Big)
}
\, .
\label{eq:frn}
\end{equation}
In this approximation, the cut-off frequency $\omega _{co}$
is given by
\begin{equation}
\omega _{co} (\chi )^2  = \omega _p^2
\Bigg(
1-\dfrac{5}{2} \chi 
\Bigg)
\, .
\label{eq:frco}
\end{equation}
While the exact Hamiltonians $\mathcal{H}$ and $\tilde{\mathcal{H}}$ are equivalent in the
sense that they give the same light rays, just with different parameterizations,
linearization of $\mathcal{H}$ and linearization of $\tilde{\mathcal{H}}$ are non-equivalent, i.e.,
they give different light rays. This is mathematically obvious because the two
Hamiltonians are related by a factor that involves $\chi$. 

The question arises which of the two approximations is better. To that end we
compare the approximative formulas (\ref{eq:linco}) and (\ref{eq:frco}) for
the cut-off frequency with the exact formula (\ref{eq:exactco}), 
see Fig. \ref{fig:cutoff}. Of course, all three formulas give the same limit 
$\omega _{co} (\chi ) \to \omega _p$ for $\chi \to 0$ 
(i.e., for $T \to 0$). We see that the formula that arises
from linearizing $\tilde{\mathcal{H}}$ is better than the
one that arises from linearizing $\mathcal{H}$, although both
approximation formulas are good for sufficiently low temperatures.

\begin{figure}[h!]
  \centering
  \includegraphics[width=0.6\textwidth]{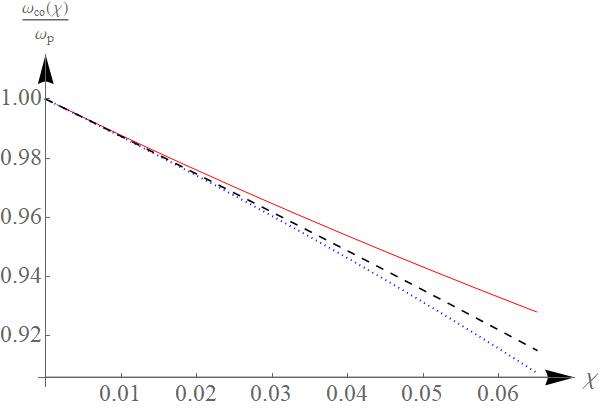}
  \caption{Cut-off frequency $\omega _{co} (\chi )$ according to the
  exact formula (\protect\ref{eq:exactco}) (solid, red), according to the 
  approximation (\protect\ref{eq:linco}) (dotted, blue) and according to the 
  approximation (\protect\ref{eq:frco}) (dashed, black).  
  }
  \label{fig:cutoff}
\end{figure}

This observation is corroborated by Fig. \ref{fig:nomega} where the 
index of refraction $n$ is plotted as a function of $\omega$ for the 
temperature of $\chi = 1/20$. Again, we see
that the formula based on the linearization of $\tilde{\mathcal{H}}$ is a better
approximation than the formula based on the linearization of $\mathcal{H}$.

\begin{figure}[h!]
  \centering
  \includegraphics[width=0.6\textwidth]{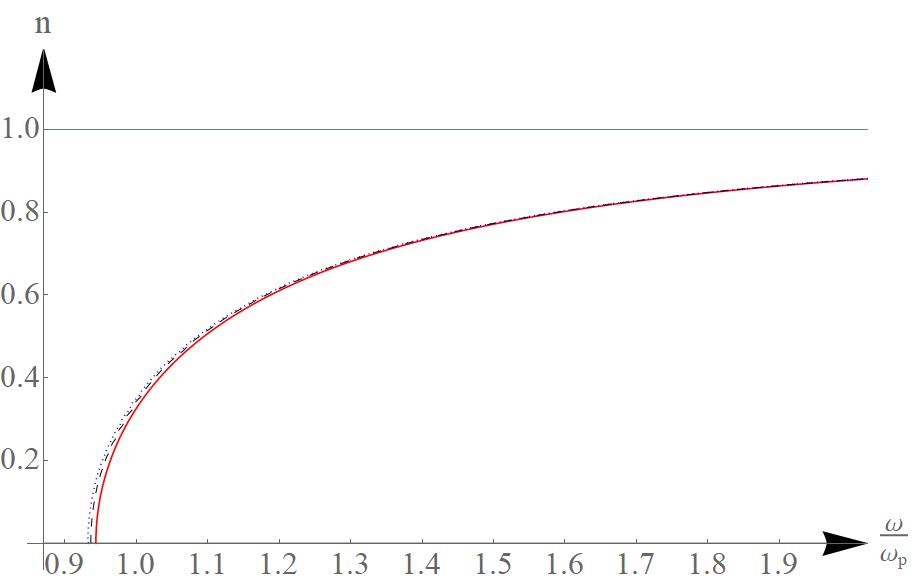}
  \caption{Index of refraction $n$ as a function of frequency $\omega$ for 
  $\chi = 1/20$ according to the
  exact formula (\protect\ref{eq:extrans}) (solid, red), according to the 
  approximation (\protect\ref{eq:linn}) (dotted, blue) and according to the 
  approximation (\protect\ref{eq:frn}) (dashed, black).  
  }
  \label{fig:nomega}
\end{figure}

We have derived here the formula (\ref{eq:frn}) for the index of refraction 
of a warm plasma on Minkowski spacetime, with a plasma of constant density 
and constant temperature which is at rest in an inertial system. 
In the body of the paper we use this formula at each cotangent space 
of a general-relativistic spacetime, where the four-velocity of 
the electron fluid is an arbitrary timelike vector field $V^{\mu}$,
$\omega = - V^{\mu} p_{\mu}$ and $\omega _p$ and $\chi$ are 
functions of the spacetime coordinates. This generalization is justified, 
provided that the principle of minimal coupling is applicable, i.e., 
provided that one does not postulate hypothetical couplings of the
plasma equations to the spacetime curvature.


\section{The energy-momentum tensor of the electron fluid}
\label{sec:energymom}

We consider the same situation and we use the same notation
as in Appendix \ref{sec:warmderiv}. 

The (kinetic) energy-momentum tensor of the electron fluid in the 
unperturbed state is 
\begin{equation}
T_{\mu \nu}^{(0)} =
c \int f^{(0)} ( \mathrm{p} ) \, 
\dfrac{p_{\mu} \, p_{\nu} \, d ^3 \vec{p}}{\sqrt{m^2c^2+ \mathrm{p}^2}}
\, .
\end{equation}
The integration over the angle coordinates can be carried
out:
\begin{equation}
T_{00}^{(0)} =
c \int _0^{\infty}  \int _0 ^{\pi} \int _0 ^{2 \pi} 
f^{(0)} ( \mathrm{p} ) \, \sqrt{m^2c^2+ \mathrm{p}^2}
\, 
\mathrm{sin} \, \vartheta \, \mathrm{p}^2 d \varphi \, d \vartheta 
\, d \mathrm{p}
= 4 \, \pi \, c \int _0^{\infty}   
f^{(0)} ( \mathrm{p} ) \, \sqrt{m^2c^2+ \mathrm{p}^2}
\, 
\mathrm{p}^2 \, d \mathrm{p}
\, ,
\end{equation}
\[
\big( T_{0i}^{(0)} \big) =
c \int _0^{\infty}  \int _0 ^{\pi} \int _0 ^{2 \pi} 
f^{(0)} ( \mathrm{p} ) \, 
\begin{pmatrix}
\mathrm{sin} \, \vartheta \, \mathrm{cos} \, \varphi
\\
\mathrm{sin} \, \vartheta \, \mathrm{sin} \, \varphi
\\
\mathrm{cos} \, \vartheta 
\end{pmatrix}
\, 
\mathrm{sin} \, \vartheta \, \mathrm{p}^3 d \varphi \, d \vartheta 
\, d \mathrm{p}
=
c \int _0^{\infty}  \int _0 ^{\pi}  
f^{(0)} ( \mathrm{p} ) \, 
\begin{pmatrix}
0
\\
0
\\
2 \, \pi \, \mathrm{cos} \vartheta 
\end{pmatrix}
\, 
\mathrm{sin} \, \vartheta \, \mathrm{p}^3 \, d \vartheta 
\, d \mathrm{p}
\]
\begin{equation}
=
2 \, \pi \, c \int _0^{\infty}  \int _{-1} ^{1}  
f^{(0)} ( \mathrm{p} ) \, 
\mathrm{p}^3  
\, u \, du \, d \mathrm{p}
\,
\begin{pmatrix}
0
\\
0
\\
1 
\end{pmatrix}
=
2 \, \pi \, c \int _0^{\infty}    
f^{(0)} ( \mathrm{p} ) \, 
\mathrm{p}^3  
\, d \mathrm{p}
\, \dfrac{u^2}{2} \Bigg| _{-1}^1 \, 
\begin{pmatrix}
0
\\
0
\\
1 
\end{pmatrix}
=
\begin{pmatrix}
0
\\
0
\\
0 
\end{pmatrix}
\end{equation}
\[
\big( T_{ij}^{(0)} \big) =
c \int _0^{\infty}  \int _0 ^{\pi} \int _0 ^{2 \pi} 
f^{(0)} ( \mathrm{p} ) \, 
\begin{pmatrix}
\mathrm{sin} ^2\vartheta \, \mathrm{cos} ^2 \varphi
&
\mathrm{sin}^2 \vartheta \, \mathrm{sin} \, \varphi \, \mathrm{cos} \, \varphi
&
\mathrm{sin} \vartheta \, \mathrm{cos} \, \vartheta \, \mathrm{cos} \, \varphi
\\
\mathrm{sin}^2 \vartheta \, \mathrm{sin} \, \varphi \, \mathrm{cos} \, \varphi
&
\mathrm{sin}^2 \vartheta \, \mathrm{sin} ^2 \varphi 
&
\mathrm{sin} \,  \vartheta \, \mathrm{sin} \, \varphi \, \mathrm{cos} \, \vartheta
\\
\mathrm{sin} \vartheta \, \mathrm{cos} \, \vartheta \, \mathrm{cos} \, \varphi
&
\mathrm{sin} \,  \vartheta \, \mathrm{sin} \, \varphi \, \mathrm{cos} \, \vartheta
&
\mathrm{cos} ^2 \vartheta 
\end{pmatrix}
\, 
\dfrac{
\mathrm{sin} \, \vartheta \, \mathrm{p}^4 d \varphi \, d \vartheta 
\, d \mathrm{p}
}{
\sqrt{m^2c^2+\mathrm{p}^2}
}
\]
\[
=
c \int _0^{\infty}  \int _0 ^{\pi}  
f^{(0)} ( \mathrm{p} ) \, 
\begin{pmatrix}
\pi \, \mathrm{sin} ^2\vartheta 
&
0
&
0
\\
0
&
\pi \, \mathrm{sin}^2 \vartheta  
&
0
\\
0
&
0
&
2 \, \pi \, \mathrm{cos} ^2 \vartheta 
\end{pmatrix}
\, 
\dfrac{
\mathrm{sin} \, \vartheta \, \mathrm{p}^4 \, d \vartheta 
\, d \mathrm{p}
}{
\sqrt{m^2c^2+\mathrm{p}^2}
}
\]
\[
=
\pi \, c  \, \int _0^{\infty}  \int _{-1} ^1  
f^{(0)} ( \mathrm{p} ) \, 
\begin{pmatrix}
1-u^2 
&
0
&
0
\\
0
&
1 - u^2
&
0
\\
0
&
0
&
2 \, u^2 
\end{pmatrix}
\, 
\dfrac{
 \mathrm{p}^4 \, d u 
\, d \mathrm{p}
}{
\sqrt{m^2c^2+\mathrm{p}^2}
}
\]
\[
=
\pi \, c  \int _0^{\infty}    
f^{(0)} ( \mathrm{p} ) \, 
\dfrac{
 \mathrm{p}^4 \, d \mathrm{p}
}{
\sqrt{m^2c^2+\mathrm{p}^2}
}
\, 
\begin{pmatrix}
u-u^3/3 
&
0
&
0
\\
0
&
u - u^3/3
&
0
\\
0
&
0
&
2 \, u^3/3 
\end{pmatrix}
\Bigg| _{-1}^1
\]
\begin{equation}
=
\dfrac{4 \, \pi \, c }{3}
 \, \int _0^{\infty}    
f^{(0)} ( \mathrm{p} ) \, 
\dfrac{
 \mathrm{p}^4 \, d \mathrm{p}
}{
\sqrt{m^2c^2+\mathrm{p}^2}
}
\big( \delta _{ij} \big)
\end{equation}
This is the energy-momentum tensor of a perfect fluid,
with energy density
\begin{equation}
\epsilon ^{(0)}
= 4 \, \pi \, c \int _0^{\infty}   
f^{(0)} ( \mathrm{p} ) \, \sqrt{m^2c^2+ \mathrm{p}^2}
\, 
\mathrm{p}^2 \, d \mathrm{p}
\end{equation}
and pressure 
\begin{equation}
P ^{(0)}
= \dfrac{4 \, \pi \, c}{3} \int _0^{\infty}   
f^{(0)} ( \mathrm{p} ) 
\, 
\dfrac{
\mathrm{p}^4 \, d \mathrm{p}
}{
\sqrt{m^2c^2+ \mathrm{p}^2}
}
\, .
\end{equation}
For the Juettner distribution (\ref{eq:Juettner}), the 
integration over $\mathrm{p}$ can be carried out with
the substitution (\ref{eq:defw}):
\[
\epsilon ^{(0)}
= \dfrac{\nu ^{(0)} \, \beta}{m^3 c^2 K_2 (\beta )}
\int _0^{\infty}   
e^{- \beta \, \sqrt{1+ \mathrm{p}^2/(m^2c^2)}}
\sqrt{m^2 c^2 + \mathrm{p}^2} \, \mathrm{p}^2
d \mathrm{p} 
= \dfrac{\nu ^{(0)} \, m \, c^2 \, \beta}{K_2 (\beta )}
\int _0^{\infty}   
e^{- \beta \, \mathrm{cosh} \, w}
\mathrm{sinh} ^2 w \, \mathrm{cosh} ^2 w \, d w
\]
\begin{equation}
= \dfrac{\nu ^{(0)} \, m \, c^2 \, \beta}{K_2 (\beta )}
\int _0^{\infty}   
e^{- \beta \, \mathrm{cosh} \, w}
\Big( \mathrm{sinh} ^2 w + \mathrm{sinh} ^4 w \Big) d w
=
\nu ^{(0)} \, m \, c^2  \Big( \dfrac{K_1 ( \beta )}{K_2 (\beta )}
+ \dfrac{3}{ \beta } \Big) \, ,
\end{equation}
\begin{equation}
P^{(0)}
= 
\dfrac{\nu ^{(0)} \, \beta}{3 \, m^3 \, c^2 \, K_2 (\beta )}
\int _0^{\infty}   
e^{- \beta \, \sqrt{1+ \mathrm{p}^2/(m^2c^2)}}
\,
\dfrac{\mathrm{p}^4 d \mathrm{p}}{\sqrt{m^2 c^2 + \mathrm{p}^2}}
= \dfrac{\nu ^{(0)} \, m \, c^2 \, \beta}{3 \, K_2 (\beta )}
\int _0^{\infty}   
e^{- \beta \, \mathrm{cosh} \, w}
\, \mathrm{sinh} ^4 w \, d w \, ,
\end{equation}
and, with $\chi = \beta ^{-1}$,
\begin{equation}
\epsilon ^{(0)}
=
\nu ^{(0)} \, m \, c^2  \Big( 1 + 
\dfrac{3}{ 2} \, \chi + O (\chi ^2 ) \Big)
\, ,
\label{eq:density}
\end{equation}
\begin{equation}
P^{(0)} 
=
\nu ^{(0)} \, m \, c^2 \, \chi
\, .
\label{eq:pressure}
\end{equation}
Here we have used (\ref{eq:defKs}) and (\ref{eq:asyKs}).
Note that (\ref{eq:pressure}) is exact (as long as the electron fluid is 
collisionless), so the electron fluid satisfies the ideal-gas equation.
Also note that the pressure vanishes for $\chi \to  0$ which confirms 
the known fact that a cold plasma is pressure-less. More precisely, for 
$ \chi \to 0$ the electron
fluid is a dust whose energy density is just the mass density of
the electrons up to a factor of $c^2$.

If we neglect the $O ( \chi ^2 )$ terms in (\ref{eq:density}), 
we get an approximation for the energy-momentum tensor of the electron fluid 
that is valid for sufficiently low temperatures.
When expressing the number density $\nu ^{(0)}$ by the plasma frequency 
$\omega _p$, energy density and pressure of the electron fluid, respectively, read
\begin{equation}
    \epsilon ^{(0)} = \dfrac{m^2 \omega _p^2}{\mu _0 q^2}
     \Big( 1 + \dfrac{3}{2} \, \chi \Big) \, , \quad 
    P ^{(0)} = \dfrac{m^2 \omega _p^2}{\mu _0 q^2} \, \chi \, .
\label{eq:eos}
\end{equation}
So in this approximation the electron fluid is a perfect
fluid with the equation of state $P^{(0)} = K \, \epsilon ^{(0)}$, where
$K$ is determined by the temperature as $K = \chi / (1+3 \chi /2)$.

\section{Axially symmetric spacetime -- Alternative formulation of deflection angle}\label{app_C}

In this appendix, we relate our results to the deflection angle in a warm plasma
to the corresponding formula for light rays in an arbitrary dispersive and 
isotropic medium of Ref.~\cite{Bezdekova-2023}. We also expand the discussion 
about the validity of the deflection-angle formula in the ergosphere in a cold 
plasma provided in Ref.~\cite{Perlick-Tsupko-2024} from the Kerr case to an 
arbitrary axially symmetric and stationary spacetime.

This requires some adaptation of notation. In the present paper we found it
convenient to represent an axially symmetric and stationary spacetime in the tetrad form
of Eq.~(\ref{eq:axistat}). This allows us to write the inverse metric immediately 
without the need of calculating a determinant, it puts the conservation laws 
(\ref{eq:Con1}) and (\ref{eq:Con2}) into a convenient form and it makes it
manifest whether an expression is well-behaved inside an ergoregion. By contrast, 
in previous studies \citep[e.g.,][]{Bezdekova-2022,Bezdekova-2023} the metric was
written in the following form:
\begin{equation}\label{ax_metr_alt}
g_{\mu \nu} dx^{\mu} dx^{\nu} =-\tilde{A}c^2dt^2  + \tilde{B}dr^2+2\tilde{P}c\,dtd\varphi
+\tilde{D}d\vartheta^2+\tilde{C} d\varphi^2.
\end{equation}
Again, the metric coefficients $\tilde{A}$, $\tilde{B}$, $\tilde{C}$, $\tilde{D}$, and $\tilde{P}$ 
are general functions of $r$ and $\vartheta$. 
Here we have added tildes to distinguish from the notation of the present paper and we 
have restored factors of $c$.

For calculating the deflection angle we restrict to the equatorial plane, $\vartheta=\pi/2$, 
assuming that light rays stay in this plane if they start tangentially to it. The metric 
coefficients are then functions of $r$ only.

Assuming that the warm plasma is stationary, the only non-zero component of its four-velocity equals
\begin{equation}\label{def_Vt_stat}
V^t\partial_t=\sqrt{\dfrac{1}{\tilde{A}}}\partial_t,
\end{equation}
and hence
\begin{equation}
\omega=-\frac{1}{c}p_\mu V^{\mu}=-\sqrt{\dfrac{1}{c^2\tilde{A}}}p_t.
\end{equation}

The corresponding Hamiltonian with a stationary warm plasma thus reads
\begin{equation}\label{Hamiltonian_warm}
\mathcal{H}(x^\alpha,p_{\alpha})=\frac{1}{2}\left[\frac{p_{r}^2}{\tilde{B}}+ \frac{p_{\varphi}^2c^2\tilde{A}-p_t^2\tilde{C}+2p_tp_{\varphi}c\tilde{P}}{c^2\tilde{A}\tilde{C}+c^2\tilde{P}^2}+\frac{\omega_p^2p_t^2}{p_t^2+\omega_p^2c^2\tilde{A}\chi}\left(1-\frac{3}{2}\chi\right)\right].
\end{equation}

The deflection angle can then easily be derived from a general formula presented in \cite{Bezdekova-2023}, see Eqs. (17) with (14) there, for a stationary dispersive isotropic medium in a spacetime given by the metric (\ref{ax_metr_alt}), leading to
\begin{equation}\label{def_angl_stat}
\alpha= 2\int_{R}^{\infty}\sqrt{\frac{\tilde{A}(r)\tilde{B}(r)}{\tilde{A}(r)\tilde{C}(r)+\tilde{P}^2(r)}}\left(\frac{h^2(r)}{\left(\frac{\tilde{P}(R)}{c\tilde{A}(R)}-\frac{\tilde{P}(r)}{c\tilde{A}(r)}\pm h(R)\right)^2}-1\right)^{-1/2}dr-\pi,
\end{equation}
where
\begin{gather}\label{h_ax_warm}
h^2(r)=\frac{\tilde{A}(r)\tilde{C}(r)+\tilde{P}^2(r)}{c^2\tilde{A}^2(r)}\left(1-
\frac{\omega_p^2c^2\tilde{A}(r)}{p_t^2+\omega_p^2c^2\tilde{A}(r)\chi}\left(1-\frac{3}{2}\chi\right)\right).
\end{gather}
As formula (\ref{def_Vt_stat}) cannot hold in the ergoregion, also the deflection-angle formula (\ref{def_angl_stat}) in the warm-plasma approximation holds only outside the ergoregion. Hence, we have to notice the following differences between this formula and the formula that follows from integrating (\ref{def_angl_axis}) over the ray trajectory: The former assumes a stationary warm plasma and it is valid only outside the ergoregion, the latter assumes a co-rotating warm plasma and it is valid also inside the ergoregion.

In the case of a cold plasma, $\chi = 0$, where (\ref{h_ax_warm}) corresponds to the formula (16) derived 
in \cite{Bezdekova-2023}, the difference vanishes and the two representations of the deflection angle become
equivalent. The reason is that then the Hamiltonian and, thus, the deflection angle becomes independent of 
the velocity of the medium, i.e., it is irrelevant if we assume (\ref{def_Vt_stat}) or some other equation
for $V^{\mu}$. Note that the constant of motion $\omega _0 = - p_t /c$ is well defined for all rays, inside
and outside the ergoregion; it is just the interpretation of $\omega _0 \big( \tilde{A}(r)) \big)^{-1/2}$ 
as the frequency measured by a stationary observer which makes sense only outside the ergoregion. Note that 
it is true that even for $\chi = 0$ the expression (\ref{def_angl_stat}) together with (\ref{h_ax_warm}) involves
the square-root of $\tilde{A}(r)$, so it is not obvious that this formula gives real results inside the
ergoregion. However, this can be made manifest in the following way.
We
rearrange the deflection angle formula by introducing a new function
\begin{equation}\label{def_gpm}
   g_\pm(R)=\frac{\tilde{P}(R)}{c\tilde{A}(R)}\pm h(R),
\end{equation}
with the function $h(R)$ from (\ref{h_ax_warm}) specified for $\chi = 0$, i.e.,  
\begin{gather}
h(R)=\left[\frac{\tilde{A}(R)\tilde{C}(R)+\tilde{P}^2(R)}{c^2\tilde{A}^2(R)}\left(1-
\frac{\omega_p^2c^2}{p_t^2}\tilde{A}(R)\right)\right]^{1/2}.
\end{gather}
Having this relation at hand, the deflection-angle formula (\ref{def_angl_stat}) can be restructured into the form
\begin{equation}
\alpha= 2\int_{R}^{\infty}\sqrt{\frac{\tilde{B}(r)}{\tilde{A}(r)\tilde{C}(r)+\tilde{P}^2(r)}}
\frac{c\tilde{A}(r)\,g_\pm(R)-\tilde{P}(r)}{\sqrt{\tilde{C}(r)-(\tilde{A}(r)\tilde{C}(r)+\tilde{P}^2(r))
\frac{\omega_p^2c^2}{p_t^2}+cg_\pm(R)\left(2\tilde{P}(r)-c\tilde{A}(r)\,g_\pm(R)\right)}
}dr-\pi.
\end{equation}
Note that both $\tilde{B}(r)>0$ and $\tilde{A}(r)\tilde{C}(r)+\tilde{P}^2(r)>0$.   This is a generalization of the formula presented in \cite{Perlick-Tsupko-2024} for the Kerr metric and a cold plasma, where the problem of rays coming to infinity from the ergosphere was discussed in more detail.
 
\end{widetext}

\bibliography{bibliography}

@PREAMBLE{
 "\providecommand{\noopsort}[1]{}" 
 # "\providecommand{\singleletter}[1]{#1}%" 
}

@ARTICLE{Bezdekova-2022,
       author = {{Bezd{\v{e}}kov{\'a}}, Barbora and {Perlick}, Volker and {Bi{\v{c}}{\'a}k}, Ji{\v{r}}{\'\i}},
        title = "{Light propagation in a plasma on an axially symmetric and stationary spacetime: Separability of the Hamilton-Jacobi equation and shadow}",
      journal = {J. Math. Phys.},
         year = 2022,
        month = sep,
       volume = {63},
       number = {9},
          eid = {092501},
        pages = {092501},
          doi = {10.1063/5.0106433},
       adsurl = {https://ui.adsabs.harvard.edu/abs/2022JMP....63i2501B},
}

@article{Bezdekova-2023,
  author    = {Bezd\v{e}kov\'{a}, Barbora and Bi\v{c}\'{a}k, Ji\v{r}\'{i}},
  title     = {Light deflection in plasma in the {H}artle-{T}horne metric and in other axisymmetric spacetimes with a quadrupole moment},
  journal   = {Phys. Rev. D},
  volume    = {108},
  number    = {8},
  pages     = {084043},
  year      = {2023},
  publisher = {APS},
    doi = {10.1103/PhysRevD.108.084043},
}

@article{bezdekova2024,
       author = {{Bezd{\v{e}}kov{\'a}}, Barbora and {Tsupko}, Oleg Yu and {Pfeifer}, Christian},
        title = "{Deflection of light rays in a moving medium around a spherically symmetric gravitating object}",
      journal = {Phys. Rev. D},
       volume = {109},
       number = {12},
        pages = {124024},
         year = {2024},
    publisher = {APS},
          doi = {10.1103/PhysRevD.109.124024}
}

@article{pfeifer2025,
       author = {Pfeifer, Christian and Bezd{\v{e}}kov{\'a}, Barbora and Tsupko, Oleg Yu},
        title = {Light deflection in axially symmetric stationary spacetimes filled with a moving medium},
      journal = {Phys. Rev. D},
         year = {2025},
       volume = {112},
        issue = {10},
        pages = {104064},
     numpages = {21},
        month = {Nov},
    publisher = {American Physical Society},
          doi = {10.1103/n7jc-x388},
          url = {https://link.aps.org/doi/10.1103/n7jc-x388}
}

@Inbook{Bittencourt2004,
       author = "Bittencourt, J. A.",
        title = "Waves in Hot Magnetized Plasmas",
    bookTitle = "Fundamentals of Plasma Physics",
         year = "2004",
    publisher = "Springer",
      address = "New York, NY",
        pages = "515--559",
         isbn = "978-1-4757-4030-1",
          doi = "10.1007/978-1-4757-4030-1_19",
          url = "https://doi.org/10.1007/978-1-4757-4030-1_19"
}

@article{genzel2010,
       author = {Genzel, Reinhard and Eisenhauer, Frank and Gillessen, Stefan},
        title = {The Galactic Center massive black hole and nuclear star cluster},
      journal = {Rev. Mod. Phys.},
       volume = {82},
       number = {4},
        pages = {3121--3195},
         year = {2010},
    publisher = {APS}
}

@article{bagenal1992,
       author = {Bagenal, Fran},
        title = {Giant planet magnetospheres},
      journal = {Ann. Rev. Earth Planet. Sci.},
       volume = {20},
        pages = {289--328},
         year = {1992}
}

@article{czechowski1995,
       author = {Czechowski, A and Grzedzielski, S},
        title = {A cold plasma layer at the heliopause},
      journal = {Adv. Space Res.},
       volume = {16},
       number = {9},
        pages = {321--325},
         year = {1995},
    publisher = {Elsevier}
}

@article{freeman1985,
       author = {Freeman, John W and Lopez, Ramon E},
        title = {The cold solar wind},
      journal = {J. Geophys. Res.: Space Physics},
       volume = {90},
       number = {A10},
        pages = {9885--9887},
         year = {1985},
    publisher = {Wiley Online Library}
}

@ARTICLE{BK-Tsupko-2009,
       author = {{Bisnovatyi-Kogan}, G.~S. and {Tsupko}, O. Yu.},
        title = "{Gravitational radiospectrometer}",
      journal = {Gravitation and Cosmology},
         year = 2009,
        month = mar,
       volume = {15},
       number = {1},
        pages = {20-27},
          doi = {10.1134/S020228930901006X},
       adsurl = {https://ui.adsabs.harvard.edu/abs/2009GrCo...15...20B},
}

@ARTICLE{BK-Tsupko-2010,
       author = {{Bisnovatyi-Kogan}, G.~S. and {Tsupko}, O. Yu.},
        title = "{Gravitational lensing in a non-uniform plasma}",
      journal = {Mon. Not. Roy. Astron. Soc.},
         year = 2010,
        month = jun,
       volume = {404},
       number = {4},
        pages = {1790-1800},
          doi = {10.1111/j.1365-2966.2010.16290.x},
       adsurl = {https://ui.adsabs.harvard.edu/abs/2010MNRAS.404.1790B},
}

@article{clemmow56,
       author = {{Clemmow}, P. C. and {Willson}, A. J.},
        title = {The dispersion equation in plasma oscillations},
      journal = {Proc. Roy. Soc. London. Ser. A. Math. Phys. Sci.},
       volume = {237},
       number = {1208},
        pages = {117--131},
         year = {1956},
    publisher = {The Royal Society London}
}

@ARTICLE{Er-Mao-2014,
       author = {{Er}, Xinzhong and {Mao}, Shude},
        title = "{Effects of plasma on gravitational lensing}",
      journal = {Mon. Not. Roy. Astron. Soc.},
         year = 2014,
        month = jan,
       volume = {437},
       number = {3},
        pages = {2180-2186},
          doi = {10.1093/mnras/stt2043},
       adsurl = {https://ui.adsabs.harvard.edu/abs/2014MNRAS.437.2180E},
}

@article{imre62,
       author = {{Imre}, K.},
        title = {Oscillations in a Relativistic Plasma},
      journal = {Phys. Fluids},
       volume = {5},
       number = {4},
        pages = {459--466},
         year = {1962}
}

@ARTICLE{Muhleman-1970,
       author = {{Muhleman}, D.~O. and {Ekers}, R.~D. and {Fomalont}, E.~B.},
        title = "{Radio interferometric test of the general relativistic light bending near the sun}",
      journal = {\prl},
         year = 1970,
        month = jun,
       volume = {24},
       number = {24},
        pages = {1377-1380},
          doi = {10.1103/PhysRevLett.24.1377},
       adsurl = {https://ui.adsabs.harvard.edu/abs/1970PhRvL..24.1377M},
}

@ARTICLE{Muhleman-1966,
       author = {{Muhleman}, D.~O. and {Johnston}, I.~D.},
        title = "{Radio propagation in the solar gravitational field}",
      journal = {\prl},
         year = 1966,
        month = aug,
       volume = {17},
       number = {8},
        pages = {455-458},
          doi = {10.1103/PhysRevLett.17.455},
       adsurl = {https://ui.adsabs.harvard.edu/abs/1966PhRvL..17..455M},
}

@BOOK{Perlick-2000,
       author = {{Perlick}, Volker},
        title = "{Ray Optics, Fermat's Principle, and Applications to General Relativity}",
         year = 2000,
        address = "Berlin-Heidelberg-New York",
        publisher={Springer},
        doi = {10.1007/3-540-46662-2},
       adsurl = {https://ui.adsabs.harvard.edu/abs/2000rofp.book.....P},
}

@ARTICLE{Perlick-Tsupko-BK-2015,
       author = {{Perlick}, Volker and {Tsupko}, Oleg Yu. and {Bisnovatyi-Kogan}, Gennady S.},
        title = "{Influence of a plasma on the shadow of a spherically symmetric black hole}",
         journal = {\prd},
         year = 2015,
        month = nov,
       volume = {92},
       number = {10},
          eid = {104031},
        pages = {104031},
          doi = {10.1103/PhysRevD.92.104031},
       adsurl = {https://ui.adsabs.harvard.edu/abs/2015PhRvD..92j4031P},
}

@ARTICLE{Perlick-Tsupko-2017,
       author = {{Perlick}, Volker and {Tsupko}, Oleg Yu.},
        title = "{Light propagation in a plasma on Kerr spacetime: Separation of the Hamilton-Jacobi equation and calculation of the shadow}",
      journal = {\prd},
         year = 2017,
        month = may,
       volume = {95},
       number = {10},
          eid = {104003},
        pages = {104003},
          doi = {10.1103/PhysRevD.95.104003},
       adsurl = {https://ui.adsabs.harvard.edu/abs/2017PhRvD..95j4003P},
}

@ARTICLE{Perlick-Tsupko-2022,
       author = {{Perlick}, Volker and {Tsupko}, Oleg Yu.},
        title = "{Calculating black hole shadows: Review of analytical studies}",
      journal = {Phys. Rep.},
         year = 2022,
        month = feb,
       volume = {947},
        pages = {1-39},
          doi = {10.1016/j.physrep.2021.10.004},
       adsurl = {https://ui.adsabs.harvard.edu/abs/2022PhR...947....1P},
}

@article{Perlick-Tsupko-2024,
  author = {{Perlick}, Volker and {Tsupko}, Oleg Yu.},
  title = "{Light propagation in a plasma on Kerr spacetime. II. Plasma imprint on photon orbits}",
  journal = {Phys. Rev. D},
  volume = {109},
  issue = {6},
  pages = {064063},
  numpages = {26},
  year = {2024},
  month = {Mar},
  publisher = {American Physical Society},
  doi = {10.1103/PhysRevD.109.064063},
  url = {https://link.aps.org/doi/10.1103/PhysRevD.109.064063}
}

@ARTICLE{Sareny-2019,
       author = {{S{\'a}ren{\'y}}, Matej and {Balek}, Vladim{\'\i}r},
        title = "{Effect of black hole-plasma system on light beams}",
      journal = {Gen. Rel. Grav.},
         year = 2019,
        month = nov,
       volume = {51},
       number = {11},
          eid = {141},
        pages = {141},
          doi = {10.1007/s10714-019-2629-8},
       adsurl = {https://ui.adsabs.harvard.edu/abs/2019GReGr..51..141S},
}

@ARTICLE{Schulze-Koops-Perlick-Schwarz-2017,
       author = {{Schulze-Koops}, Karen and {Perlick}, Volker and {Schwarz}, Dominik J.},
        title = "{Sachs equations for light bundles in a cold plasma}",
      journal = {Class. Quantum Grav.},
         year = 2017,
        month = nov,
       volume = {34},
       number = {21},
          eid = {215006},
        pages = {215006},
          doi = {10.1088/1361-6382/aa8d46},
       adsurl = {https://ui.adsabs.harvard.edu/abs/2017CQGra..34u5006S},
}

@BOOK{Synge-1960,
       author = {{Synge}, J.~L.},
        title = "{Relativity: The General Theory}",
         year = 1960,
        address = "Amsterdam",
        publisher={North-Holland Publishing Company},
       adsurl = {https://ui.adsabs.harvard.edu/abs/1960rgt..book.....S},
}

@article{Tsupko-2021,
  title = {Deflection of light rays by a spherically symmetric black hole in a dispersive medium},
  author = {Tsupko, Oleg Yu.},
  journal = {Phys. Rev. D},
  volume = {103},
  issue = {10},
  pages = {104019},
  numpages = {11},
  year = {2021},
  month = {May},
  publisher = {American Physical Society},
  doi = {10.1103/PhysRevD.103.104019},
  url = {https://link.aps.org/doi/10.1103/PhysRevD.103.104019}
}

@ARTICLE{Tsupko-BK-2013,
       author = {{Tsupko}, Oleg Yu. and {Bisnovatyi-Kogan}, Gennady S.},
        title = "{Gravitational lensing in plasma: Relativistic images at homogeneous plasma}",
      journal = {\prd},
         year = 2013,
        month = jun,
       volume = {87},
       number = {12},
          eid = {124009},
        pages = {124009},
          doi = {10.1103/PhysRevD.87.124009},
       adsurl = {https://ui.adsabs.harvard.edu/abs/2013PhRvD..87l4009T},
}

@ARTICLE{Tsupko-BK-2020-microlensing,
       author = {{Tsupko}, Oleg Yu and {Bisnovatyi-Kogan}, Gennady S.},
        title = "{Hills and holes in the microlensing light curve due to plasma environment around gravitational lens}",
      journal = {Mon. Not. Roy. Astron. Soc.},
         year = 2020,
        month = feb,
       volume = {491},
       number = {4},
        pages = {5636-5649},
          doi = {10.1093/mnras/stz3365},
       adsurl = {https://ui.adsabs.harvard.edu/abs/2020MNRAS.491.5636T},
}

@ARTICLE{Rogers2015,
       author = {{Rogers}, Adam},
        title = "{Frequency-dependent effects of gravitational lensing within plasma}",
      journal = {Mon. Not. Roy. Astron. Soc.},
         year = 2015,
       volume = {451},
       number = {1},
        pages = {4536–4544},
}

@article{Michel1972,
    author = "Michel, F. Curtis",
    title = "{Accretion of matter by condensed objects}",
    journal = "Astrophys. Space Sci.",
    volume={15},
    pages = "153--160",
    year = "1972"
}

@article{AkiyamaEtAl2021,
    author = "Akiyama et al., K. ",
    title = "{First M87 Event Horizon Telescope results. 
              VIII. Magnetic field structure near the event horizon}",
    doi = "10.3847/2041-8213/abe4de",
    journal = "Astrophys. J.",
    volume={910},
    pages = "L13",
    year = "2021"
}

@article{LanzuisiEtAl2019,
    author = "Lanzuisi et al., G.",
    title = "{NuSTAR measurement of coronal temperature in two luminous,
               high-redshift quasars}",
    doi = "10.3847/2041-8213/ab15dc",
    journal = "Astrophys. J.",
    volume={875},
    pages = "L20",
    year = "2019"
}

@article{JiangEtAl2019,
   author    = "Jiang et al., J.",
   title     = "High density reflection spectroscopy – {II}. {T}he density of the inner 
                black hole accretion disc in {AGN}",
   journal   = "Mon. Not. Roy. Astron. Soc.",
   volume    = "489",
   year      = "2019",
   pages     = "3436",
}

@article{MallickEtAl2022,
   author    = "Mallick et al., L.",
   title     = "High-density disc reflection spectroscopy of low-mass active galactic nuclei",
   journal   = "Mon. Not. Roy. Astron. Soc.",
   volume    = "513",
   year      = "2022",
   pages     = "4361",
}

@article{IssaounEtAl2019,
   author    = "Issaoun et al., S.",
   title     = "The size, shape, and scattering of {S}agittarius {$A^*$} at 86 {GHz}: 
                {F}irst {VLBI} with {ALMA}",
   journal   = "Astrophys J.",
   volume    = "871",
   year      = "2019",
   pages     = "30",
}

\end{document}